\newcommand*{\ATLASLATEXPATH}{}
\pdfoutput=1
\documentclass[UKenglish, texlive=2016, cernpreprint]{\ATLASLATEXPATH atlasdoc}

\usepackage[subfigure,subcaption]{\ATLASLATEXPATH atlaspackage}

\usepackage{\ATLASLATEXPATH atlasbiblatex}

\usepackage{\ATLASLATEXPATH atlasphysics}

\usepackage{rotating}
\usepackage{multirow}

\DeclareOldFontCommand{\bf}{\normalfont\bfseries}{\mathbf}
\DeclareOldFontCommand{\rm}{\normalfont\rmfamily}{\mathrm}
\DeclareOldFontCommand{\sf}{\normalfont\sffamily}{\mathsf}
\DeclareOldFontCommand{\tt}{\normalfont\ttfamily}{\mathtt}
\DeclareOldFontCommand{\bf}{\normalfont\bfseries}{\mathbf}
\DeclareOldFontCommand{\it}{\normalfont\itshape}{\mathit}
\DeclareOldFontCommand{\sl}{\normalfont\slshape}{\@nomath\sl}
\DeclareOldFontCommand{\sc}{\normalfont\scshape}{\@nomath\sc}

\usepackage{TOPQ_2017_03-defs}

 \AtlasTitle{Measurement of the top quark mass in the $t\bar{t}\to$ lepton+jets
 channel from $\sqrt{s}=8$ TeV ATLAS data and combination with previous results}
 \author{The ATLAS Collaboration\\}
 \AtlasRefCode{TOPQ-2017-03}

 \PreprintIdNumber{CERN-EP-2018-238}
 \AtlasDate{\today}
 \AtlasJournalRef{Eur. Phys. J. C 79 (2019) 290}
 \AtlasDOI{10.1140/epjc/s10052-019-6757-9}
\AtlasAbstract{%
 The top quark mass is measured using a template method in the \ttbarlj\ channel
 (lepton is $e$ or $\mu$) using ATLAS data recorded in 2012 at the LHC.
 The data were taken at a proton--proton centre-of-mass energy of
 $\sqrts=8$~\TeV\ and correspond to an integrated luminosity of
 \atlumo~\ifb.
 The \ttbarlj\ channel is characterized by the presence of a charged lepton, a
 neutrino and four jets, two of which originate from bottom quarks~($b$).
 Exploiting a three-dimensional template technique, the top quark mass is
 determined together with a global jet energy scale factor and a relative
 $b$-to-light-jet energy scale factor.
 The mass of the top quark is measured to be
 $\mt=\XZ{\EightLpJVal}{\EightLpJSta}{\EightLpJSys}$~\GeV.
 A combination with previous ATLAS \mt\ measurements gives
 $\mt= \XZ{\CombVal}{\CombSta}{\CombSys}$~\GeV.
}

\hypersetup{pdftitle={ATLAS document},pdfauthor={The ATLAS Collaboration}}

\begin{document}
\maketitle
\clearpage
%
\section{Introduction}
\label{sect:intro}
 The mass of the top quark \mt is an important parameter of the Standard
 Model~(SM).
 Precise measurements of \mt\ provide crucial information for global fits of
 electroweak parameters~\cite{LEPewkfits, Baak2014,PDG2014} which help to assess
 the internal consistency of the SM and probe its extensions. In addition, the
 value of \mt\ affects the stability of the SM Higgs potential, which has
 cosmological implications~\cite{Degrassi2012, HiggsStab1, HiggsStab2}.
 
 Many measurements of \mt\ in each \ttbar\ decay channel were performed by
 the Tevatron and LHC collaborations.
 The most precise measurements per experiment in the \ttbarlj\ channel are
 $\mt=\mtljcdf$~\GeV\ by CDF~\cite{CDF-1207},
 $\mt=\mtljdnu$~\GeV\ by D0~\cite{D00-1401},
 $\mt=\mtljatlold$~\GeV\ by ATLAS~\cite{TOPQ-2013-02} and
 $\mt=\mtljcms$~\GeV\ by CMS~\cite{CMS-1401}.
 Combinations are performed, by either the individual experiments, or by several
 Tevatron and LHC experiments~\cite{ATLAS:2014wva}.
 In these combinations, selections of measurements from all \ttbar\ decay
 channels are used.
 The latest combinations per experiment are
 $\mt=\mtcdf$~\GeV\ by CDF~\cite{CDF-1402},
 $\mt=\mtdnu$~\GeV\ by D0~\cite{Abazov:2017ktz},
 $\mt=\mtatlold$~\GeV\ by ATLAS~\cite{TOPQ-2016-03} and
 $\mt=\mtcms$~\GeV\ by CMS~\cite{CMS-1401}.

 In this paper, an ATLAS measurement of \mt\ in the \ttbarlj\ channel is
 presented. The result is obtained from \pp\ collision data recorded in 2012 at
 a \cme\ of $\sqrts=8$~\TeV\ with an integrated luminosity of about
 $\atlumo$~\invfb.
 The analysis exploits the decay $\ttbar\to\Wplus\Wminus\bbbar\to\ell\nu
 q\bar{q}^\prime\bbbar$, which occurs when one \Wboson\ boson decays into a
 charged lepton~($\ell$ is $e$ or $\mu$ including $\tau\to e,\mu$ decays) and a
 neutrino~($\nu$), and the other into a pair of quarks.
 In the analysis presented here, \mt\ is obtained from the combined sample of
 events selected in the electron+jets and muon+jets final states.
 Single-top-quark events with the same reconstructed final states contain
 information about the top quark mass and are therefore included as signal
 events.

 The measurement uses a template method, where simulated distributions are
 constructed for a chosen quantity sensitive to the physics parameter under
 study using a number of discrete values of that parameter.
 These templates are fitted to functions that interpolate between different
 input values of the physics parameter while fixing all other parameters of the
 functions.
 In the final step, an unbinned likelihood fit to the observed data distribution
 is used to obtain the value of the physics parameter that best describes the
 data.
 In this procedure, the experimental distributions are constructed such that
 fits to them yield unbiased estimators of the physics parameter used as input
 in the signal Monte Carlo~(MC) samples. Consequently, the top quark mass
 determined in this way corresponds to the mass definition used in the MC
 simulation.
 Because of various steps in the event simulation, the mass measured in this way
 does not necessarily directly coincide with mass definitions within a given
 renormalization scheme, e.g.~the top quark pole mass.
 Evaluating these differences is a topic of theoretical
 investigations~\cite{Buckley:2011ms,MITP2014,Butenschoen:2016lpz,HOA-1801,RAV-1801}.

 The measurement exploits the three-dimensional template fit technique presented
 in \Ref{\cite{TOPQ-2013-02}}.
 To reduce the uncertainty in \mt\ stemming from the uncertainties in the jet
 energy scale~(\JES) and the additional \bjet\ energy scale~(\bJES), \mt\ is
 measured together with the jet energy scale factor~(\JSF) and the relative
 $b$-to-light-jet energy scale factor~(\bJSF).
 Given the larger data sample than used in \Ref{\cite{TOPQ-2013-02}}, the
 analysis is optimized to reject combinatorial background arising from incorrect
 matching of the observed jets to the daughters arising from the top quark
 decays, thereby achieving a better balance of the statistical and systematic
 uncertainties and reducing the total uncertainty.
 Given this new measurement, an update of the ATLAS combination of
 \mt\ measurements is also presented.

 This document is organized as follows.  After a short description of the ATLAS
 detector in \Sect{\ref{sect:atlas}}, the data and simulation samples are
 discussed in \Sect{\ref{sect:mc}}. Details of the event selection are given in
 \Sect{\ref{sect:evsel}}, followed by the description of the reconstruction of
 the three observables used in the template fit in \Sect{\ref{sect:evreco}}.
 The optimization of the event selection using a multivariate analysis approach
 is presented in \Sect{\ref{sect:bdtsel}}.
 The template fits are introduced in \Sect{\ref{sect:templates}}.
 The evaluation of the systematic uncertainties and their statistical
 uncertainties are discussed in \Sect{\ref{sect:unc}}, and the measurement of
 \mt\ is given in \Sect{\ref{sect:result}}.
 The combination of this measurement with previous ATLAS results is discussed in
 \Sect{\ref{sect:comb}} and compared with measurements of other experiments.
 The summary and conclusions are given in \Sect{\ref{sect:conclusion}}.
 Additional information about the optimization of the event selection and on
 specific uncertainties in the new measurement of \mt\ in the \ttbarlj\ channel
 are given in \App{\ref{sect:addlpj}}, while \App{\ref{sect:addcom}} contains
 information about various combinations performed, together with comparisons
 with results from other experiments.

\newcommand{\AtlasCoordFootnote}{
 ATLAS uses a right-handed coordinate system with its origin at the nominal
 interaction point (IP) in the centre of the detector and the $z$-axis along the
 beam pipe.  The $x$-axis points from the IP to the centre of the LHC ring, and
 the $y$-axis points upwards.  Cylindrical coordinates $(r,\phi)$ are used in
 the transverse plane, $\phi$ being the azimuthal angle around the $z$-axis.
 The pseudorapidity is defined in terms of the polar angle $\theta$ as $\eta =
 -\ln \tan(\theta/2)$.  Angular distance is measured in units of $\Delta R
 \equiv \sqrt{(\Delta\eta)^{2} + (\Delta\phi)^{2}}$.}

\section{The ATLAS experiment}
\label{sect:atlas}
 The ATLAS experiment~\cite{PERF-2007-01} at the LHC is a multipurpose particle
 detector with a forward--backward symmetric cylindrical geometry and a near
 $4\pi$ coverage in the solid angle.\footnote{\AtlasCoordFootnote} It consists
 of an inner tracking detector surrounded by a thin superconducting solenoid
 providing a \SI{2}{\tesla} axial magnetic field, electromagnetic and hadronic
 calorimeters, and a muon spectrometer.
 The inner tracking detector covers the pseudorapidity range $|\eta| < 2.5$.  It
 consists of silicon pixel, silicon microstrip, and transition radiation
 tracking detectors.
 Lead/liquid-argon (LAr) sampling calorimeters provide electromagnetic (EM)
 energy measurements with high granularity.
%
 A hadronic (steel/scintillator-tile) calorimeter covers the central
 pseudorapidity range ($|\eta| < 1.7$). The endcap and forward regions are
 instrumented with LAr calorimeters for both the EM and hadronic energy
 measurements up to $|\eta| = 4.9$.
 The muon spectrometer surrounds the calorimeters and is based on three large
 air-core toroid superconducting magnets with eight coils each. Its bending
 power is \num{2.0} to \SI{7.5}{\tesla\metre}. It includes a system of precision
 tracking chambers and fast detectors for triggering.

 A three-level trigger system was used to select events. The first-level trigger
 is implemented in hardware and used a subset of the detector information to
 reduce the accepted rate to at most \SI{75}{\kilo\hertz}. This is followed by
 two software-based trigger levels that together reduced the accepted event rate
 to \SI{400}{\hertz} on average depending on the data-taking conditions during
 2012.

\section{Data and simulation samples}
\label{sect:mc}
 The analysis is based on \pp\ collision data recorded by the ATLAS detector in
 2012 at a \cme\ of $\sqrts=8$~\TeV.
 The integrated luminosity is $\atlumo$~\invfb\ with an uncertainty of
 \atlumounc~\cite{DAPR-2013-01}.
 The modelling of top quark pair~(\ttbar) and single-top-quark signal events, as
 well as most background processes, relies on MC simulations.
 For the simulation of \ttbar\ and single-top-quark events, the
 \PowhegBox\ v1~\cite{NAS-0401,FRI-0701,Alioli:2010xd} program was used.
 Within this framework, the simulations of the \ttbar~\cite{FRI-0702} and
 single-top-quark production in the $s$- and $t$-channels~\cite{ALI-0901} and
 the $Wt$-channel~\cite{REE-1101} used matrix elements at next-to-leading
 order~(NLO) in the strong coupling constant \alphas with the
 NLO~CT10~\cite{LAI-1001} parton distribution function~(PDF) set and the
 \hdamp\ parameter\footnote{The \hdamp\ parameter controls the transverse
   momentum \pt\ of the first additional emission beyond the leading-order
   Feynman diagram in the parton shower and therefore regulates the
   high-\pt\ emission against which the \ttbar\ system recoils.} set to
 infinity.
 Using \mt\ and the top quark transverse momentum \pt\ for the underlying
 leading-order Feynman diagram, the dynamic factorization and renormalization
 scales were set to $\sqrt{\mt^2 + \pt^2}$.
 The \Pythia~(v6.425) program~\cite{SJO-0601} with the P2011C~\cite{Skands} set
 of tuned parameters~(tune) and the corresponding CTEQ6L1
 PDFs~\cite{Pumplin2002} provided the parton shower, hadronization and
 underlying-event modelling.

 For \mt\ hypothesis testing, the \ttbar\ and single-top-quark event samples
 were generated with five different assumed values of \mt\ in the range from
 $167.5$ to $177.5$~\GeV\ in steps of $2.5$~\GeV.
 The integrated luminosity of the simulated \ttbar\ sample with
 $\mt=172.5$~\GeV\ is about $360$~\invfb.
 Each of these MC samples is normalized according to the best available
 cross-section calculations. For $\mt=172.5$~\GeV, the \ttbar\ cross-section is
 $\sigma_{\ttbar}=253^{+13}_{-15}$~\pb, calculated at next-to-next-to-leading
 order~(NNLO) with next-to-next-to-leading logarithmic soft gluon
 terms~\cite{CAC-1101,PRL-109-132001,JHEP-1212,JHEP-1301,Czakon:2013goa} with
 the {\sc Top++}~2.0 program~\cite{CZA-1101}.
 The PDF- and \alphas-induced uncertainties in this cross-section were
 calculated using the PDF4LHC prescription~\cite{PDF4LHC} with the MSTW2008
 $68\%$ CL~NNLO PDF~\cite{MAR-0901,MAR-0902},
 CT10~NNLO~PDF~\cite{LAI-1001,CT10NNLO} and NNPDF2.3~5f~FFN
 PDF~\cite{Ball:2012cx} and were added in quadrature with the uncertainties
 obtained from the variation of the factorization and renormalization scales by
 factors of 0.5 and 2.0.
 The cross-sections for single-top-quark production were calculated at NLO and
 are $\sigma_\mathrm{t}=87.8\,^{+3.4}_{-1.9}$~\pb~\cite{Kidonakis:2011wy},
 $\sigma_{Wt}=22.4\,\pm\,1.5$~\pb~\cite{Kidonakis:2010ux} and
 $\sigma_\mathrm{s}=5.6\,\pm\,0.2$~\pb~\cite{Kidonakis:2010tc}
 in the $t$-, the $Wt$- and the $s$-channels, respectively.

 The \Alpgen~(v2.13) program~\cite{MAN-0301} interfaced to the
 \Pythiasix\ program was used for the simulation of the production of \Wbos\ or
 \Zboson\ bosons in association with jets.
 The CTEQ6L1 PDFs and the corresponding AUET2 tune~\cite{ATL-PHYS-PUB-2011-008}
 were used for the matrix element and parton shower settings.
 The \Wj\ and \Zj\ events containing heavy-flavour~(HF) quarks~($Wbb$+jets,
 $Zbb$+jets, $Wcc$+jets, $Zcc$+jets, and $Wc$+jets) were generated separately
 using leading-order (LO) matrix elements with massive bottom and \cquark{s}.
 Double-counting of HF quarks in the matrix element and the parton shower
 evolution was avoided via a HF overlap-removal procedure that used the
 \dR\ between the additional heavy quarks as the criterion. If the \dR\ was
 smaller than 0.4, the parton shower prediction was taken, while for larger
 values, the matrix element prediction was used.
 The \Zj\ sample is normalized to the inclusive NNLO
 calculation~\cite{Melnikov:2006kv}. Due to the large uncertainties in the
 overall \Wj\ normalization and the flavour composition, both are estimated
 using data-driven techniques as described in Section~\ref{sect:bkg}.
 Diboson production processes~($WW$, $WZ$ and $ZZ$) were simulated using the
 \Alpgen\ program with CTEQ6L1 PDFs interfaced to the
 \HERWIG~(v6.520)~\cite{Corcella:2000bw} and \Jimmy~(v4.31)~\cite{SAMPLES-JIMMY}
 programs. The samples are normalized to their predicted cross-sections at
 NLO~\cite{Campbell:1999ah}.

 All samples were simulated taking into account the effects of multiple soft
 \pp\ interactions~(pile-up) that are present in the 2012 data. These
 interactions were modelled by overlaying simulated hits from events with
 exactly one inelastic collision per bunch crossing with hits from minimum-bias
 events produced with the \Pythia~(v8.160) program~\cite{SJO-0801} using the A2
 tune~\cite{ATL-PHYS-PUB-2012-003} and the MSTW2008~LO~PDF.
 The number of additional interactions is Poisson-distributed around the mean
 number of inelastic \pp\ interactions per bunch crossing $\mu$. For a given
 simulated hard-scatter event, the value of $\mu$ depends on the instantaneous
 luminosity and the inelastic \pp\ cross-section, taken to be
 73~mb~\cite{DAPR-2013-01}.
 Finally, the simulation sample is reweighted such as to match the
 pile-up observed in data.

 A simulation~\cite{SOFT-2010-01} of the ATLAS detector response based on
 \Geantfour~\cite{AGO-0301} was performed on the MC events.
 This simulation is referred to as full simulation.
 The events were then processed through the same reconstruction software as the
 data.
 A number of samples used to assess systematic uncertainties were produced
 bypassing the highly computing-intensive full \Geantfour\ simulation of the
 calorimeters.
 They were produced with a faster version of the
 simulation~\cite{ATL-PHYS-PUB-2010-013}, which retained the full simulation of
 the tracking but used a parameterized calorimeter response based on resolution
 functions measured in full simulation samples.
 This simulation is referred to as fast simulation.

\section{Object reconstruction, background estimation and event preselection}
\label{sect:evsel}
 The reconstructed objects resulting from the top quark pair decay are electron
 and muon candidates, jets and missing transverse momentum~(\met).
 In the simulated events, corrections are applied to these objects based on
 detailed data-to-simulation comparisons for many different processes, so as to
 match their performance in data.
%
\subsection{Object reconstruction}
\label{sect:obsel}
 Electron candidates~\cite{PERF-2016-01} are required to have a transverse
 energy of $\et>25$~\GeV\ and a pseudorapidity of the corresponding EM cluster
 of $\absetaclus < 2.47$ with the transition region $1.37<\absetaclus<1.52$
 between the barrel and the endcap calorimeters excluded.
 Muon candidates~\cite{PERF-2014-05} are required to have transverse momentum
 $\pt>25$~\GeV\ and $\vert\eta\vert<2.5$.
 To reduce the contamination by leptons from HF decays inside jets or from
 photon conversions, referred to collectively as non-prompt~(NP) leptons, strict
 isolation criteria are applied to the amount of activity in the vicinity of the
 lepton candidate~\cite{Rehermann2011,PERF-2014-05,PERF-2016-01}.

 Jets are built from topological clusters of calorimeter
 cells~\cite{PERF-2014-07} with the \antikt\ jet clustering
 algorithm~\cite{Cacciari:2008gp} using a radius parameter of $R=0.4$.
 The clusters and jets are calibrated using the local cluster weighting~(LCW)
 and the global sequential calibration~(GSC) algorithms,
 respectively~\cite{PERF-2011-03,PERF-2012-01,ATLAS-CONF-2015-037}.
 The subtraction of the contributions from pile-up is performed via the jet area
 method~\cite{PERF-2014-03}.
 Jets are calibrated using an energy- and $\eta$-dependent simulation-based
 scheme with in situ corrections based on data~\cite{PERF-2012-01}.
 Jets originating from pile-up interactions are identified via their jet vertex
 fraction~(JVF), which is the \pt\ fraction of associated tracks stemming from
 the primary vertex.
 The requirement $\mathrm{JVF}>0.5$ is applied solely to jets with
 $\pt<50$~\GeV\ and $\abseta<2.4$~\cite{PERF-2014-03}.
 Finally, jets are required to satisfy $\pt>25$~\GeV\ and $\abseta<2.5$.

 Muons reconstructed within a $\dR=0.4$ cone around the axis of a jet with
 $\pt>25$~\GeV\ are excluded from the analysis.
 In addition, the closest jet within a $\dR=0.2$ cone around an electron
 candidate is removed, and then electrons within a $\dR=0.4$ cone around any of
 the remaining jets are discarded.

 The identification of jets containing reconstructed $b$-hadrons, called \btag,
 is used for event reconstruction and background suppression.
 In the following, irrespective of their origin, jets tagged by the
 \btag\ algorithm are referred to as \btagged\ jets, whereas those not tagged
 are referred to as untagged jets.
 Similarly, whether they are tagged or not, jets containing $b$-hadrons in
 simulation are referred to as \bjets\ and those containing only lighter-flavour
 hadrons from $u, d, c, s$-quarks, or originating from gluons, are collectively
 referred to as \ljets.
 The working point of the neural-network-based MV1
 \btag\ algorithm~\cite{PERF-2012-04} corresponds to an average
 \btag\ efficiency of 70$\%$ for \bjet{s}\ in simulated \ttbar\ events and
 rejection factors of 5 for jets containing a $c$-hadron and 140 for jets
 containing only lighter-flavour hadrons.
 To match the \btag\ performance in the data, \pt- and $\eta$-dependent scale
 factors, obtained from dijet and \ttbarll\ events, are applied to MC jets
 depending on their generated quark flavour, as described in
 Refs.~\cite{PERF-2012-04,ATLAS-CONF-2014-004,ATLAS-CONF-2014-046}.

 The missing transverse momentum \met\ is the absolute value of the vector
 $\metvec$ calculated from the negative vectorial sum of all transverse
 momenta. The vectorial sum takes into account all energy deposits in the
 calorimeters projected onto the transverse plane. The clusters are corrected
 using the calibrations that belong to the associated physics object. Muons are
 included in the calculation of the \met\ using their momentum reconstructed in
 the inner tracking detectors~\cite{PERF-2014-04}.
%
\subsection{Background estimation}
\label{sect:bkg}
 The contribution of events falsely reconstructed as \ttbarlj\ events due to the
 presence of objects misidentified as leptons~(fake leptons) and NP leptons
 originating from HF decays, is estimated from data using the
 matrix-method~\cite{ATLAS-CONF-2014-058}.
 The technique employed uses $\eta$- and \pt-dependent efficiencies for
 NP/fake-leptons and prompt-leptons.
 They are measured in a background-enhanced control region with low \met\ and
 from events with dilepton masses around the \Zboson\ boson
 peak~\cite{TOPQ-2010-01}, respectively.
 For the \Wj\ background, the overall normalization is estimated from data.
 The estimate is based on the charge-asymmetry method~\cite{TOPQ-2011-08},
 relying on the fact that at the LHC more \Wplus\ than \Wminus\ bosons are
 produced.
 In addition, a data-driven estimate of the $Wb\bar{b}$, $Wc\bar{c}$, $Wc$ and
 $W$+light-jet fractions is performed in events with exactly two jets and at
 least one $b$-tagged jet. Further details are given in
 Ref.~\cite{TOPQ-2012-18}.
 The \Zj\ and diboson background processes are normalized to their predicted
 cross-sections as described in Section~\ref{sect:mc}.
%
\subsection{Event preselection}
\label{sect:presel}
 Triggering of events is based solely on the presence of a single electron or
 muon, and no information from the hadronic final state is used.
 A logical {\it OR} of two triggers is used for each of the \ttbarej\ and
 \ttbarmj\ channels. The triggers with the lower thresholds of 24~\GeV\ for
 electrons or muons select isolated leptons. The triggers with the higher
 thresholds of 60~\GeV\ for electrons and 36~\GeV\ for muons do not include an
 isolation requirement.
 The further selection requirements closely follow those in
 \Ref{\cite{TOPQ-2013-02}} and are
\begin{itemize}
 \item Events are required to have at least one primary vertex with at least
   five associated tracks. Each track needs to have a minimum \pt\ of
   0.4~\GeV. For events with more than one primary vertex, the one with the
   largest $\sum p_{\rm T}^2$ is chosen as the vertex from the hard scattering.
 \item The event must contain exactly one reconstructed charged lepton, with
   $\ET> 25$~\GeV\ for electrons and $\pt > 25$~\GeV\ for muons, that matches
   the charged lepton that fired the corresponding lepton trigger.
 \item In the \ttbarmj\ channel, $\met>20$~\GeV\ and $\met+\mWt>60$~\GeV\ are
   required.\footnote{Here \mWt\ is the transverse mass of the \Wboson\ boson,
     defined as $\sqrt{2\,\ptlepton\,\met
       \left(1-\cos\phi(\ell,\metvec)\right)}$, where $\metvec$ provides an
     estimate of the transverse momentum of the neutrino.}
 \item In the \ttbarej\ channel, more stringent requirements on \met\ and
   \mWt\ are applied because of the higher level of NP/fake-lepton background.
   The requirements are $\met > 30$~\GeV\ and $\mWt>30$~\GeV.
 \item The presence of at least four jets with $\pt>25$~\GeV\ and
   $\vert\eta\vert<2.5$ is required.
 \item The presence of exactly two \btagged\ jets is required.
\end{itemize}
 The resulting event sample is statistically independent of the ones used for
 the measurement of \mt\ in the \ttbarll\ and \ttbarjj\ channels at
 $\sqrts=8$~\TeV~\cite{TOPQ-2016-03, TOPQ-2015-03}.
 The observed number of events in the data after this preselection and the
 expected numbers of signal and background events corresponding to the same
 integrated luminosity as the data are given in~\Tab{\ref{tab:selections}}.
 For all predictions, the uncertainties are estimated as the sum in quadrature
 of the statistical uncertainty, the uncertainty in the integrated luminosity
 and all systematic uncertainties assigned to the measurement of \mt\ listed in
 \Sect{\ref{sect:unc}}, except for the PDF and pile-up uncertainties, which are
 small.
 The normalization uncertainties listed below are included for the predictions
 shown in this section, but due to their small effect on the measured top quark
 mass they are not included in the final measurement.

 For the signal, the \xsecunc\ uncertainty in the \ttbar\ cross-section
 introduced in \Sect{\ref{sect:mc}} and a \stopxsecunc\ uncertainty in the
 single-top-quark cross-section are used. The latter uncertainty is obtained
 from the cross-section uncertainties given in \Sect{\ref{sect:mc}} and the
 fractions of the various single-top-quark production processes after the
 selection requirements.
 The background uncertainties contain uncertainties of \Xjunc\ in the
 normalization of the diboson and \Zj\ production processes. These uncertainties
 are calculated using Berends--Giele scaling~\cite{BERENDS199132}.
 Assuming a top quark mass of $\mt=172.5$~\GeV, the predicted number of events
 is consistent within uncertainties with the number observed in the data.
%
\begin{table}[tb!]
\caption{The observed numbers of events in data after the event preselection and
  the \mvabased\ selection~(see \protect\Sect{\ref{sect:bdtsel}}).
  In addition, the expected numbers of signal events for $\mt=172.5$~\GeV\ and
  background events corresponding to the same integrated luminosity as the data
  are given.
  The uncertainties in the predicted number of events take into account the
  statistical and systematic sources explained in \Sect{\ref{sect:presel}}.
 Two significant digits are used for the uncertainties in the predicted events.
\label{tab:selections}
}
\begin{center}
\small
\begin{tabular}{|l|r@{$\,\pm\,$}r|r@{$\,\pm\,$}r|}
\hline
Selection & \multicolumn{2}{c|}{Preselection}
          & \multicolumn{2}{c|}{\mvabased\ selection}\\ \hline
    Data  & \multicolumn{2}{c|}{96105} & \multicolumn{2}{c|}{38054} \\\hline
\ttbar\ signal                & 85000 & 10000 & 36100 &  5500 \\
Single-top-quark signal       &  4220 &   360 &   883 &    85 \\\hline
NP/fake leptons~(data-driven) &   700 &   700 &   9.2 &   9.2 \\
\Wj~(data-driven)             &  2800 &   700 &   300 &   100 \\
\Zj\                          &   430 &   230 &    58 &    33 \\
$WW/WZ/ZZ$                    &    63 &    32 &   7.0 &   5.2 \\\hline
Signal+background             & 93000 & 10000 & 37300 &  5500 \\\hline
Expected background fraction  & 0.043 & 0.012 & 0.010 & 0.003 \\
Data / (Signal+background)    &  1.03 & 0.12  & 1.02  & 0.15  \\
\hline
\end{tabular}
\end{center}
\end{table}
%
\clearpage
%
\section{Reconstruction of the three observables}
\label{sect:evreco}
 As in \Ref{\cite{TOPQ-2013-02}}, a full kinematic reconstruction of the event
 is done with a likelihood fit using the
 \KLFitter\ package~\cite{KLFitterNIM,TOPQ-2011-15}.  The \KLFitter\ algorithm
 relates the measured kinematics of the reconstructed objects to the
 leading-order representation of the \ttbar\ system decay using
 $\ttbar\to\ell\nu\blep\,\qone\qtwo\bhad$.
 In this procedure, the measured jets correspond to the quark decay products of
 the \Wboson\ boson, \qone\ and \qtwo, and to the \bquarks, \blep\ and \bhad,
 produced in the semi-leptonic and hadronic top quark decays, respectively.

 The event likelihood is the product of Breit--Wigner~(BW) distributions for the
 \Wboson\ bosons and top quarks and transfer functions~(TFs) for the energies of
 the reconstructed objects that are input to \KLFitter.
 The $W$ boson BW distributions use the world combined values of the $W$ boson
 mass and decay width from Ref.~\cite{PDG2014}.
 A common mass parameter \mtr\ is used for the BW distributions describing the
 semi-leptonically and hadronically decaying top quarks and is fitted
 event-by-event. The top quark width varies with \mtr\ according to the SM
 prediction~\cite{PDG2014}. The TFs are derived from the
 \PowhegPythia\ \ttbar\ signal MC simulation sample at an input mass of
 $\mt=172.5$~\GeV.
 They represent the experimental resolutions in terms of the probability that
 the observed energy at reconstruction level is produced by a given parton-level
 object for the leading-order decay topology and in the fit constrain the
 variations of the reconstructed objects.

 The input objects to the event likelihood are the reconstructed charged lepton,
 the missing transverse momentum and up to six jets.
 These are the two \btagged\ jets and the four untagged jets with the highest
 \pt.
 The $x$- and $y$-components of the missing transverse momentum are starting
 values for the neutrino transverse-momentum components, and its longitudinal
 component~$p_{\nu,z}$ is a free parameter in the kinematic likelihood fit. Its
 starting value is computed from the $W\to\ell\nu$ mass constraint. If there are
 no real solutions for $p_{\nu,z}$, a starting value of zero is used. If there
 are two real solutions, the one giving the largest likelihood value is taken.

 Maximizing the event-by-event likelihood as a function of \mtr\ establishes the
 best assignment of reconstructed jets to partons from the \ttbarlj\ decay. The
 maximization is performed by testing all possibilities for assigning
 \btagged\ jets to \bquark\ positions and untagged jets to light-quark
 positions. 
 With the above settings of the reconstruction algorithm, compared with
 the settings\footnote{
 In \Ref{\cite{TOPQ-2013-02}} only four input jets were used. In addition,
 \btag\ efficiencies and rejection factors were used to favour permutations for
 which a \btagged\ jet is assigned to a \bquark\ position and penalise those
 where a \btagged\ jet is assigned to a light-quark position. However, the
 latter permutations were still accepted whenever they resulted in the largest
 likelihood.}
 used in \Ref{\cite{TOPQ-2013-02}}, a larger fraction of correct assignments of
 reconstructed jets to partons from the \ttbarlj\ decay is achieved.
 The performance of the reconstruction algorithm is discussed in
 \Sect{\ref{sect:bdtsel}}.

 The value of \mtr\ obtained from the kinematic likelihood fit is used as the
 observable primarily sensitive to the underlying \mt.
 The invariant mass of the hadronically decaying \Wboson\ boson~\mWr, which is
 sensitive to the \JES, is calculated from the assigned jets of the chosen
 permutation.
 Finally, an observable called \rbqr, designed to be sensitive to the \bJES, is
 computed as the scalar sum of the transverse momenta of the two \btagged\ jets
 divided by the scalar sum of the transverse momenta of the two jets associated
 with the hadronic \Wboson\ boson decay:
%
\begin{linenomath*}
  \begin{align}
    \rbqr & = \frac{p_\mathrm{T}^{b_{\rm had}} + p_\mathrm{T}^{b_{\rm lep}}}
          {p_\mathrm{T}^{\qone}    + p_\mathrm{T}^{\qtwo}}\,.
    \nonumber
  \end{align}
\end{linenomath*}
%
 The values of \mWr\ and \rbqr\ are computed from the jet four-vectors as given
 by the jet reconstruction instead of using the values obtained in the kinematic
 likelihood fit. This ensures the maximum sensitivity to the jet calibration for
 \ljets\ and \bjets.

\begin{figure*}[tbp!]
\centering
\subfigure[Transverse momentum of the lepton]
          {\includegraphics[width=0.42\textwidth]{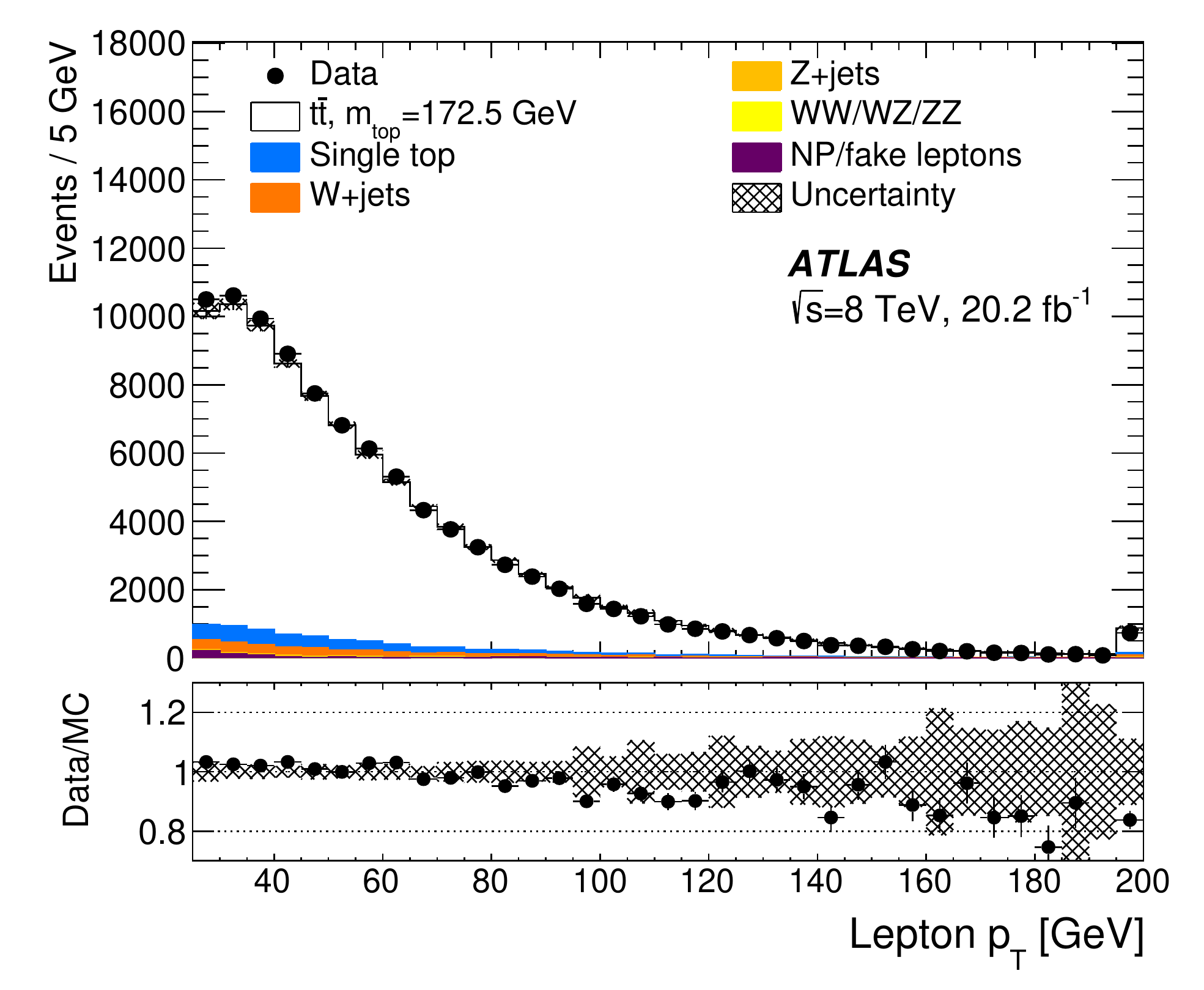}\label{fig:fig_01a}}
\subfigure[Average transverse momentum of jets]
          {\includegraphics[width=0.42\textwidth]{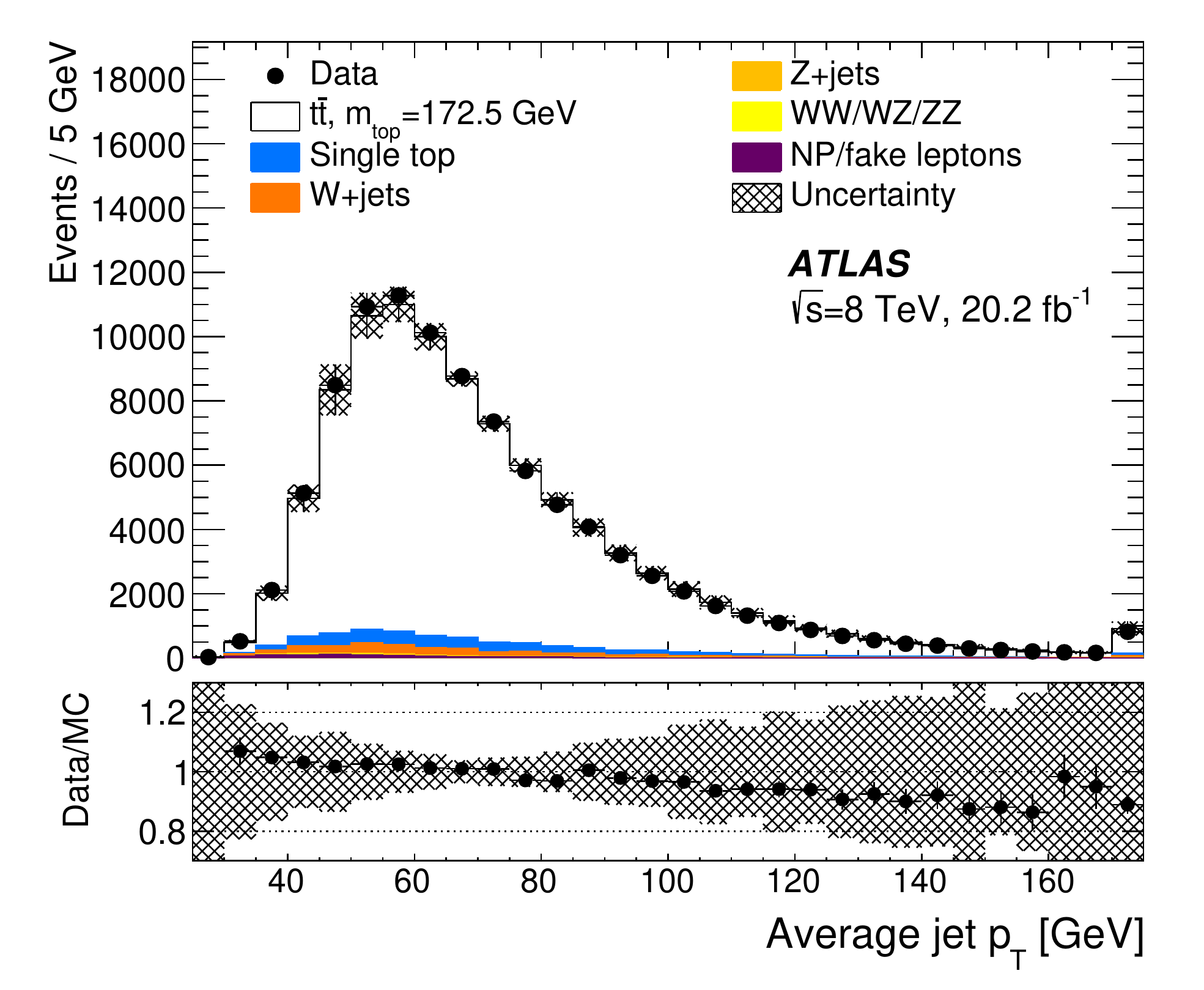}\label{fig:fig_01b}}
\hfill
\subfigure[Transverse momentum of the hadronically decaying top quark]
          {\includegraphics[width=0.42\textwidth]{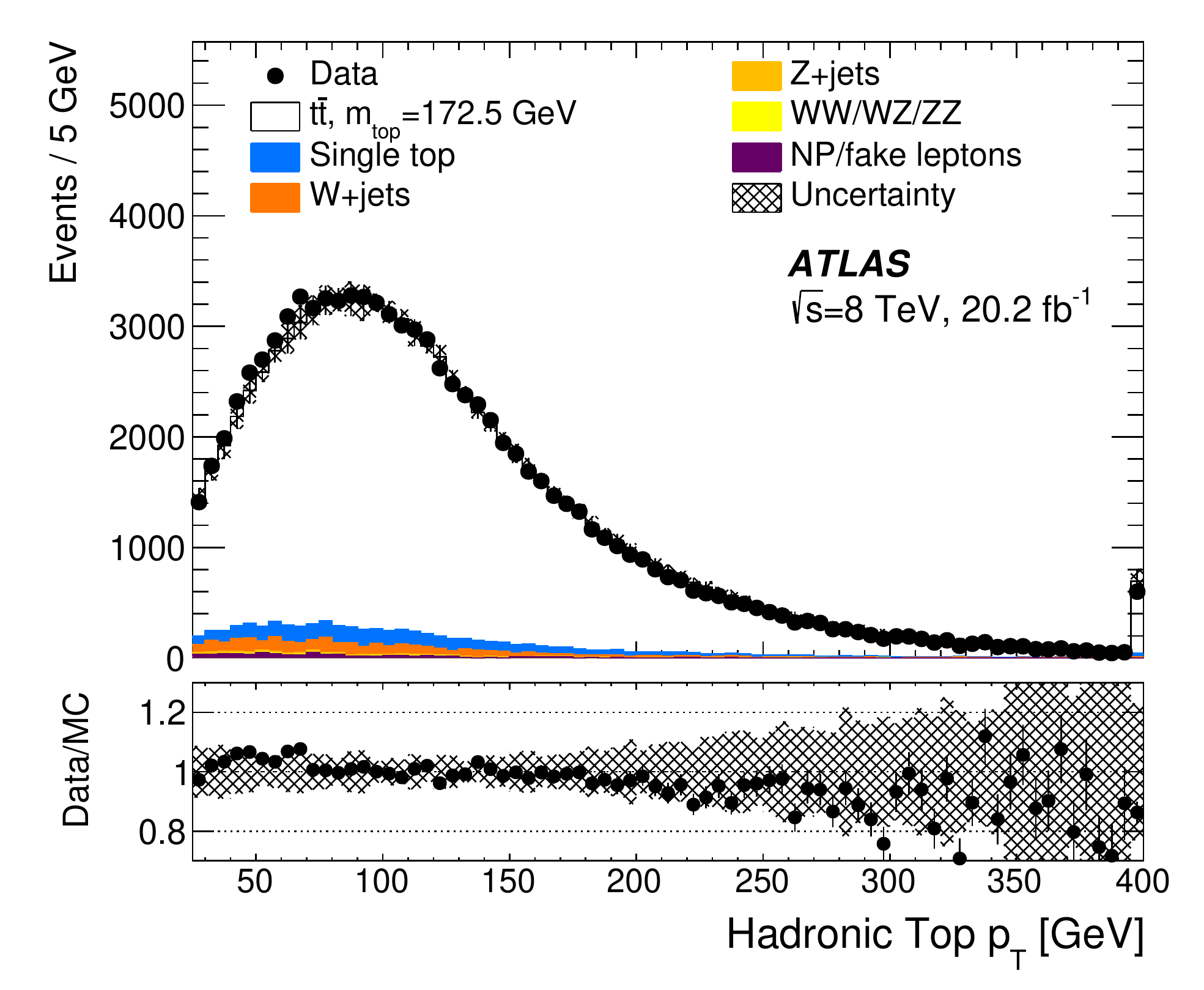}\label{fig:fig_01c}}
\subfigure[Transverse momentum of the \ttbar\ system]
          {\includegraphics[width=0.42\textwidth]{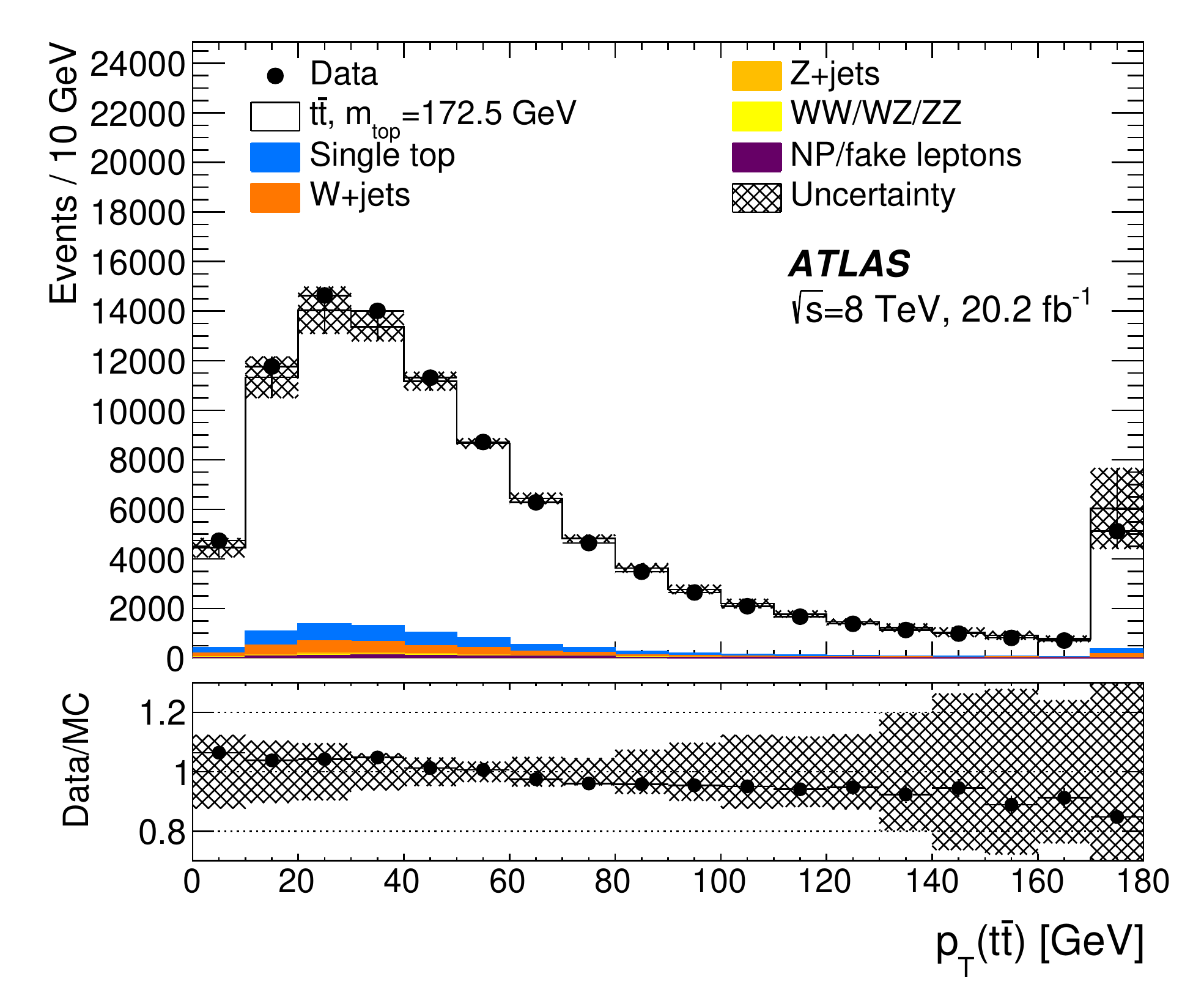}\label{fig:fig_01d}}
\hfill
\subfigure[Logarithm of the event likelihood]
          {\includegraphics[width=0.42\textwidth]{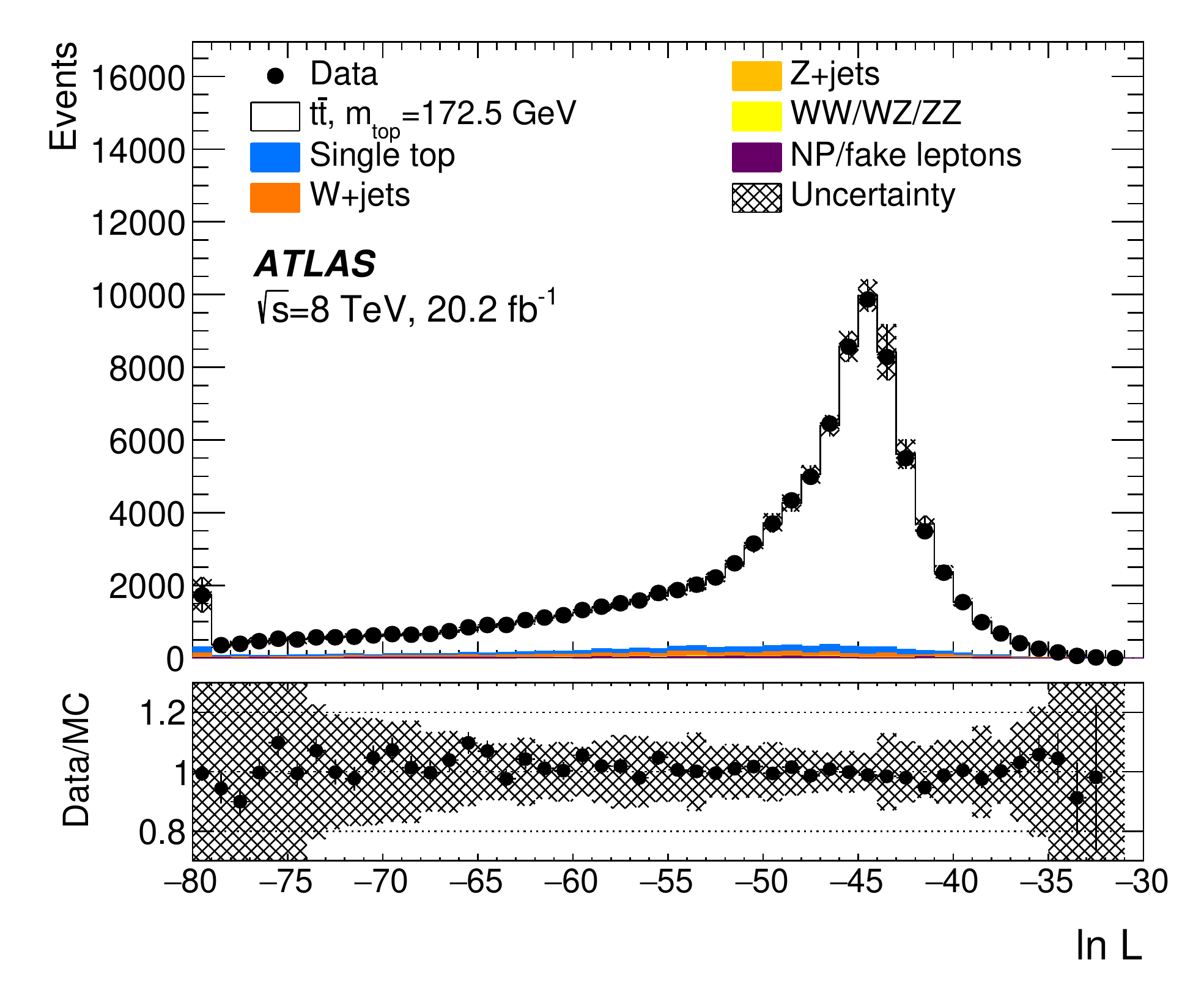}\label{fig:fig_01e}}
\subfigure[\dR\ of jets from the \Wboson\ boson decay]
          {\includegraphics[width=0.42\textwidth]{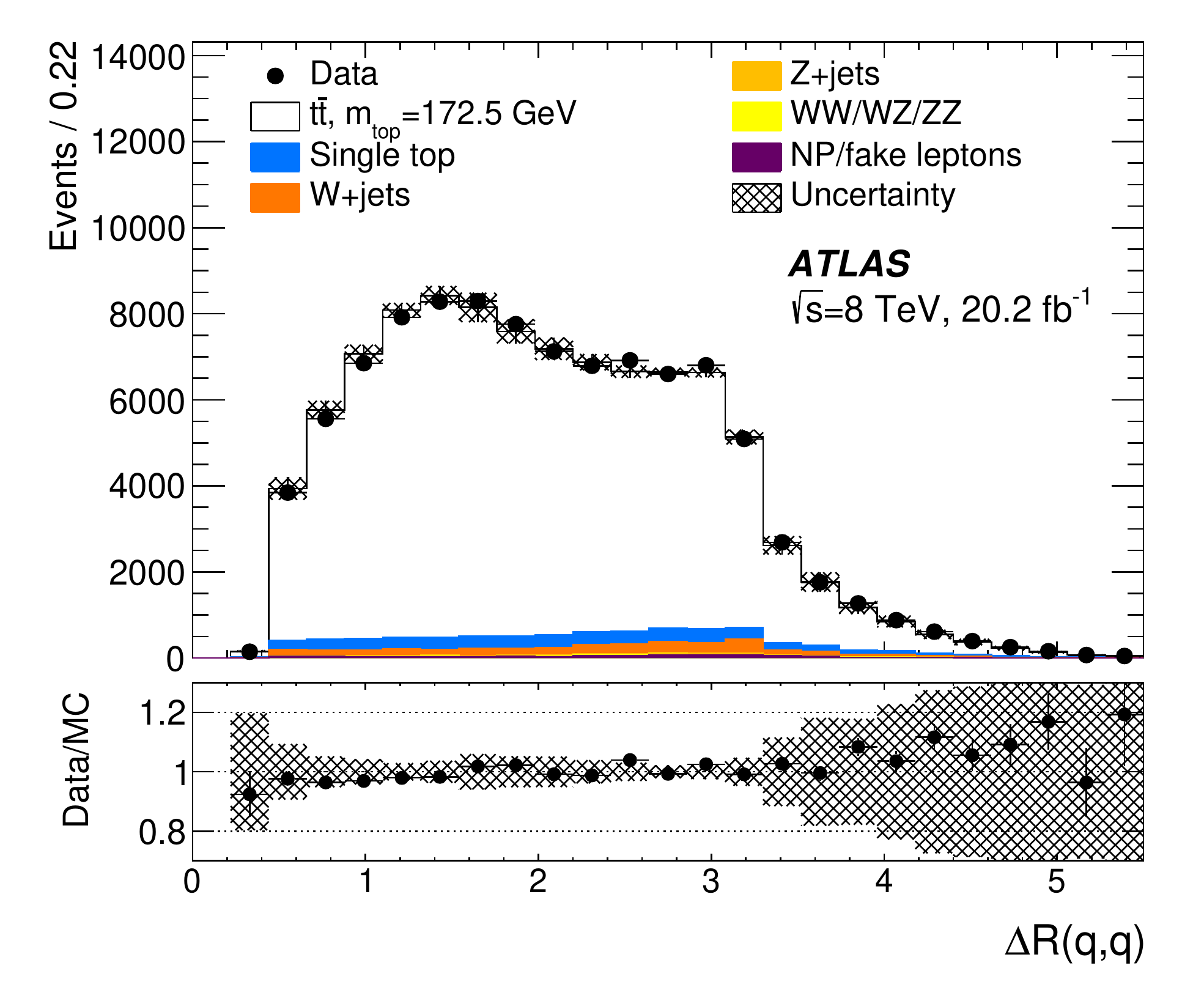}\label{fig:fig_01f}}
\caption{Distributions for the events passing the preselection.
 The data are shown together with the signal-plus-background prediction,
 normalized to the number of events observed in the data.
 The hatched area is the uncertainty in the prediction as described in the
 text.
 The rightmost bin contains all entries with values above the lower edge of this
 bin, similarly the leftmost bin contains all entries with values below the
 upper edge of this bin.
 Figure~(a) shows the transverse momentum of the lepton,
 figure~(b) shows the average transverse momentum of the jets,
 figure~(c) shows the transverse momentum of the hadronically decaying top
 quark,
 figure~(d) shows the transverse momentum of the \ttbar\ system,
 figure~(e) shows the logarithm of the event likelihood of the best permutation
 and
 figure~(f) shows the distance \dR\ of the two untagged jets \qone\ and
 \qtwo\ from the hadronically decaying \Wboson\ boson.
\label{fig:fig_01}
}
\end{figure*}

 Some distributions of the observed event kinematics after the event
 preselection and for the best permutation are shown in \Fig{\ref{fig:fig_01}}.
 Given the good description of the observed number of events by the prediction
 shown in \Sect{\ref{sect:presel}} and that the measurement of \mt\ is mostly
 sensitive to the shape of the distributions, the comparison of the data with
 the predictions is based solely on the distributions normalized to the number
 of events observed in data.
 The systematic uncertainty assigned to each bin is calculated from the sum in
 quadrature of all systematic uncertainties discussed in
 \Sect{\ref{sect:presel}}.
 Within uncertainties, the predictions agree with the observed distributions in
 \Fig{\ref{fig:fig_01}}, which shows the transverse momentum of the lepton, the
 average transverse momentum of the jets, the transverse momentum of the
 hadronically decaying top quark~\pthad, the transverse momentum of the \ttbar
 system, the logarithm of the event likelihood of the best permutation and the
 distance \dR\ of the two untagged jets \qone\ and \qtwo\ assigned to the
 hadronically decaying \Wboson\ boson.
 The distributions of transverse momenta predicted by the simulation, e.g.~the
 \pthad\ distribution shown in \Fig{\ref{fig:fig_01c}}, show a slightly
 different trend than observed in data, with the data being softer.  This
 difference is fully covered by the uncertainties.
 This trend was also observed in \Ref{\cite{TOPQ-2016-03}} for the
 \ptlb\ distribution in the \ttbarll\ channel and in the measurement of the
 differential \ttbar\ cross-section in the lepton+jets
 channel~\cite{TOPQ-2015-06}.
%

 In anticipation of the template parameterization described in
 Section~\ref{sect:templates}, the following restrictions on the three
 observables are applied: $125 \leq \mtr \leq 200~\GeV$, $55 \leq \mWr \leq
 110~\GeV$, and $0.3 \leq \rbqr \leq 3$.
 Since in this analysis only the best permutation is considered, events that do
 not pass these requirements are rejected.
 This removes events in the tails of the three distributions, which are
 typically poorly reconstructed with small likelihood values and do not contain
 significant information about \mt.
 The resulting templates have simpler shapes, which are easier to model
 analytically with fewer parameters.
 The preselection with these additional requirements is referred to as the {\it
   standard selection} to distinguish it from the boosted decision tree (BDT)
 optimization for the smallest total uncertainty in \mt, discussed in the next
 section.
%
\clearpage 
\section{Multivariate analysis and BDT event selection}
\label{sect:bdtsel}
 For the measurement of \mt, the event selection is refined enriching the
 fraction of events with correct assignments of \recolevel\ objects to their
 \genlevel\ counterparts which should be better measured and therefore lead to
 smaller uncertainties.
 The optimization of the selection is based on the multivariate BDT algorithm
 implemented in the \TMVA\ package~\cite{Hocker:2007ht}.
 The reconstruction-level objects are matched to the closest parton-level object
 within a \dR\ of $0.1$ for electrons and muons and $0.3$ for jets.
 A matched object is defined as a reconstruction-level object that falls within
 the relevant \dR\ of any parton-level object of that type, and a correct match
 means that this generator-level object is the one it originated from.
 Due to acceptance losses and reconstruction inefficiencies, not all
 reconstruction-level objects can successfully be matched to their parton-level
 counterparts.
 If any object cannot be unambiguously matched, the corresponding event is
 referred to as {\it unmatched}.
 The efficiency for correctly matched events \effcm\ is the fraction of
 correctly matched events among all the matched events, and the
 \selPurity\ \purcm\ is the fraction of correctly matched events among all
 selected events, regardless of whether they could be matched or not.

\begin{figure*}[tbp!]
\centering
\subfigure[Logarithm of the event likelihood]
          {\includegraphics[width=0.48\textwidth]{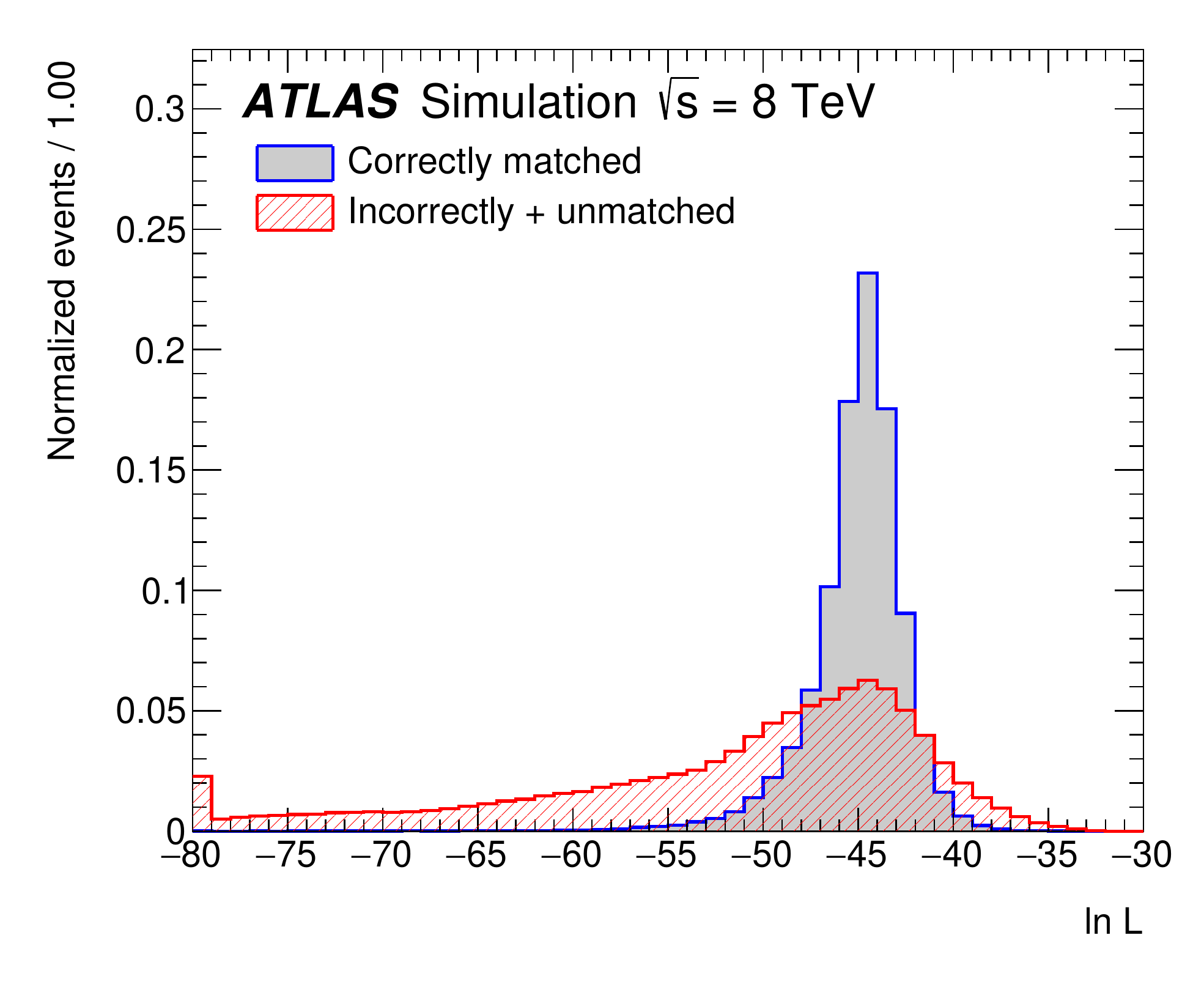}\label{fig:fig_02a}}
\subfigure[\dR\ of jets from the \Wboson\ boson decay]
          {\includegraphics[width=0.48\textwidth]{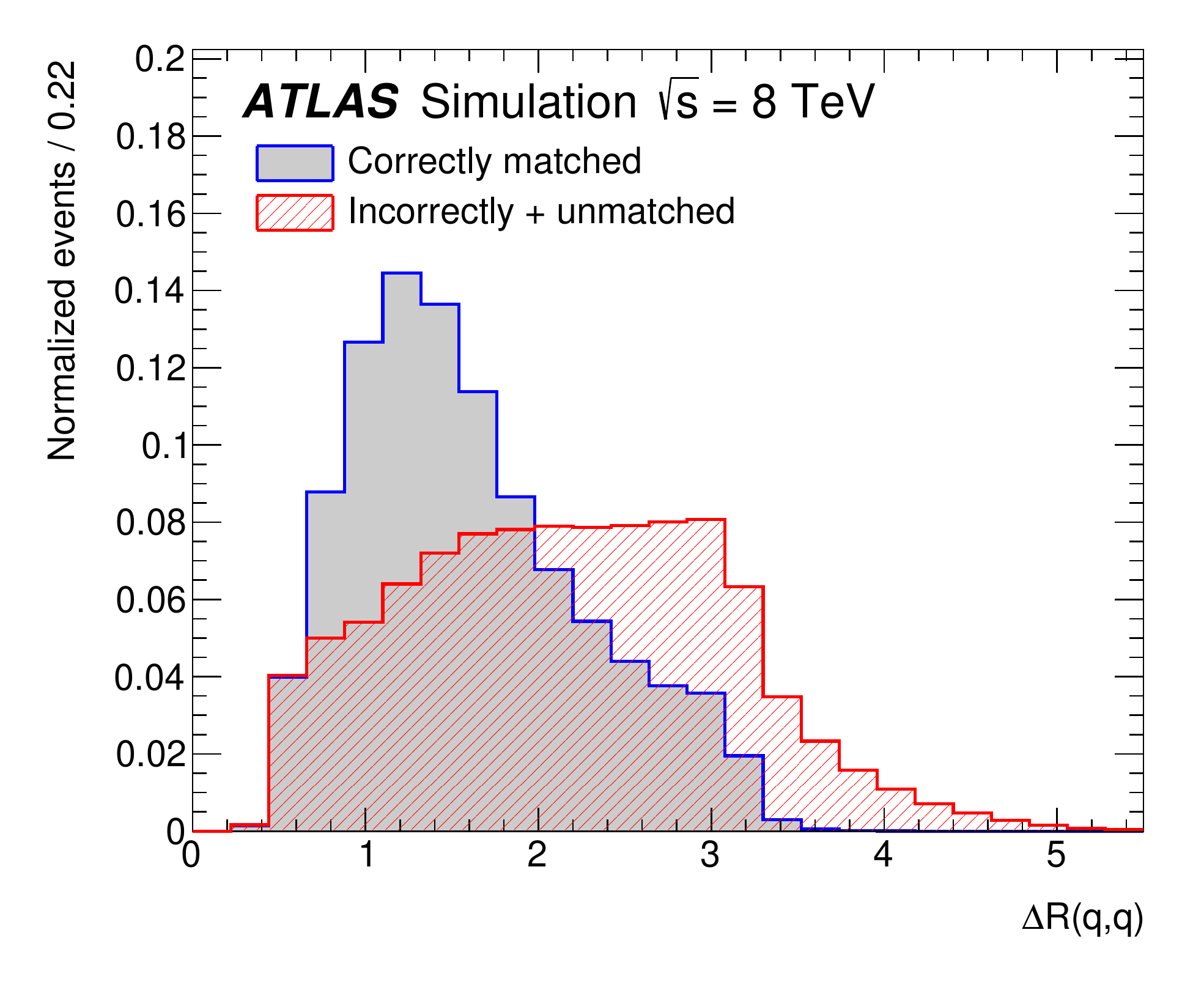}\label{fig:fig_02b}}
\hfill
\subfigure[BDT test and training samples]
          {\includegraphics[width=0.48\textwidth]{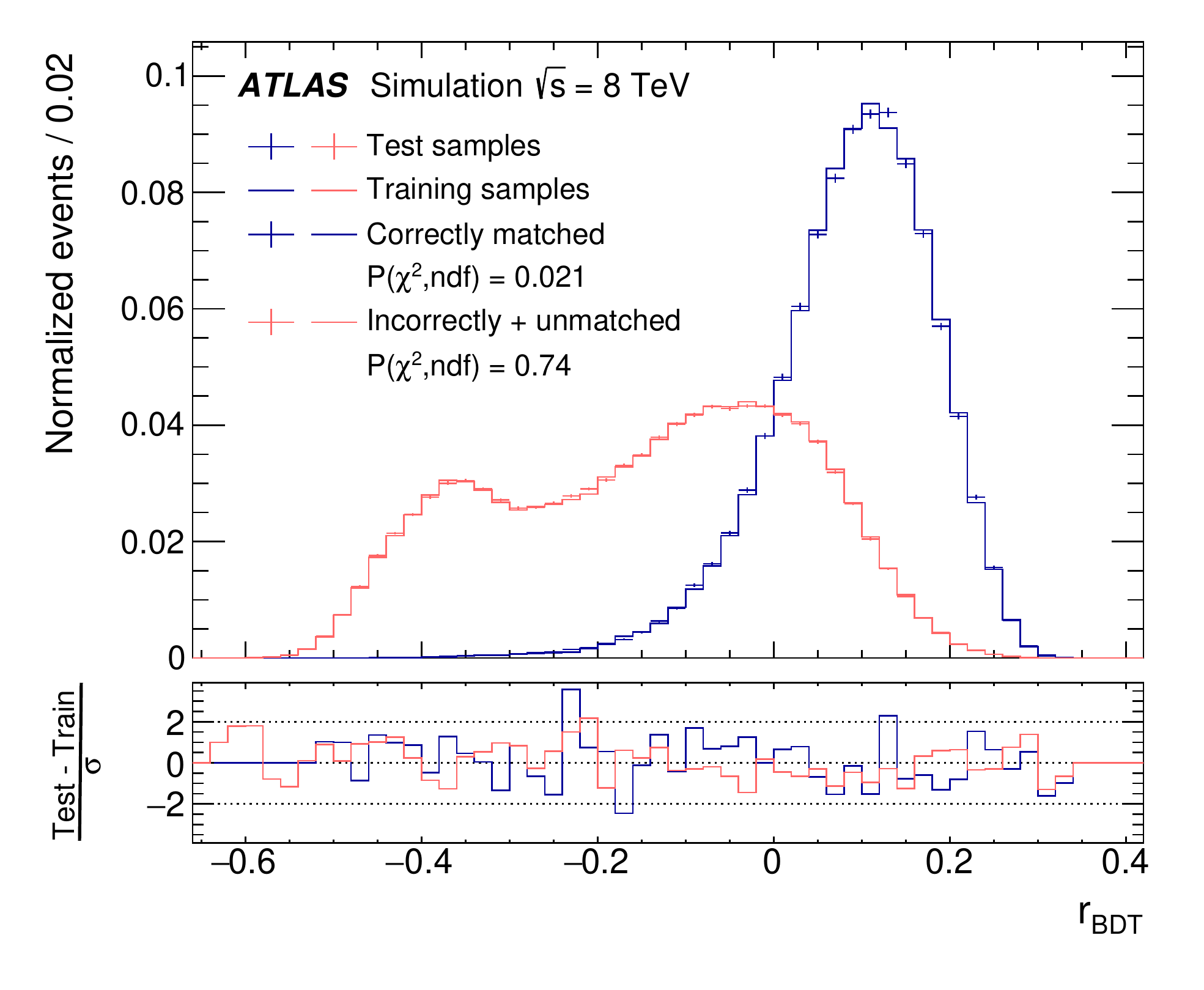}\label{fig:fig_02c}}
\subfigure[Data-to-simulation comparison for \rBDT]
          {\includegraphics[width=0.48\textwidth]{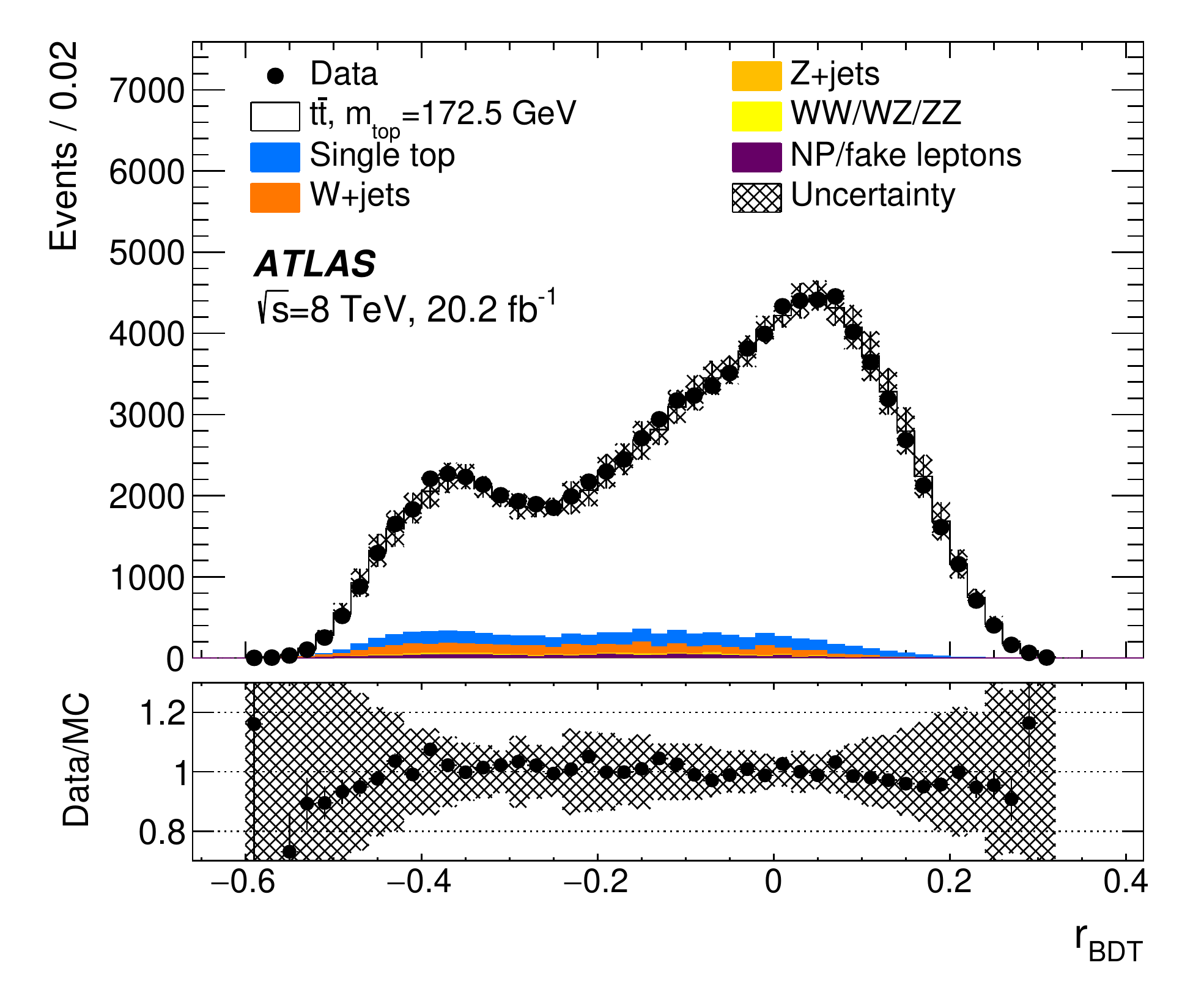}\label{fig:fig_02d}}
\caption{Input and results of the BDT training on \ttbar\ signal events for the
  preselection.
 Figure~(a) shows the logarithm of the event likelihood of the best permutation
 (\lnL) for the correctly matched events and the remainder.
 Similarly, figure~(b) shows the distribution of the \dR\ between the two
 untagged jets assigned to the \Wboson\ boson decay.
 Figure~(c) shows the distribution of the BDT output (\rBDT) for the two classes
 of events for both the training (histograms) and test samples (points with
 statistical uncertainties).
 The compatibility in terms of the \chiq\ probability is also listed.
 The distributions peaking at around $\rBDT=0.1$ are for the correctly matched
 events, the ones to the left are for incorrectly or unmatched events.
 The ratio figure shows the difference between the number of events in the
 training and test samples divided by the statistical uncertainty in this
 difference.
 Finally, figure~(d) shows the comparison of the \rBDT\ distributions observed
 in data and MC simulation.
 The hatched area includes the uncertainties as detailed in the text. The
 uncertainty bars correspond to the statistical uncertainties in the data.
 \label{fig:fig_02}
}
\end{figure*}

 The BDT algorithm is exploited to enrich the event sample in events that have
 correct jet-to-parton matching by reducing the remainder, i.e.~the sum of
 incorrectly matched and unmatched events.
 Using the preselection, the BDT algorithm is trained on the simulated
 \ttbar\ signal sample with $\mt=172.5$~\GeV.
 Many variables were studied and only those with a separation\footnote{The
   chosen definition of the separation is given in \Eqn{(1)} of the TMVA
   manual~\cite{Hocker:2007ht}.} larger than $0.1\%$ are used in the training.
 The 13 variables chosen for the final training are given in
 \Tab{\ref{tab:BDTvar}}.
 For all input variables to the BDT algorithm, good agreement between the MC
 predictions and the data is found, as shown in
 \Figs{\ref{fig:fig_01e}}{\ref{fig:fig_01f}} for the examples of the likelihood
 of the chosen permutation and the opening angle \dR\ of the two untagged jets
 associated with the \Wboson\ boson decay.
%
\begin{table}[tbp!]
\caption{The input variables to the BDT algorithm sorted by their separation.
 \label{tab:BDTvar}
}
\begin{center}
\begin{tabular}{|c|l|}
\hline
Separation & Description \\ \hline
   31$\%$  & Logarithm of the event likelihood of the best permutation, \lnL\\
   13$\%$  & \dR\ of the two untagged jets \qone\ and \qtwo\ from \\
           & the hadronically decaying \Wboson\ boson, \dRqq \\
  5.0$\%$  & \pt\ of the hadronically decaying \Wboson\ boson \\
  4.3$\%$  & \pt\ of the hadronically decaying top quark \\
  4.2$\%$  & Relative event probability of the best permutation \\
  2.0$\%$  & \pt\ of the reconstructed \ttbar\ system \\
  1.7$\%$  & \pt\ of the semi-leptonically decaying top quark \\
  1.2$\%$  & Transverse mass of the leptonically decaying \Wboson\ boson\\
  0.3$\%$  & \pt\ of the leptonically decaying \Wboson\ boson \\
  0.3$\%$  & Number of jets \\
  0.2$\%$  & \dR\ of the reconstructed \btagged\ jets \\
  0.2$\%$  & Missing transverse momentum \\
  0.1$\%$  & \pt\ of the lepton \\
\hline
\end{tabular}
\end{center}
\end{table}
%
 These two variables also have the largest separation for the correctly matched
 events and the remainder.
 The corresponding distributions for the two event classes are shown in
 \Figs{\ref{fig:fig_02a}}{\ref{fig:fig_02b}}.
%
%
 These figures show a clear separation of the correctly matched events and the
 remainder.
 Half the simulation sample is used to train the algorithm and the other half to
 assess its performance.
 The significant difference between the distributions of the output
 value~\rBDT\ of the BDT classifier between the two classes of events in
 \Fig{\ref{fig:fig_02c}} shows their efficient separation by the BDT algorithm.
 In addition, reasonable agreement is found for the \rBDT\ distributions in the
 statistically independent test and training samples.
 The \rBDT\ distributions in simulation and data in \Fig{\ref{fig:fig_02d}}
 agree within the experimental uncertainties.
 The above findings justify the application of the BDT approach to the data.
%
\begin{figure}[tbp!]
\centering
\includegraphics[width=0.92\textwidth]{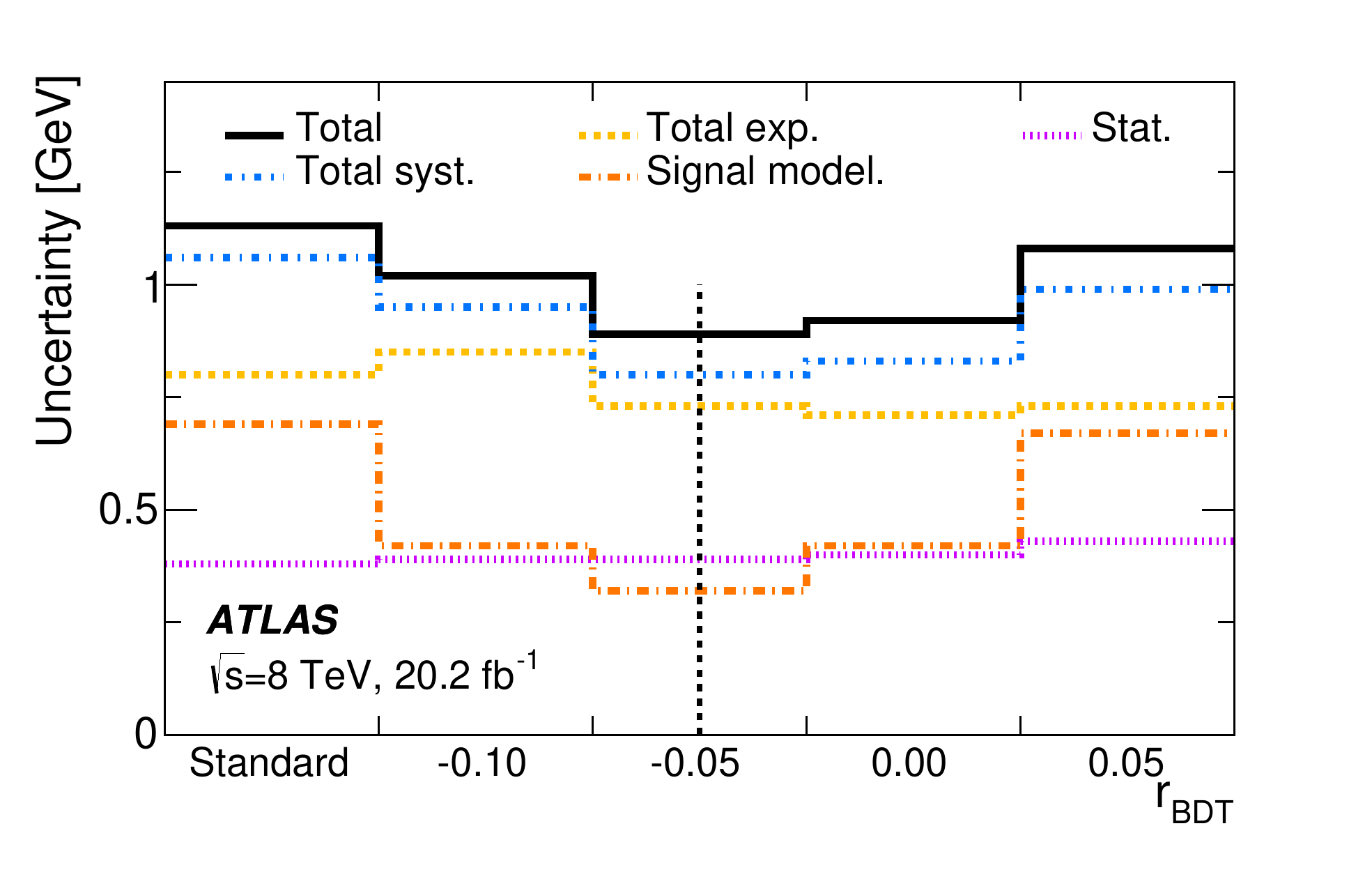}
\caption{Various classes of \mt\ uncertainties as a function of the minimum
  requirement on the BDT output \rBDT\ and for the standard selection.
 The total uncertainty (solid line) is the sum in quadrature of the statistical
 (dotted line) and total systematic uncertainty (short dash-dotted line). The
 total systematic uncertainty consists of the total experimental (dashed line)
 and total signal-modelling uncertainty (long dash-dotted line). The
 uncertainties in the background estimate are included in the total experimental
 uncertainty.
 The minimum requirement on \rBDT\ defining the \mvabased\ selection is
 indicated by the vertical black dashed line. All uncertainties are included
 except for the method and the pile-up uncertainties.
\label{fig:BDT_optimisation}
}
\end{figure}

 The full \mt\ analysis detailed in \Sect{\ref{sect:unc}} is performed, except
 for the evaluation of the small method and pile-up uncertainties described in
 \Sect{\ref{sect:unc}}, for several minimum requirements on \rBDT\ in the range
 of $[-0.10, 0.05]$ in steps of 0.05 to find the point with smallest total
 uncertainty.
 The total uncertainty in \mt\ together with the various classes of uncertainty
 sources as a function of \rBDT\ evaluated in the \mvabased\ optimization are
 shown in \Fig{\ref{fig:BDT_optimisation}}.
 The minimum requirement $\rBDT=-0.05$ provides the smallest total uncertainty
 in \mt.
 The resulting numbers of events for this BDT selection are given in
 \Tab{\ref{tab:selections}}.
 Compared with the preselection, \effcm\ is increased from \effpre\ to \effBDT,
 albeit at the expense of a significant reduction in the number of selected
 events. The purity \purcm\ is increased from \purpre\ to \purBDT.
 In addition, the intrinsic resolution in \mt\ of the remaining event sample is
 improved, i.e.~the statistical uncertainty in \mt\ in
 \Fig{\ref{fig:BDT_optimisation}} is almost constant as a function of \rBDT; in
 particular, it does not scale with the square root of the number of events
 retained.
 For the signal sample with $\mt=172.5$~\GeV, the template fit functions for the
 standard selection and the \mvabased\ selection, together with their ratios,
 are shown in \Fig{\ref{fig:Resolution}} in \App{\ref{sect:addlpj}}.
 The shape of the signal modelling uncertainty derives from a sum of
 contributions with different shapes. The curves from the signal Monte Carlo
 generator and colour reconnection uncertainties decrease, the one from the
 underlying event uncertainty is flat, the one from the initial- and final-state
 QCD radiation has a valley similar to the sum of all contributions, and finally
 the one from the hadronization uncertainty rises.
 
\begin{figure*}[tbp!]
\centering
\subfigure[\Wboson\ boson transverse mass]
          {\includegraphics[width=0.48\textwidth]{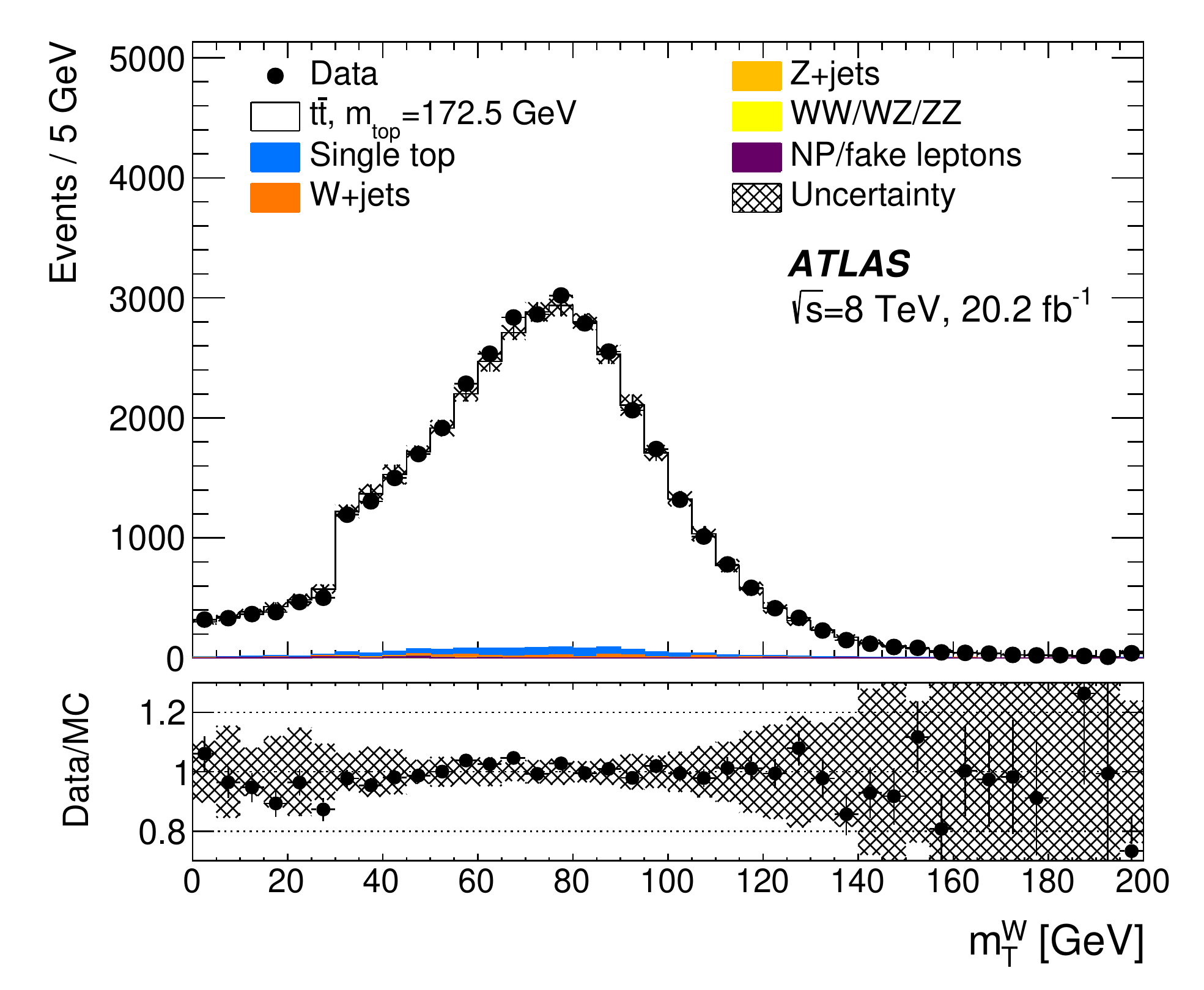}\label{fig:fig_03a}}
\subfigure[Reconstructed top quark mass]
          {\includegraphics[width=0.48\textwidth]{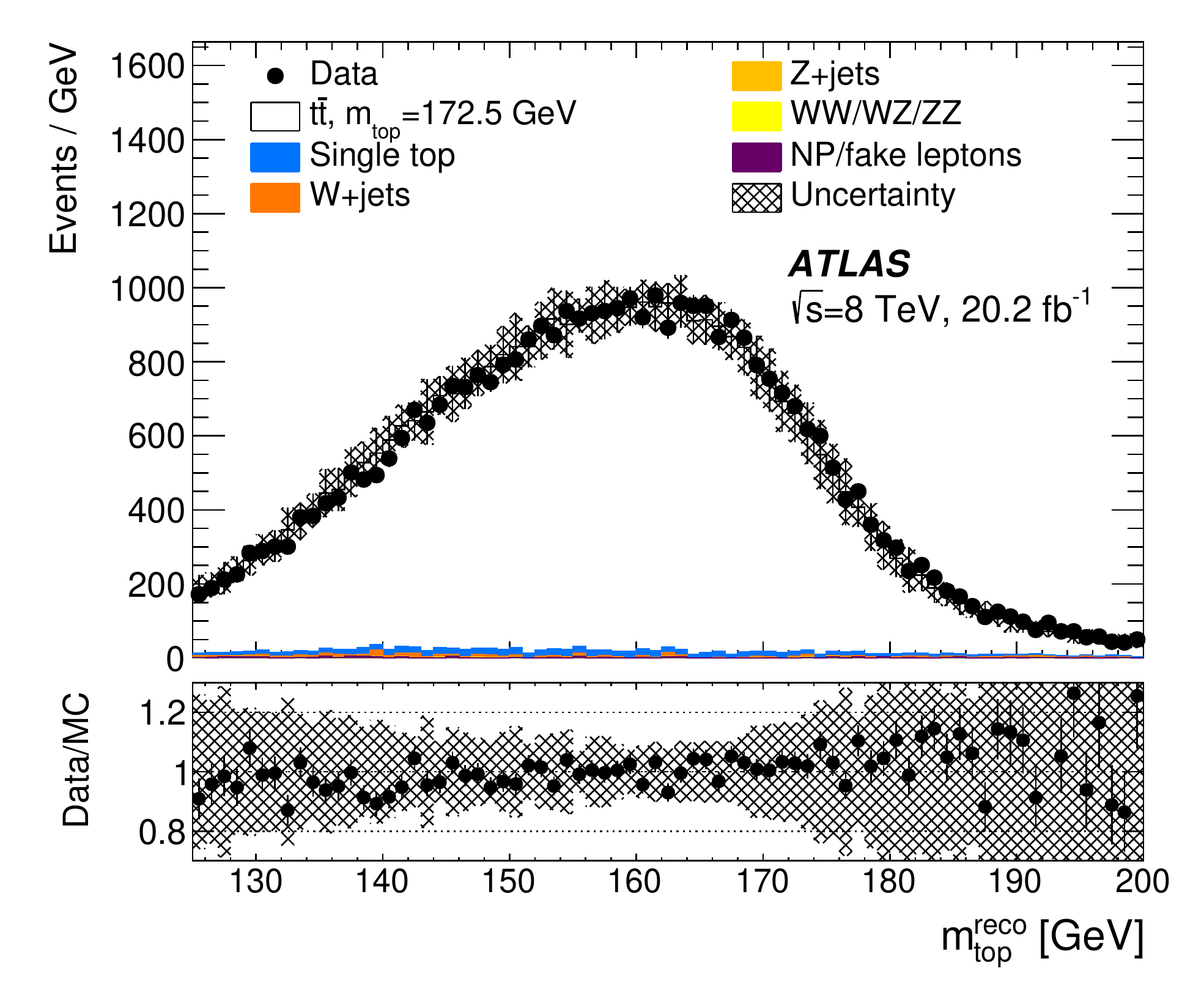}\label{fig:fig_03b}}
\hfill
\subfigure[Reconstructed \Wboson\ boson mass]
          {\includegraphics[width=0.48\textwidth]{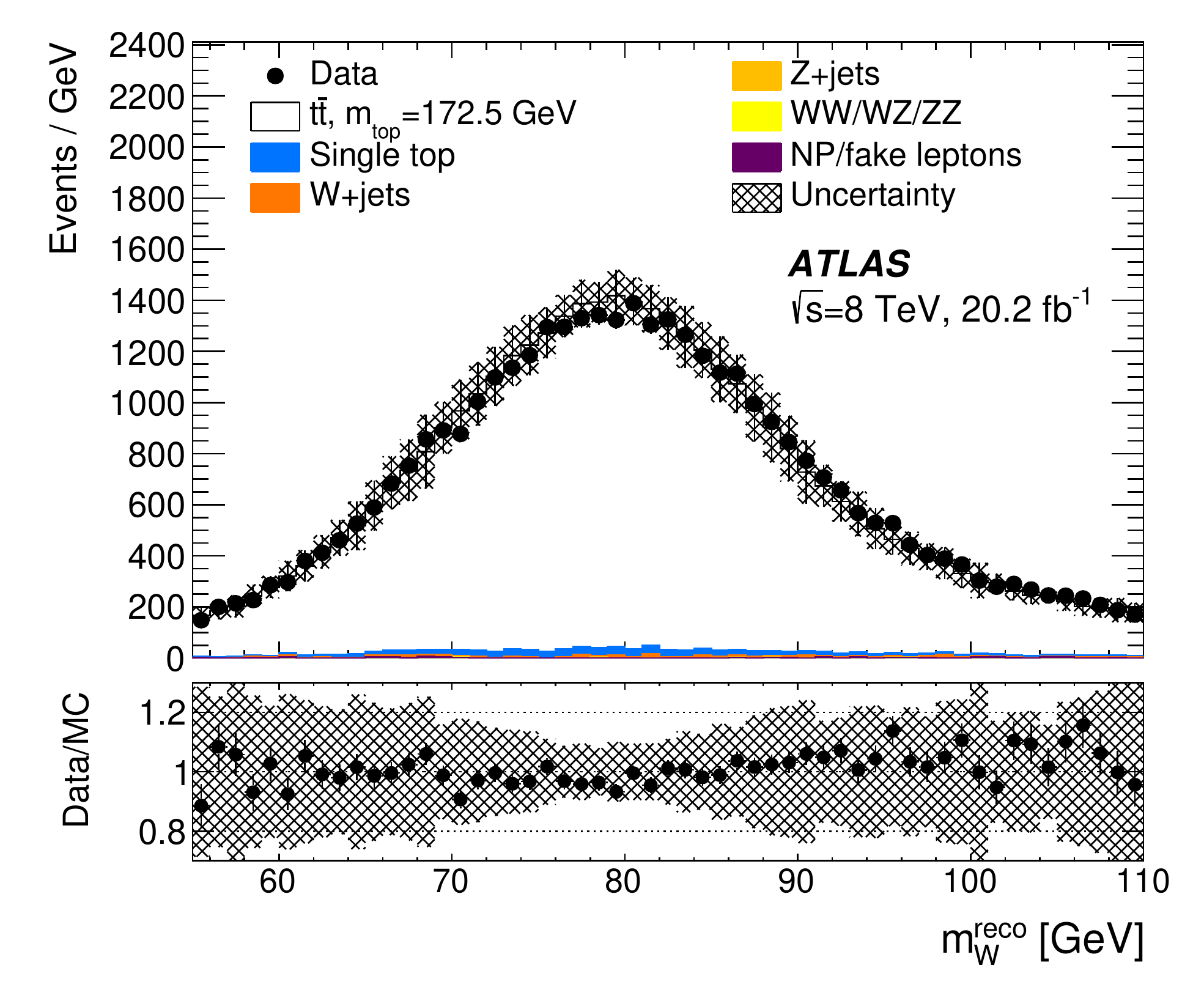}\label{fig:fig_03c}}
\subfigure[Reconstructed ratio of jet transverse momenta]
          {\includegraphics[width=0.48\textwidth]{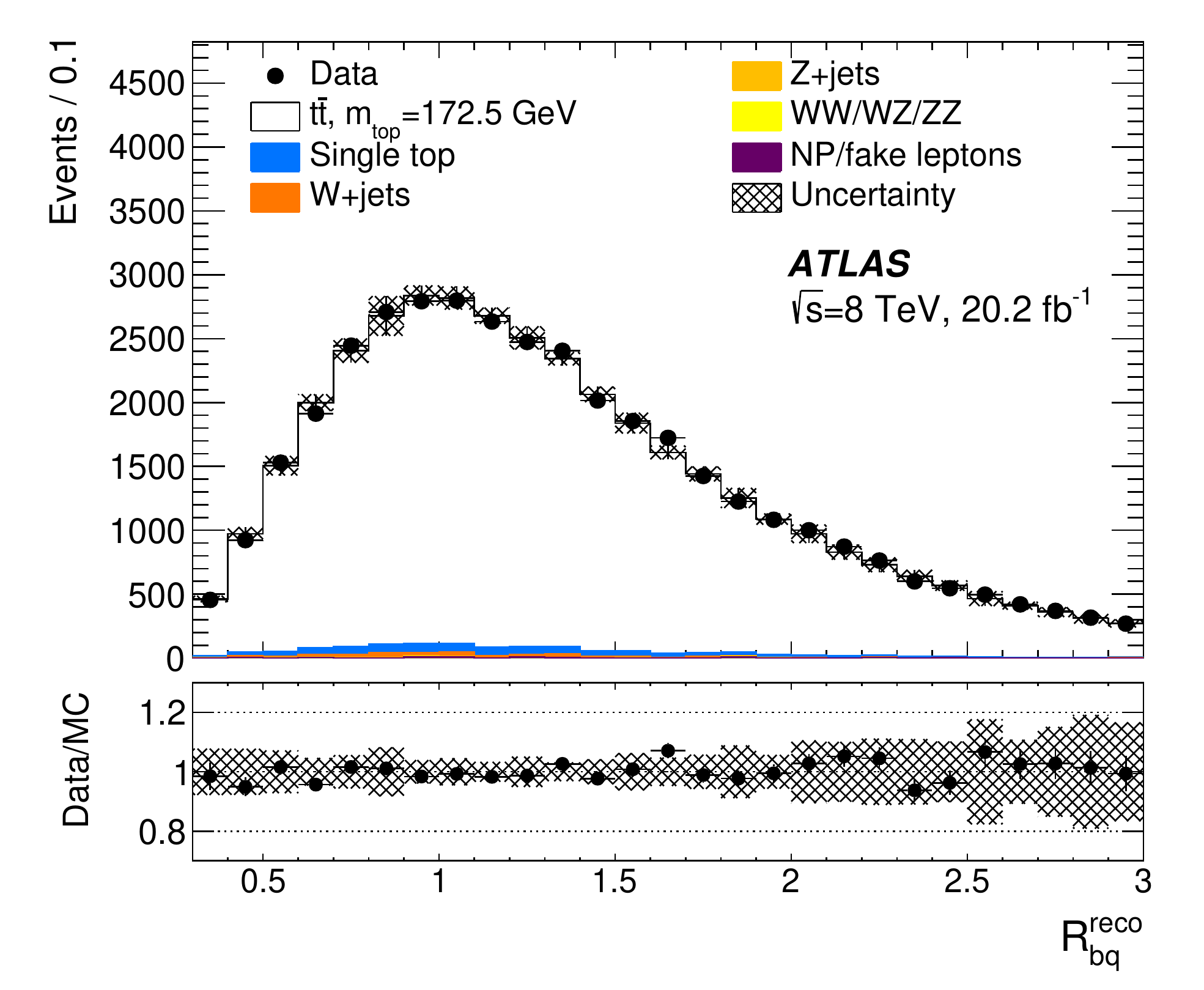}\label{fig:fig_03d}}
\caption{Distributions for the events passing the BDT selection.
 The data are shown, together with the signal-plus-background prediction
 normalized to the number of events observed in the data.
 The hatched area is the uncertainty in the prediction described in the
 text.
 The rightmost bin contains all entries with values above the lower edge of this
 bin, similarly the leftmost bin contains all entries with values below the
 upper edge of this bin.
 Figure~(a) shows the \Wboson\ boson transverse mass for the semi-leptonic top
 quark decay.
 The remaining figures show the three observables used for the determination of
 \mt, where figure~(b) shows the reconstructed top quark mass \mtr, figure~(c)
 shows the reconstructed invariant mass of the \Wboson\ boson \mWr and
 figure~(d) shows the reconstructed ratio of jet transverse momenta \rbqr.
 The three distributions are shown within the ranges of the template fit.
 \label{fig:fig_03}
}
\end{figure*}

 Some distributions of the observed event kinematics after the
 \mvabased\ selection are shown in \Fig{\ref{fig:fig_03}}.
 Good agreement between the MC predictions and the data is found, as seen for
 the preselection in \Fig{\ref{fig:fig_01}}.
 The examples shown are the observed \Wboson\ boson transverse mass for the
 semi-leptonically decaying top quark in \Fig{\ref{fig:fig_03a}} and the three
 observables of the \mt\ analysis~(within the ranges of the template fit) in
 \Figrange{\ref{fig:fig_03b}}{\ref{fig:fig_03d}}.
 The sharp edge observed at 30~\GeV\ in \Fig{\ref{fig:fig_03a}} originates from
 the different selection requirements for the \Wboson\ boson transverse mass in
 the electron+jets and muon+jets final states.
%

\clearpage
\section{Template fit}
\label{sect:templates}
 This analysis uses a three-dimensional template fit technique which determines
 \mt\ together with the jet energy scale factors \JSF\ and \bJSF.
 The aim of the multi-dimensional fit to the data is to measure \mt\ and, at the
 same time, to absorb the mean differences between the jet energy scales
 observed in data and MC simulated events into jet energy scale factors.
 By using \JSF\ and \bJSF, most of the uncertainties in \mt\ induced by
 \JES\ and \bJES\ uncertainties are transformed into additional statistical
 components caused by the higher dimensionality of the fit.
 This method reduces the total uncertainty in \mt\ only for sufficiently large
 data samples. In this case, the sum in quadrature of the additional statistical
 uncertainty in \mt\ due to the \JSF~(or \bJSF) fit and the residual
 \JES-induced~(or \bJES-induced) systematic uncertainty is smaller than the
 original \JES-induced~(or \bJES-induced) uncertainty in \mt.
 This situation was already realized for the $\sqrts=7$~\TeV\ data
 analysis~\cite{TOPQ-2013-02} and is even more advantageous for the much larger
 data sample of the $\sqrts=8$~\TeV\ data analysis.
 Since \JSF\ and \bJSF\ are global factors, they do not completely absorb the
 \JES\ and \bJES\ uncertainties which have \pt- and $\eta$-dependent components.

 For simultaneously determining \mt, \JSF\ and \bJSF, templates are constructed
 from the MC samples.
 Templates of \mtr\ are constructed with several input \mt\ values used in the
 range 167.5--177.5~\GeV\ and for the sample at $\mt=172.5$~\GeV\ also with
 independent input values for \JSF\ and \bJSF\ in the range 0.96--1.04 in steps
 of 0.02.
 Statistically independent MC samples are used for different input values of
 \mt.
 The templates with different values of \JSF\ and \bJSF\ are constructed by
 scaling the energies of the jets appropriately.
 In this procedure, \JSF\ is applied to all jets, while \bJSF\ is solely applied
 to \bjets\ according to the generated quark flavour.
 The scaling is performed after the various correction steps of the jet
 calibration but before the event selection. This procedure results in different
 events passing the \mvabased\ selection from one energy scale variation to
 another.
 However, many events are in all samples, resulting in a large statistical
 correlation of the samples with different jet scale factors.
 Similarly, templates of \mWr\ and \rbqr\ are constructed with the above listed
 input values of \mt, \JSF\ and \bJSF.

 Independent signal templates are derived for the three observables for all
 \mt-dependent samples, consisting of the \ttbar\ signal events and
 single-top-quark production events.
 This procedure is adopted because single-top-quark production carries
 information about the top quark mass, and in this way, \mt-independent
 background templates can be used.
 The signal templates are simultaneously fitted to the sum of a Gaussian and two
 Landau functions for \mtr, to the sum of two Gaussian functions for \mWr and to
 the sum of two Gaussian and one Landau function for \rbqr.
 This set of functions leads to an unbiased estimate of \mt, but is not unique.
 For the background, the \mtr\ distribution is fitted to a Landau function,
 while both the \mWr\ and the \rbqr\ distributions are fitted to the sum of two
 Gaussian functions.

\begin{figure*}[tbp!]
\centering
\vspace{-0.3cm}
\subfigure[\mtr\ with three input \mt]
          {\includegraphics[width=0.47\textwidth]{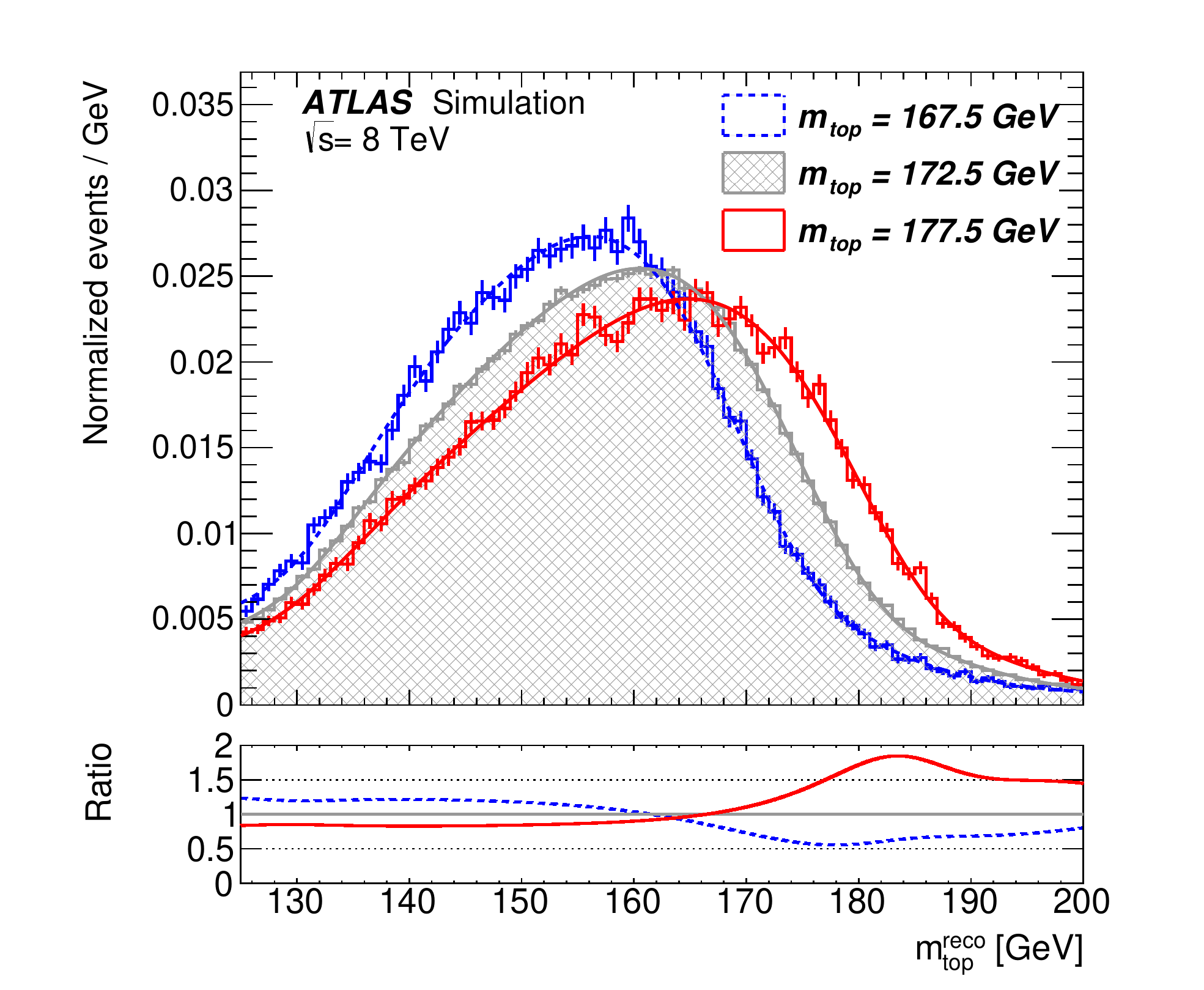}
            \label{fig:fig_04a}}
\subfigure[\mtr\ with three input \JSF]
          {\includegraphics[width=0.47\textwidth]{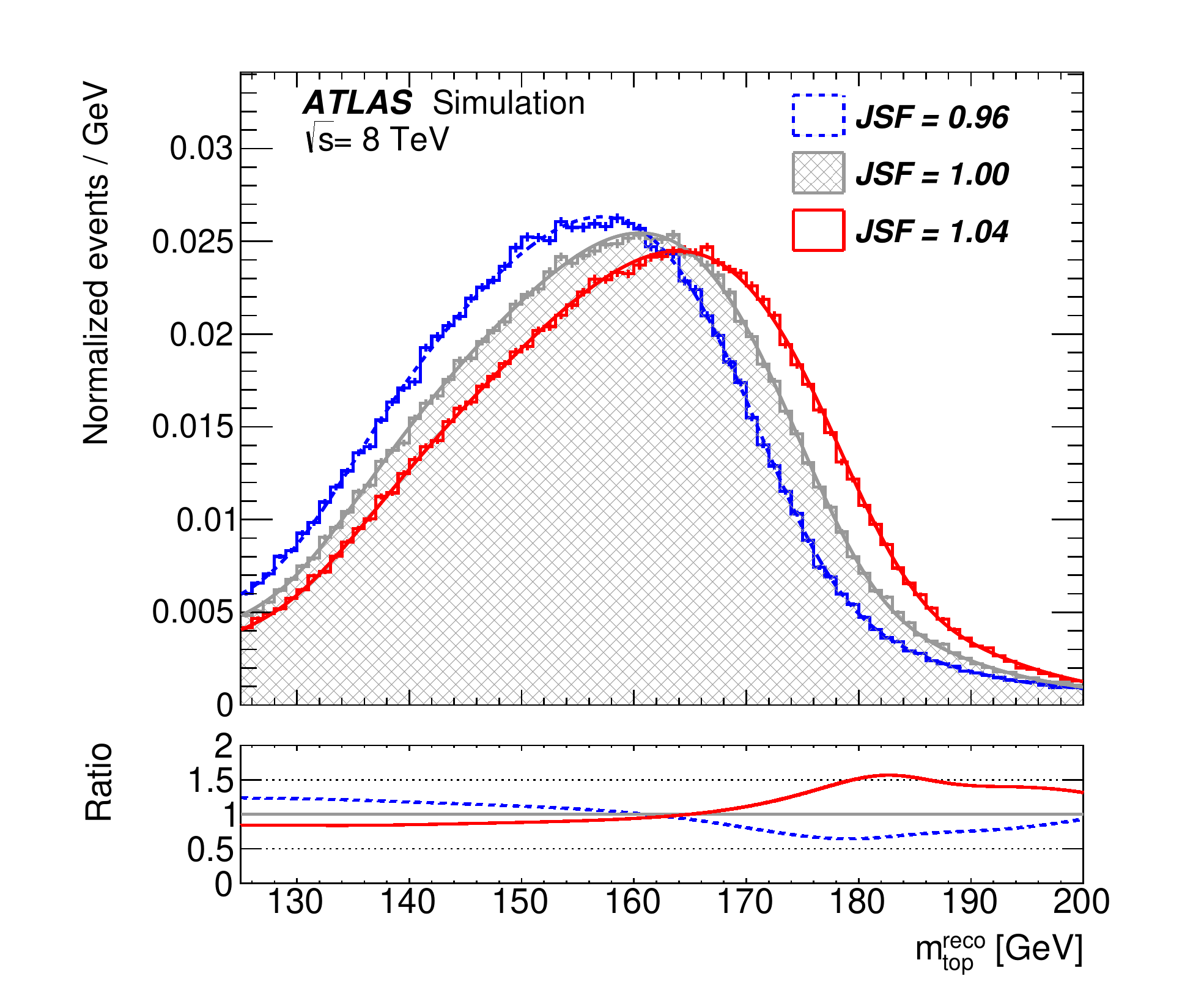}
            \label{fig:fig_04b}}
\hfill
\vspace{-0.3cm}
\subfigure[\mtr\ with three input \bJSF]
          {\includegraphics[width=0.47\textwidth]{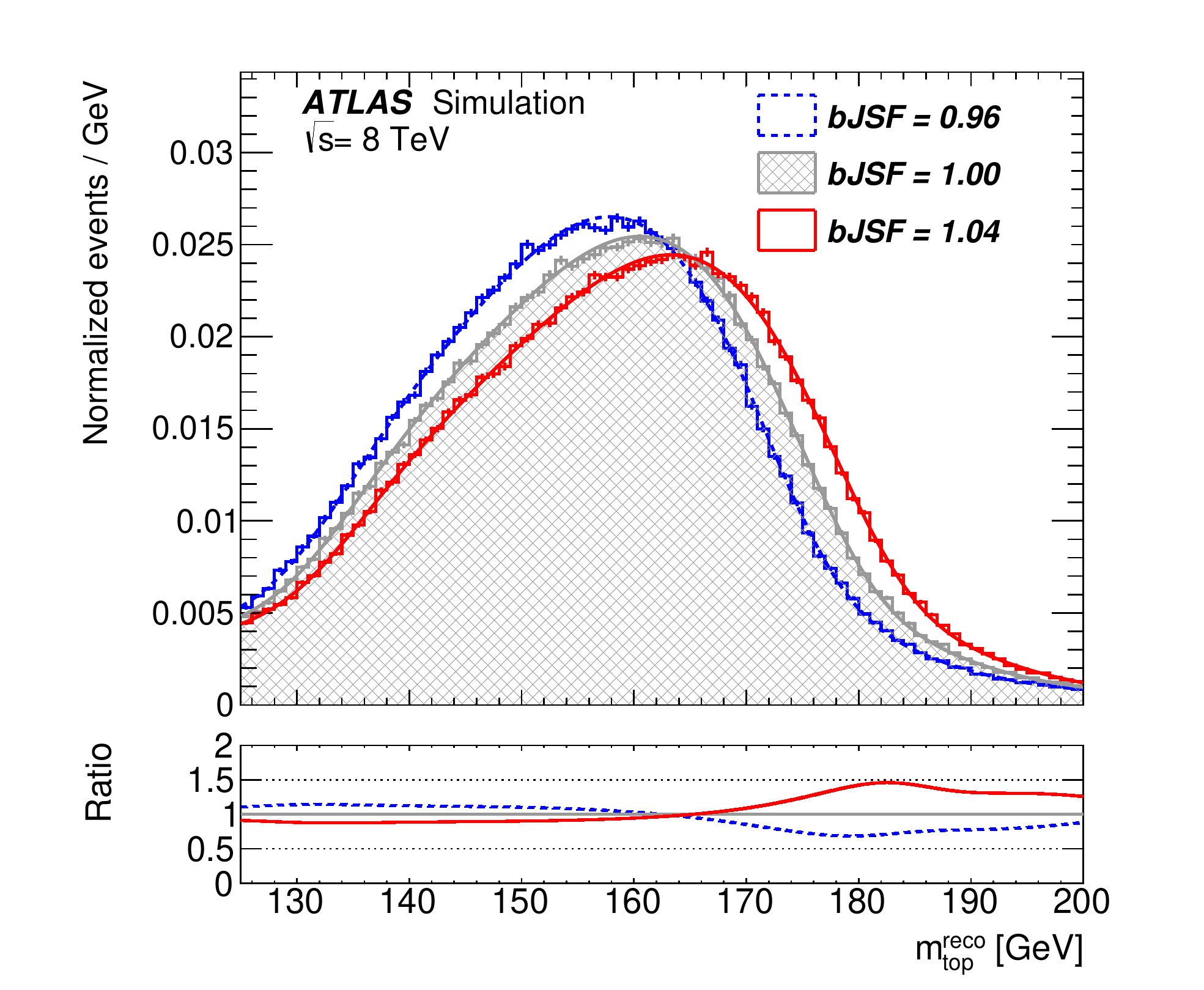}
            \label{fig:fig_04c}}
\subfigure[\mWr\ with three input \JSF]
          {\includegraphics[width=0.47\textwidth]{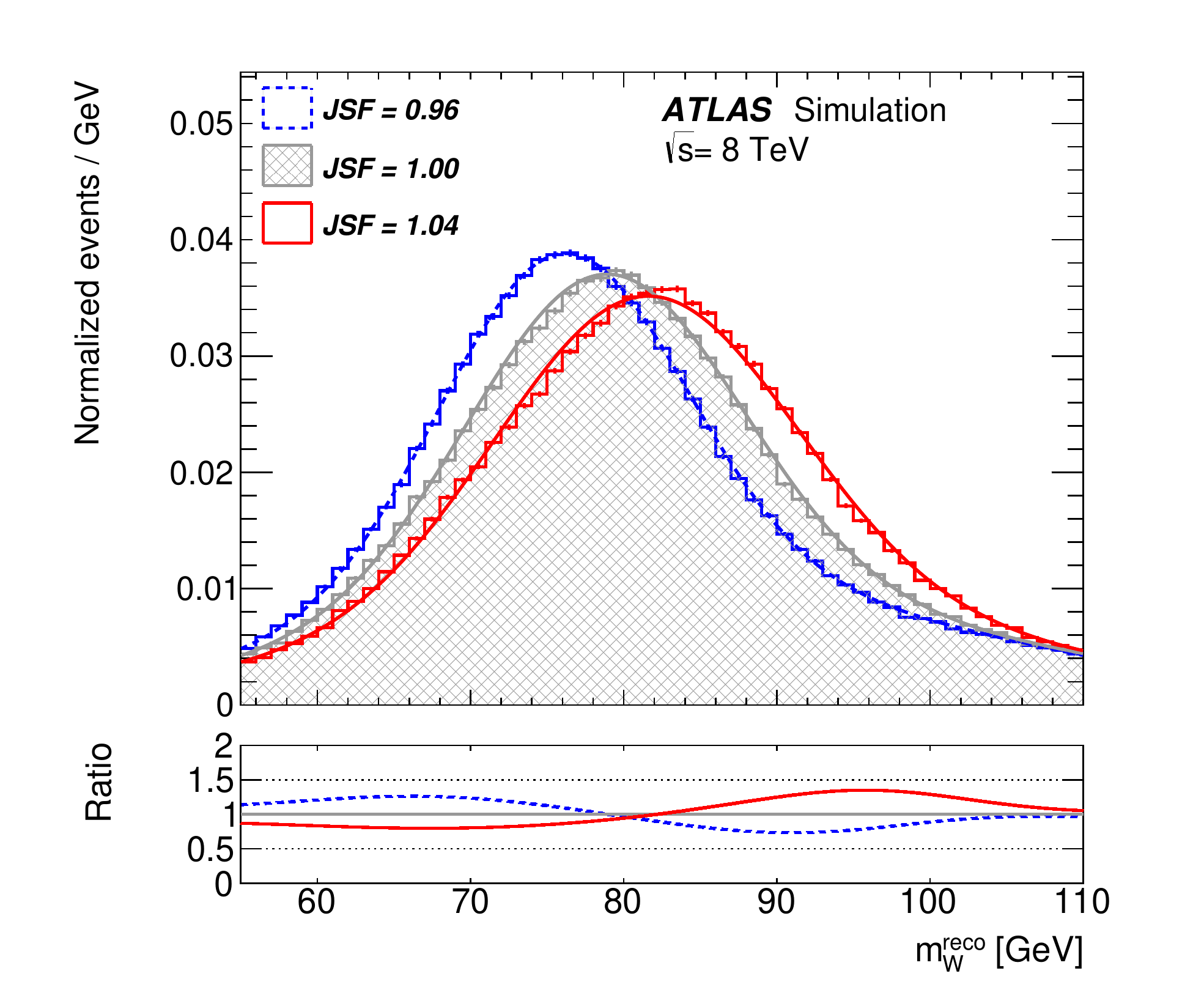}
            \label{fig:fig_04d}}
\hfill
\subfigure[\rbqr\ with three input \mt]
          {\includegraphics[width=0.47\textwidth]{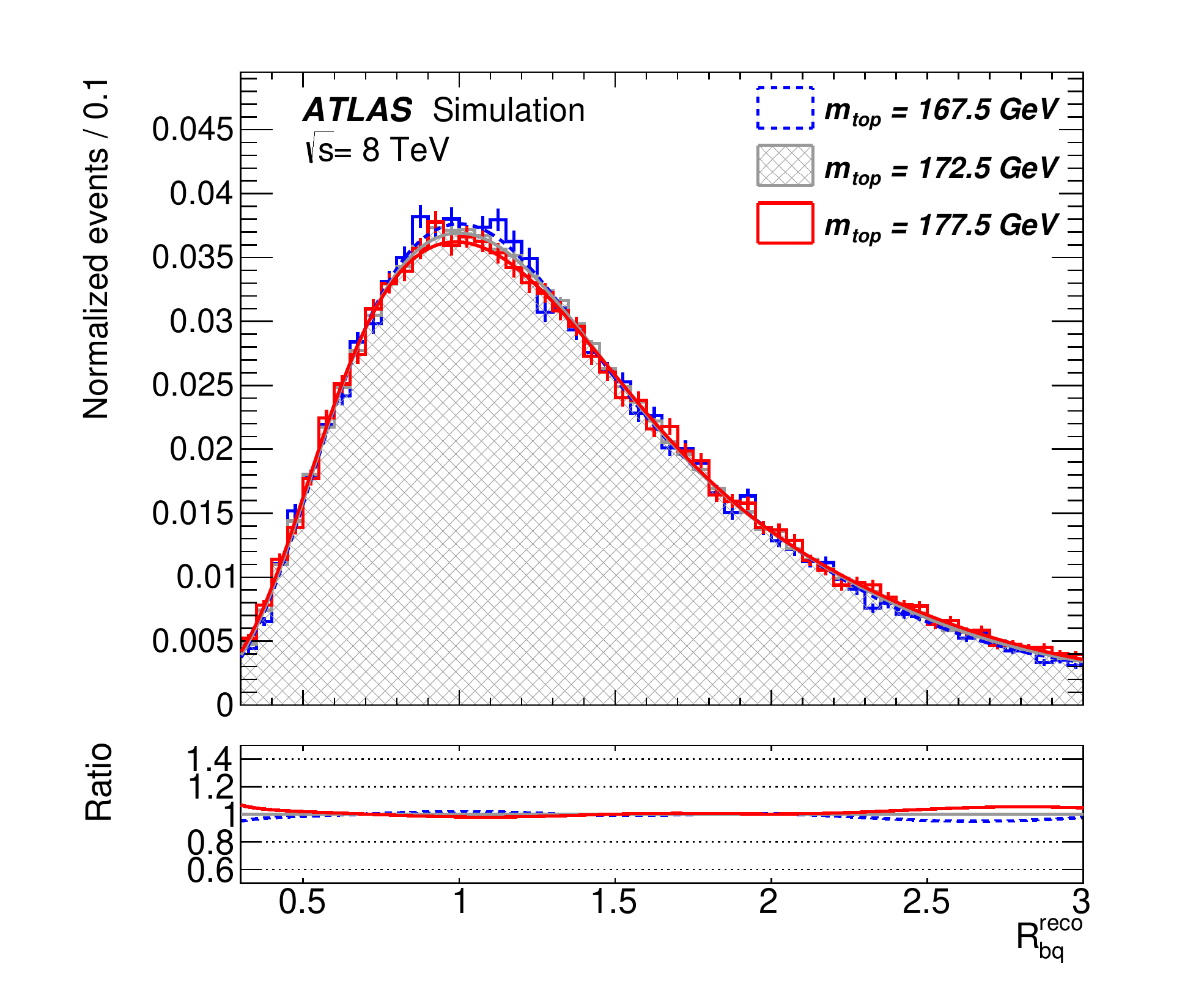}
            \label{fig:fig_04e}}
\subfigure[\rbqr\ with three input \bJSF]
          {\includegraphics[width=0.47\textwidth]{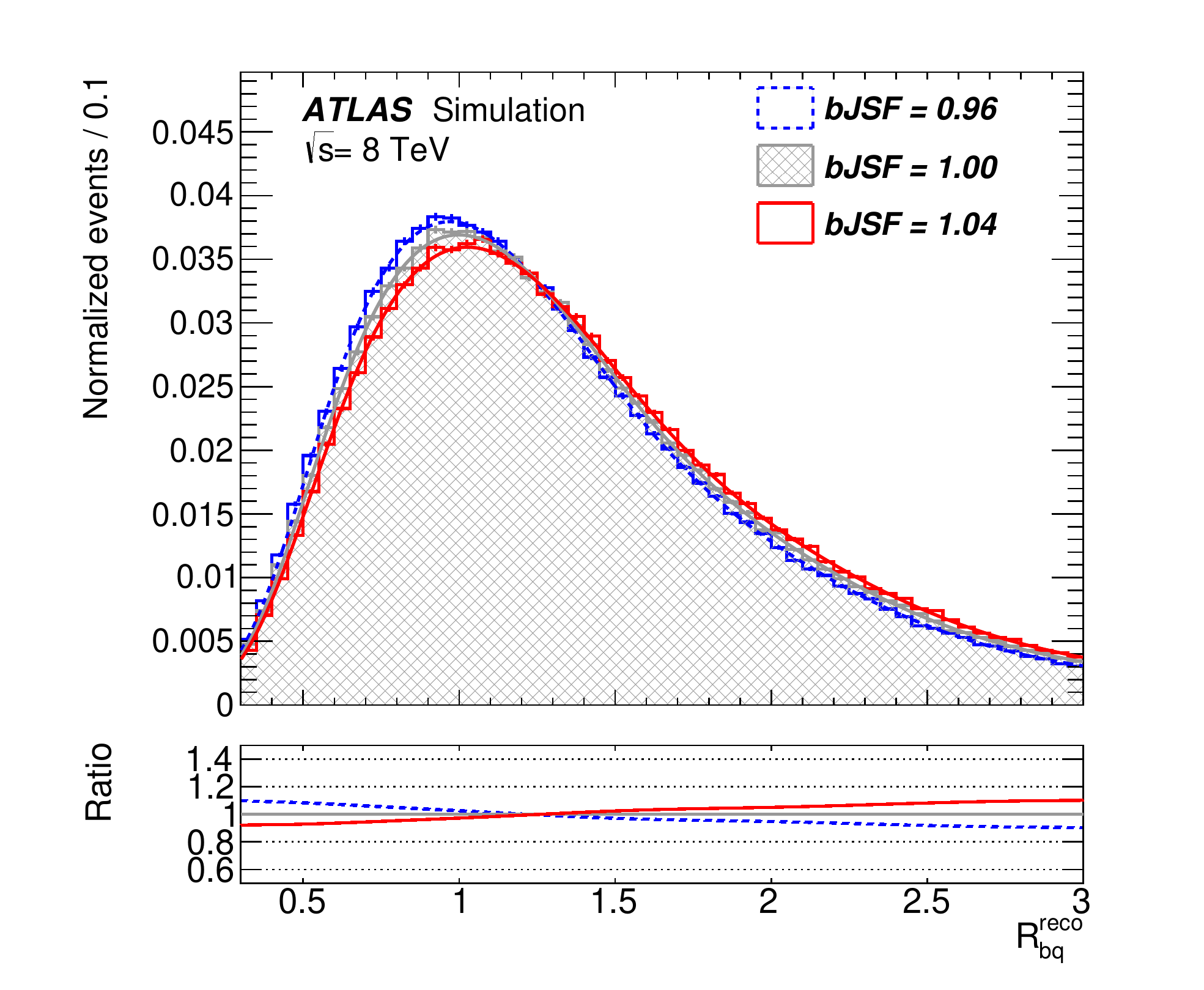}
            \label{fig:fig_04f}}
\caption{Template parameterizations for signal events, composed of \ttbar\ and
  single-top-quark production events.
 Figures~(a)--(c) show the sensitivity of \mtr\ to \mt, \JSF and \bJSF,
 figure~(d) shows the sensitivity of \mWr\ to \JSF and
 figures~(e) and (f) show the sensitivity of \rbqr\ to \mt\ and \bJSF.
 Each template is overlaid with the corresponding probability density
 function from the combined fit to all templates described in the text.
 The ratios shown are calculated relative to the probability density function of
 the central sample with $\mt = 172.5$~\GeV, $\JSF = 1$ and $\bJSF = 1$.
 \label{fig:fig_04}
}
\end{figure*}

 In \Figrange{\ref{fig:fig_04a}}{\ref{fig:fig_04c}}, the sensitivity of \mtr\ to
 the fit parameters \mt, \JSF\ and \bJSF\ is shown by the superposition of the
 signal templates and their fits for three input values per varied parameter.
 In a similar way, the sensitivity of \mWr\ to \JSF\ is shown in
 \Fig{\ref{fig:fig_04d}}.
 The dependences of \mWr\ on the input values of \mt\ and \bJSF\ are negligible
 and are not shown.
 Consequently, to increase the size of the simulation sample, the fit is
 performed on the sum of the \mWr\ distributions of the samples with different
 input top quark masses.
 Finally, the sensitivity of \rbqr\ to the input values of \mt\ and \bJSF\ is
 shown in \Figs{\ref{fig:fig_04e}}{\ref{fig:fig_04f}}.
 The dependence of \rbqr\ on \JSF~(not shown) is much weaker than the dependence
 on \bJSF.
%

 For the signal, the parameters of the fitting functions for \mtr\ depend
 linearly on \mt, \JSF\ and \bJSF. The parameters of the fitting functions for
 \mWr\ depend linearly on \JSF. Finally, the parameters of the fitting functions
 for \rbqr\ depend linearly on \mt, \JSF\ and \bJSF.
 For the background, the dependences of the parameters of the fitting functions
 are identical to those for the signal, except that they do not depend on
 \mt\ and that those for \rbqr\ do not depend on \JSF.

 Signal and background probability density functions \Ptopsig\ and \Ptopbkg\ for
 the \mtr, \mWr\ and \rbqr\ distributions are used in an unbinned likelihood fit
 to the data for all events, $i=1,\dots N$.
 The likelihood function maximized is
%
\begin{linenomath*}
  \begin{align}
    \Like_\mathrm{shape}^{\elljets}(\mt, \JSF, \bJSF, \fbkg)  &=
    \prod_{i=1}^{N} \Ptop(\mtri\,\vert\,\mt, \JSF, \bJSF, \fbkg) \nonumber \\
    &\quad\quad\times\,\PW(\mWri\,\vert\,\JSF, \fbkg) \nonumber \\
    &\quad\quad\times\,\Prbq(\rbqri\,\vert\,\mt, \JSF, \bJSF, \fbkg),
    \label{Eq:LikeLJ}
  \end{align}
\end{linenomath*}
%
\noindent with
%
\begin{linenomath*}
  \begin{align*}
    \Ptop(\mtri\,\vert\,\mt, \JSF, \bJSF, \fbkg) &=
    (1-\fbkg)\cdot\Ptopsig(\mtri\,\vert\,\mt, \JSF, \bJSF) + \\
    &\quad\quad\quad\,\,\,
    \fbkg\cdot\Ptopbkg(\mtri\,\vert\,\JSF, \bJSF)\,\,, \\
    \PW(\mWri\,\vert\,\JSF, \fbkg) &= 
    (1-\fbkg)\cdot\PWsig(\mWri\,\vert\,\JSF) + \\
    & \quad\quad\quad\,\,\, \fbkg\cdot\PWbkg(\mWri\,\vert\,\JSF)\,\,, \textrm{ and} \\
    \Prbq(\rbqri\,\vert\,\mt, \JSF, \bJSF, \fbkg)
    &= (1-\fbkg)\cdot\Prbqsig(\rbqri\,\vert\,\mt, \JSF, \bJSF) + \\
    & \quad\quad\quad\,\,\, \fbkg\cdot\Prbqbkg(\rbqri\,\vert\, \bJSF)\,\,
  \end{align*}
\end{linenomath*}
%
 where the fraction of background events is denoted by \fbkg.
 The parameters determined by the fit are \mt, \JSF\ and \bJSF, while \fbkg\ is
 fixed to its expectation shown in \Tab{\ref{tab:selections}}. 
 It was verified that the correlations between \mtr, \mWr\ and \rbqr\ of
 $\rho(\mtr, \mWr)= \rhomtrmwr$, $\rho(\mtr, \rbqr)= \rhomtrrbqr$, and
 $\rho(\mWr, \rbqr)= \rhomWrrbqr$, are small enough that formulating the
 likelihood in \Eqn{(\ref{Eq:LikeLJ})} as a product of three one-dimensional
 likelihoods does not bias the result.

 Pseudo-experiments are used to verify the internal consistency of the fitting
 procedure and to obtain the expected statistical uncertainty for the data. For
 each set of parameter values, \NPexpUnc\ pseudo-experiments are performed, each
 corresponding to the integrated luminosity of the data.
 To retain the correlation of the three observables for the three-dimensional
 fit, individual events are used. Because this exceeds the number of available
 MC events, results are corrected for oversampling~\cite{BAR-0001}.
 The results of pseudo-experiments for different input values of \mt\ are
 obtained from statistically independent samples, while the results for
 different \JSF\ and \bJSF\ are obtained from statistically correlated samples
 as explained above.
 For each fitted quantity and each variation of input parameters, the residual,
 i.e.~the difference between the input value and the value obtained by the fit,
 is compatible with zero.
 The three expected statistical uncertainties are
%
\begin{linenomath*}
\begin{alignat*}{5}
\sigmaX{\mathrm{stat}}(\mt) &={}& &\Expstatuncmeamtop &{}\pm{}& \Expstatuncrmsmtop~\GeV, \\
\sigmaX{\mathrm{stat}}(\JSF) &={}& &\ExpstatuncmeaJSF &{}\pm{}& \ExpstatuncrmsJSF\,, \textrm{ and}\\
\sigmaX{\mathrm{stat}}(\bJSF) &={}& &\ExpstatuncmeabJSF &{}\pm{}& \ExpstatuncrmsbJSF\,, 
\end{alignat*}
\end{linenomath*}
%
 where the values quoted are the mean and RMS of the distribution of the
 statistical uncertainties in the fitted quantities from pseudo-experiments.
 The widths of the pull distributions are below unity for \mt\ and the two jet
 scale factors, which results in an overestimation of the uncertainty in \mt\ of
 up to 7$\%$.
 Since this leads to a conservative estimate of the uncertainty in \mt, no
 attempts to mitigate this feature are made.

\section{Uncertainties affecting the \boldmath$\mt$ determination}
\label{sect:unc}
%
\begin{table*}[tb!]
\caption{Systematic uncertainties in \mt.
 The measured values of \mt\ are given together with the statistical and
 systematic uncertainties in \GeV\ for the standard and the BDT event
 selections.
 For comparison, the result in the \ttbarlj\ channel at $\sqrts=7$~\TeV\ from
 \Ref{\protect\cite{TOPQ-2013-02}} is also listed.
 For each systematic uncertainty listed, the first value corresponds to the
 uncertainty in \mt, and the second to the statistical precision in this
 uncertainty.
 An integer value of zero means that the corresponding uncertainty
 is negligible and therefore not evaluated.
 Statistical uncertainties quoted as 0.00 are smaller than 0.005.
 The statistical uncertainty in the total systematic uncertainty is calculated
 from uncertainty propagation.
 The last line refers to the sum in quadrature of the statistical and systematic
 uncertainties.
\label{tab:LpJresults8TeV}
}
\small
\begin{center}
\begin{tabular}{|l|r|r|r|}\cline{2-4}
\multicolumn{1}{c|}{}                  & $\sqrts=7$~\TeV     & \multicolumn{2}{c|}{$\sqrts=8$~\TeV} \\\hline
                       Event selection & Standard            &         Standard &              BDT  \\\hline
                    \mt\ result [\GeV] &        \SevenStaVal &     \EightStaVal &     \EightLpJVal  \\\hline
                            Statistics &        \SevenStaSta &     \EightStaSta &     \EightLpJSta  \\
        {\it $~~$-- Stat. comp. (\mt)} &      \SevenStaStamt &   \EightStaStamt &   \EightLpJStamt  \\
       {\it $~~$-- Stat. comp. (\JSF)} &     \SevenStaStaJSF &  \EightStaStaJSF &  \EightLpJStaJSF  \\
      {\it $~~$-- Stat. comp. (\bJSF)} &    \SevenStaStabJSF & \EightStaStabJSF & \EightLpJStabJSF  \\
                                Method & 0.11 $\pm$ 0.10     &  0.04 $\pm$ 0.11 & 0.13 $\pm$ 0.11   \\\hline
          Signal Monte Carlo generator & 0.22 $\pm$ 0.21     &  0.50 $\pm$ 0.17 & 0.16 $\pm$ 0.17   \\
                         Hadronization & 0.18 $\pm$ 0.12     &  0.05 $\pm$ 0.10 & 0.15 $\pm$ 0.10   \\
Initial- and final-state QCD radiation & 0.32 $\pm$ 0.06     &  0.28 $\pm$ 0.11 & 0.08 $\pm$ 0.11   \\
                      Underlying event & 0.15 $\pm$ 0.07     &  0.08 $\pm$ 0.15 & 0.08 $\pm$ 0.15   \\
                   Colour reconnection & 0.11 $\pm$ 0.07     &  0.37 $\pm$ 0.15 & 0.19 $\pm$ 0.15   \\
          Parton distribution function & 0.25 $\pm$ 0.00     &  0.08 $\pm$ 0.00 & 0.09 $\pm$ 0.00   \\\hline
              Background normalization & 0.10 $\pm$ 0.00     &  0.04 $\pm$ 0.00 & 0.08 $\pm$ 0.00   \\
                        $W$+jets shape & 0.29 $\pm$ 0.00     &  0.05 $\pm$ 0.00 & 0.11 $\pm$ 0.00   \\
                    Fake leptons shape & 0.05 $\pm$ 0.00     &                0 &               0   \\\hline
                      Jet energy scale & 0.58 $\pm$ 0.11     &  0.63 $\pm$ 0.02 & 0.54 $\pm$ 0.02   \\
Relative $b$-to-light-jet energy scale & 0.06 $\pm$ 0.03     &  0.05 $\pm$ 0.01 & 0.03 $\pm$ 0.01   \\
                 Jet energy resolution & 0.22 $\pm$ 0.11     &  0.23 $\pm$ 0.03 & 0.20 $\pm$ 0.04   \\
         Jet reconstruction efficiency & 0.12 $\pm$ 0.00     &  0.04 $\pm$ 0.01 & 0.02 $\pm$ 0.01   \\
                   Jet vertex fraction & 0.01 $\pm$ 0.00     &  0.13 $\pm$ 0.01 & 0.09 $\pm$ 0.01   \\

                                 \btag & 0.50 $\pm$ 0.00     &  0.37 $\pm$ 0.00 & 0.38 $\pm$ 0.00   \\
                               Leptons & 0.04 $\pm$ 0.00     &  0.16 $\pm$ 0.01 & 0.16 $\pm$ 0.01   \\
           Missing transverse momentum & 0.15 $\pm$ 0.04     &  0.08 $\pm$ 0.01 & 0.05 $\pm$ 0.01   \\
                           Pile-up & 0.02 $\pm$ 0.01     &  0.14 $\pm$ 0.01 & 0.15 $\pm$ 0.01   \\\hline
         Total systematic uncertainty  & \SevenStaSys $\pm$ \SevenStaSysUnc
                                       & \EightStaSys $\pm$ \EightStaSysUnc
                                       & \EightLpJSys $\pm$ \EightLpJSysUnc \\\hline
                                Total  & \SevenStaTot $\pm$ \SevenStaTotUnc
                                       & \EightStaTot $\pm$ \EightStaTotUnc
                                       & \EightLpJTot $\pm$ \EightLpJTotUnc \\\hline

\end{tabular}
\end{center}
\end{table*}
%
 This section focuses on the treatment of uncertainty sources of a systematic
 nature.
 The same systematic uncertainty sources as in \Ref{\cite{TOPQ-2013-02}} are
 investigated.
 If possible, the corresponding uncertainty in \mt\ is evaluated by varying the
 respective quantities by $\pm 1 \sigma$ from their default values, constructing
 the corresponding event sample and measuring the average \mt\ change relative
 to the result from the nominal MC sample with \NPexpUnc\ pseudo-experiments
 each, drawn from the full MC sample.
 In the absence of a $\pm 1 \sigma$ variation, e.g.~for the evaluation of the
 uncertainty induced by the choice of signal MC generator, the full observed
 difference is assigned as a symmetric systematic uncertainty and further
 treated as a variation equivalent to a $\pm 1 \sigma$ variation.
 Wherever a $\pm 1 \sigma$ variation can be performed, half the observed
 difference between the $+1\sigma$ and $-1\sigma$ variation in \mt\ is assigned
 as an uncertainty if the \mt\ values obtained from the variations lie on
 opposite sides of the nominal result.
 If they lie on the same side, the maximum observed difference is taken as a
 symmetric systematic uncertainty.
 Since the systematic uncertainties are derived from simulation or data samples
 with limited numbers of events, all systematic uncertainties have a
 corresponding statistical uncertainty, which is calculated taking into account
 the statistical correlation of the considered samples, as explained in
 \Sect{\ref{sect:syststat}}.
 The statistical uncertainty in the total systematic uncertainty is dominated by
 the limited sizes of the simulation samples.
 The resulting systematic uncertainties are given in
 \Tab{\ref{tab:LpJresults8TeV}} independent of their statistical significance.
 Further information is given in
 \Tabrange{\ref{tab:LpJresults8TeVbreak}}{\ref{tab:btagresults8TeV}} in
 \App{\ref{sect:addlpj}}.
 This approach follows the suggestion in \Ref{\cite{Barlow:2002yb}} and relies
 on the fact that, given a large enough number of considered uncertainty
 sources, statistical fluctuations average out.\footnote{In the limit of many
   small systematic uncertainties with large statistical uncertainties, this
   procedure on average leads to an overestimate of the total systematic
   uncertainty.}
 The uncertainty sources are designed to be uncorrelated with each other, and
 thus the total uncertainty is taken as the sum in quadrature of uncertainties
 from all sources.
 The individual uncertainties are compared in \Tab{\ref{tab:LpJresults8TeV}} for
 three cases: the standard selection for the $\sqrts=7$~\TeV~\cite{TOPQ-2013-02}
 and 8~\TeV\ data and the \mvabased\ selection for $\sqrts=8$~\TeV\ data.
 Many uncertainties in \mt\ obtained with the standard selection at the two
 \cmes\ agree within their statistical uncertainties such that the resulting
 total systematic uncertainties are almost identical.
 Consequently, repeating the $\sqrts=7$~\TeV\ analysis on $\sqrts=8$~\TeV\ data
 would have only improved the statistical precision.
 The picture changes when comparing the uncertainties in $\sqrts=8$~\TeV\ data
 for the standard selection and the \mvabased\ selection. In general, the
 experimental uncertainties change only slightly, with the largest reduction
 observed for the JES uncertainty.
 In contrast, a large improvement comes from the reduced uncertainties in the
 modelling of the \ttbar\ signal processes as shown in
 \Tab{\ref{tab:LpJresults8TeV}}.
 This, together with the improved intrinsic resolution in \mt, more than
 compensates for the small loss in precision caused by the increased statistical
 uncertainty.
 The individual sources of systematic uncertainties and the evaluation of their
 effect on \mt\ are described in the following.
%
%
 \subsection{Statistics and method calibration}
 Uncertainties related to statistical effects and the method calibration are
 discussed here.
 \paragraph{Statistical:}\mbox{}
 The quoted statistical uncertainty consists of three parts: a purely
 statistical component in \mt\ and the contributions stemming from the
 simultaneous determination of \JSF\ and \bJSF.
 The purely statistical component in \mt\ is obtained from a one-dimensional
 template method exploiting only the \mtr\ observable, while fixing the values
 of \JSF\ and \bJSF\ to the results of the three-dimensional analysis.
 The contribution to the statistical uncertainty in the fitted parameters due to
 the simultaneous fit of \mt\ and \JSF\ is estimated as the difference in
 quadrature between the statistical uncertainty in a two-dimensional fit to
 \mtr\ and \mWr while fixing the value of \bJSF and the one-dimensional fit to
 the data described above.
 Analogously, the contribution of the statistical uncertainty due to the
 simultaneous fit of \mt\ together with \JSF\ and \bJSF\ is defined as the
 difference in quadrature between the statistical uncertainties obtained in the
 three-dimensional and the two-dimensional fits to the data.
 This separation allows a comparison of the statistical sensitivities of the
 \mt\ estimator used in this analysis, to those of analyses exploiting a
 different number of observables in the fit.
 In addition, the sensitivity of the estimators to the global jet energy scale
 factors can be compared directly.
 These uncertainties are treated as uncorrelated uncertainties in
 \mt\ combinations. Together with the systematic uncertainty in the residual jet
 energy scale uncertainties discussed below, they directly replace the
 uncertainty in \mt\ from the jet energy scale variations present without the in
 situ determination.
 \paragraph{Method:}\mbox{}
 The residual difference between fitted and generated \mt\ when analysing a
 template from a MC sample reflects the potential bias of the method.
 Consequently, the largest observed fitted \mt\ residual and the largest
 observed statistical uncertainty in this quantity, in any of the five signal
 samples with different assumed values of \mt, is assigned as the method
 calibration uncertainty and its corresponding statistical uncertainty,
 respectively.
 This also covers effects from limited numbers of simulated events in the
 templates and potential deficiencies in the template parameterizations.
%
\subsection{Modelling of signal processes}
 The modelling of \ttbarlj\ events incorporates a number of processes that have
 to be accurately described, resulting in systematic effects, ranging from the
 \ttbar\ production to the hadronization of the showered objects.

 Thanks to the restrictive event-selection requirements, the contribution of
 non-\ttbar\ processes, comprising the single-top-quark process and the various
 background processes, is very low.
 The systematic uncertainty in \mt\ from the uncertainty in the single-top-quark
 normalization is estimated from the corresponding uncertainty in the
 theoretical cross-section given in \Sect{\ref{sect:mc}}.
 The resulting systematic uncertainty is small compared with the systematic
 uncertainty in the \ttbar\ production and is consequently neglected.
 For the modelling of the signal processes, the consequence of including
 single-top-quark variations in the uncertainty evaluation was investigated for
 various uncertainty sources and found to be negligible. Therefore, the
 single-top-quark variations are not included in the determination of the signal
 event uncertainties.
 \paragraph{Signal Monte Carlo generator:}\mbox{}
 The full observed difference in fitted \mt\ between the event samples produced
 with the \PowhegBox\ and \Mcatnlo~\cite{FRI-0201,FRI-0301} programs is quoted
 as a systematic uncertainty.
 For the renormalization and factorization scales the \PowhegBox\ sample uses
 the function given in \Sect{\ref{sect:mc}}, while the \Mcatnlo\ sample uses
 $\murf=\sqrt{\mt^2 + 0.5 (p_{{\rm T},t}^2 + p_{{\rm T}, \bar{t}}^2)}$.
 Both samples are generated with a top quark mass of $\mt=172.5$~\GeV\ with the
 CT10 PDFs in the matrix-element calculation and use the \HERWIG\ and
 \Jimmy\ programs with the ATLAS \AUETt\ tune~\cite{ATL-PHYS-PUB-2011-008}.
 \paragraph{Hadronization:}\mbox{}
 To cover the choice of parton shower and hadronization models, samples produced
 with the \PowhegBox\ program are showered with either the \Pythiasix\ program
 using the P2011C tune or the \HERWIG\ and \Jimmy\ programs using the ATLAS
 \AUETt\ tune.
 This includes different approaches in shower modelling, such as using a
 \pt-ordered parton showering in the \Pythia\ program or angular-ordered parton
 showering in the \HERWIG\ program, the different parton shower matching scales,
 as well as fragmentation functions and hadronization models, such as choosing
 the Lund string model~\cite{Andersson198331,Andersson:1998tv} implemented in
 the \Pythia\ program or the cluster fragmentation model~\cite{Webber:1983if}
 used in the \HERWIG\ program.
 The full observed difference between the samples is quoted as a systematic
 uncertainty.

 As shown in \Fig{\ref{fig:fig_01}}, the distributions of transverse momenta in
 data are slightly softer than those in the \PowhegPythia\ MC simulation
 samples.
 Similarly to what was observed in the \ttbarll\ channel for the
 \ptlb\ distribution, in the \ttbarlj\ channel the \PowhegHerwig\ sample is
 much closer to the data for several distributions of transverse momenta.
 The \pthad\ distribution is much better described by the \PowhegHerwig\ sample
 as was also observed in Ref.~\cite{TOPQ-2015-06}.
 In addition, but to a lesser extent, the \Mcatnlo\ sample used to assess the
 signal Monte Carlo generator uncertainty and the samples to assess the initial-
 and final-state QCD radiation uncertainty discussed next also lead to a softer
 distribution in simulation.
 Given this, the observed difference in the \pthad\ distribution is covered by a
 combination of the signal-modelling uncertainties given in
 \Tab{\ref{tab:LpJresults8TeV}}.

 Despite the fact that the \JES\ and \bJES\ are estimated independently using
 dijet and other non-\ttbar\ samples~\cite{PERF-2012-01}, some double-counting
 of hadronization-uncertainty-induced uncertainties in the \JES\ and \mt\ cannot
 be excluded.
 This was investigated closely for the ATLAS top quark mass measurement in the
 \ttbarlj\ channel at $\sqrts=7$~\TeV. The results in
 \Ref{\cite{ATL-PHYS-PUB-2015-042}} revealed that the amount of double-counting
 of \JES\ and hadronization effects for the \ttbarlj\ channel is small.
 \paragraph{Initial- and final-state QCD radiation~(ISR/FSR):}\mbox{}
 ISR/FSR leads to a higher jet multiplicity and different jet energies than the
 hard process, which affects the distributions of the three observables.
 The uncertainties due to ISR/FSR modelling are estimated with samples generated
 with the \PowhegBox\ program interfaced to the \Pythiasix\ program for which
 the parameters of the generation are varied to span the ranges compatible with
 the results of measurements of \ttbar\ production in association with
 jets~\cite{TOPQ-2011-21,TOPQ-2012-03,ATL-PHYS-PUB-2015-002}.
 This uncertainty is evaluated by comparing two dedicated samples that differ in
 several parameters, namely the QCD scale \LambdaQCD, the transverse momentum
 scale for space-like parton-shower evolution $Q^2_\mathrm{max}$, the
 \hdamp\ parameter~\cite{ATL-PHYS-PUB-2014-005} and the P2012~\RadLo\ and
 \RadHi\ tunes~\cite{Skands}.
 In \Ref{\cite{ATL-PHYS-PUB-2015-002}}, it was shown that a number of
 final-state distributions are better accounted for by the
 \PowhegPythia\ samples with $\hdamp=\mt$. Therefore, these samples are used for
 evaluating this uncertainty, taking half the observed difference between the up
 variation and the down variation sample.
 Because the parameterizations for the template fit to data are obtained from
 \PowhegPythia\ samples using $\hdamp=\infty$, it was verified that, considering
 the method uncertainty quoted in \Tab{\ref{tab:LpJresults8TeV}}, applying the
 same functions to the $\hdamp=\mt$ samples leads to a result compatible with
 the input top quark mass.
 \paragraph{Underlying event:}\mbox{}
 To reduce statistical fluctuations in the evaluation of this systematic
 uncertainty, the difference in underlying-event modelling is assessed by
 comparing a pair of \PowhegBox\ samples based on the same partonic events
 generated with the CT10 PDFs.
 A sample with the P2012 tune is compared with a sample with the
 P2012~\mpiHi\ tune~\cite{Skands}, with both tunes using the same CTEQ6L1
 PDFs~\cite{cteq5l} for parton showering and hadronization.
 The Perugia 2012~\mpiHi\ tune provides more semi-hard multiple parton
 interactions and is used for this comparison with identical colour reconnection
 parameters in both tunes. The full observed difference is assigned as a
 systematic uncertainty.
 \paragraph{Colour reconnection:}\mbox{}
 This systematic uncertainty is estimated using a pair of samples with the same
 partonic events as for the underlying-event uncertainty evaluation but with the
 P2012 tune and the P2012~\loCR\ tune~\cite{Skands} for parton showering and
 hadronization.
 The full observed difference is assigned as a systematic uncertainty.
 \paragraph{Parton distribution function~(PDF):}\mbox{}
 The PDF systematic uncertainty is the sum in quadrature of three contributions.
 These are the sum in quadrature of the differences in fitted \mt\ for the
 26~eigenvector variations of the CT10 PDF and two differences in \mt\ obtained
 from reweighting the central CT10 PDF set to the MSTW2008 PDF~\cite{MAR-0901}
 and the NNPDF2.3 PDF~\cite{Ball:2012cx}.
%
\subsection{Modelling of background processes}
 Uncertainties in the modelling of the background processes are taken into
 account by variations of the corresponding normalizations and shapes of the
 distributions.
 \paragraph{Background normalization:}\mbox{}
 The normalizations are varied for the data-driven background estimates
 according to their uncertainties. For the negligible contribution from diboson
 production, no normalization uncertainty is evaluated.
 \paragraph{Background shape:}\mbox{}
 For the \Wj\ background, the shape uncertainty is evaluated from the variation
 of the heavy-flavour fractions. The corresponding uncertainty is small.
 Given the very small contribution from \Zj, diboson and \fakedash\ backgrounds,
 no shape uncertainty is evaluated for these background sources.
%
\subsection{Detector modelling} 
\label{sect:detectormodelling7TeV}
 The level of understanding of the detector response and of the particle
 interactions therein is reflected in numerous systematic uncertainties.
 \paragraph{Jet energy scale~(JES):}\mbox{}
 The \JES\ is measured with a relative precision of about $1\%$ to $4\%$,
 typically falling with increasing jet \pt\ and rising with increasing jet
 $\vert\eta\vert$~\cite{ATLAS-CONF-2015-057,ATLAS-CONF-2015-017}.
 The total \JES\ uncertainty consists of more than 60 subcomponents originating
 from the various steps in the jet calibration.
 The number of these nuisance parameters is reduced with a matrix
 diagonalization of the full \JES\ covariance matrix including all nuisance
 parameters for a given category of the \JES\ uncertainty components.

 The analyses of $\sqrts=7$~\TeV\ and $\sqrts=8$~\TeV\ data make use of the
 EM+JES and LCW+GSC~\cite{ATLAS-CONF-2015-057} jet calibrations, respectively.
 The two calibrations feature different sets of nuisance parameters, and the
 LCW+GSC calibration generally has smaller uncertainties than the EM+JES
 calibration.
 While the pile-up correction for the jet calibration for $\sqrts=7$~\TeV\ data
 only depends on the number of primary vertices~(\nvtx) and the mean number of
 interactions per bunch crossing~($\mu$), a pile-up subtraction method based on
 jet area is introduced for the $\sqrts=8$~\TeV\ data.
 Terms to account for uncertainties in the pile-up estimation are added. They
 depend on the jet \pt\ and the local transverse momentum density. 
 In addition, the punch-through uncertainty, i.e.~an uncertainty for jets that
 penetrate through to the muon spectrometer, is added.
 The final reduced number of nuisance parameters for the
 $\sqrts=8$~\TeV\ analysis is 25.
 The JES-uncertainty-induced uncertainty in \mt\ is the dominant systematic
 uncertainty for all results shown in \Tab{\ref{tab:LpJresults8TeV}}.
 When only a one-dimensional fit to \mtr\ or a two-dimensional fit to \mtr\ and
 \mWr\ is done, this uncertainty is \JESUnctoned~\GeV\ or \JESUnctd~\GeV,
 respectively.
 \paragraph{Relative $b$-to-light-jet energy scale~(bJES):}\mbox{}
 The \bJES\ uncertainty is an additional uncertainty for the remaining
 differences between \bjets\ and \ljets\ after the global \JES\ is applied, and
 therefore the corresponding uncertainty is uncorrelated with the
 \JES\ uncertainty.
 An additional uncertainty of $0.2\%$ to $1.2\%$ is assigned to \bjets, with
 the lowest uncertainty for \bjets\ with high transverse
 momenta~\cite{PERF-2012-01}.
 Due to the determination of \bJSF, the \bJES\ uncertainty leads to a very small
 contribution to the uncertainty in \mt\ in \Tab{\ref{tab:LpJresults8TeV}}.
 However, performing only a two-dimensional fit to \mtr\ and \mWr\ would result
 in an uncertainty of \bJESUnctd~\GeV\ from this source.
 \paragraph{Jet energy resolution~(JER):}\mbox{}
 The JER uncertainty is determined following an eigenvector decomposition
 strategy similar to the \JES\ systematic
 uncertainties~\cite{ATLAS-CONF-2015-057,ATLAS-CONF-2015-017}. The 11 components
 take into account various effects evaluated from simulation-to-data comparisons
 including calorimeter noise terms in the forward region.
 The corresponding uncertainty in \mt\ is the sum in quadrature of the
 components of the eigenvector decomposition.
 \paragraph{Jet reconstruction efficiency~(JRE):}\mbox{}
 This uncertainty is evaluated by randomly removing $0.23\%$ of the jets with
 $\pt< 30$~\GeV\ from the simulated events prior to the event selection to
 reflect the precision with which the data-to-simulation JRE ratio is
 known~\cite{PERF-2011-03}.
 The fitted \mt\ difference between the varied sample and the nominal sample is
 taken as a systematic uncertainty.
 \paragraph{Jet vertex fraction~(JVF):}\mbox{}
 When summing the scalar \pt\ of all tracks in a jet, the JVF is the fraction
 contributed by tracks originating at the primary vertex.
 The uncertainty in \mt\ is evaluated by varying the requirement on the JVF
 within its uncertainty~\cite{PERF-2014-03}.
 \paragraph{\boldmath$\btag$:}\mbox{}
 Mismodelling of the \btag\ efficiency and mistag rate is accounted for by the
 application of jet-specific scale factors to simulated
 events~\cite{PERF-2012-04}.
 These scale factors depend on jet \pt, jet $\eta$ and the underlying quark
 flavour.
 The ones used in this analysis are derived from dijet and
 \ttbarll~\cite{PERF-2012-04} events.
 They are the same as those used for the measurement of \mt\ in the
 \ttbarll\ channel~\cite{TOPQ-2016-03}.
 Similarly to the \JES\ uncertainties, the \btag\ uncertainties are estimated by
 using an eigenvector approach, based on the \btag\ calibration
 analysis~\cite{PERF-2012-04,ATLAS-CONF-2014-004,ATLAS-CONF-2014-046}.
 They include the uncertainties in the \btag, $c/\tau$-tagging and mistagging
 scale factors.
 This uncertainty in \mt\ is derived by varying the scale factors within their
 uncertainties and adding the resulting fitted differences in quadrature.
 In this procedure, uncertainties that are considered both in the
 \btag\ calibration and as separate sources in the \mt\ analysis are taken into
 account simultaneously by applying the corresponding varied \btag\ scale
 factors together with the varied sample when assessing the corresponding
 uncertainty in \mt.
 The final uncertainty is the sum in quadrature of these independent
 components. Compared with the result from $\sqrts=7$~\TeV\ data, this
 uncertainty is reduced by about one third for both the standard and
 \mvabased\ event selections in accordance with the improvements made in the
 calibrations of the \btag\ algorithm~\cite{PERF-2012-04,ATLAS-CONF-2014-004}.
 \paragraph{Leptons:}\mbox{}
 The lepton uncertainties are related to the electron energy or muon momentum
 scale and resolution, as well as trigger, isolation and identification
 efficiencies.
 These are measured very precisely in high-purity $\jpsi\to \ellell$ and
 $Z\to\ellell$ data~\cite{PERF-2013-05,PERF-2014-05,PERF-2016-01}.
 For each component, the corresponding uncertainty is propagated to the analysis
 by variation of the respective quantity. The changes are propagated to the
 \met\ as well.
 \paragraph{Missing transverse momentum:}\mbox{}
 The remaining contribution to the missing-transverse-momentum uncertainty stems
 from the uncertainties in calorimeter-cell energies associated with
 low-\pt\ jets~($7~\GeV< \pt < 20~\GeV$) without any corresponding reconstructed
 physics object or from pile-up interactions.
 They are accounted for as described in \Ref{\cite{PERF-2014-04}}.
 The corresponding uncertainty in \mt\ is small.
 \paragraph{Pile-up:}\mbox{}
 Besides the component treated in the \JES\ uncertainty, the residual dependence
 of the fitted \mt\ on the amount of pile-up activity and a possible
 mismodelling of pile-up in MC simulation is determined.
 For this, the \mt\ dependence in bins of \nvtx\ and $\mu$ is determined
 for data and MC simulated events.
 Within the statistical uncertainties, the slopes of the linear dependences of
 \mt\ observed in data and predicted by the MC simulation are compatible. The
 same is true for \JSF\ and \bJSF.
 The final effect on the measurement is assessed by a convolution of the linear
 dependence with the respective \nvtx\ and  $\mu$ distributions observed for
 data and MC simulated events.
 The maximum of the \nvtx\ and $\mu$ effects is assigned as an uncertainty due
 to pile-up.
 The pile-up conditions differ between the $\sqrts=7$ and 8~\TeV\ data.
 For the BDT selection of $\sqrts=8$~\TeV\ data used here, the average of the
 mean number of inelastic \pp\ interactions per bunch crossing is $\meanmu=20.3$
 and the average number of reconstructed primary vertices is about $\nvtx=9.4$,
 to be compared with $\meanmu=8.8$ and $\nvtx=7.0$ for
 $\sqrts=7$~\TeV\ data~\cite{PERF-2014-03}.
 The corresponding uncertainty is somewhat larger than for $\sqrts=7$~\TeV\ data
 but still small.
%
\subsection{Statistical precision of systematic uncertainties}
\label{sect:syststat}
 The systematic uncertainties quoted in \Tab{\ref{tab:LpJresults8TeV}} carry
 statistical uncertainties themselves. In view of a combination with other
 measurements, the statistical precision $\sigma$ from a comparison of two
 samples (1 and 2) is determined for each uncertainty source based on the
 statistical correlation \rhoX{12}\ of the underlying samples using $\sigmaXq{}=
 \sigmaXq{1}+\sigmaXq{2}-2\rhoX{12}\sigmaX{1}\sigmaX{2}$.
 The statistical correlation is expressed as a function of the fraction of
 shared events of both samples
 $\rhoX{12}=\sqrt{\N{12}/\N{1}\cdot\N{12}/\N{2}}=\N{12}/\sqrt{\N{1}\cdot\N{2}}$,
 with $\N{1}$ and $\N{2}$ being the unweighted numbers of events in the two
 samples and $\N{12}$ being the unweighted number of events present in both
 samples.
 The size of the MC sample at $\mt=172.5$~\GeV\ results in a statistical
 precision in \mt\ of about 0.1~\GeV.
 Most estimations are based on the same sample with only a change in a single
 parameter, such as lepton energy scale uncertainties.
 This leads to a high correlation of the central \mt\ values and a
 correspondingly low statistical uncertainty in their difference.
 Others, which do not share the same generated events or exhibit other
 significant differences, have a lower correlation, and the corresponding
 statistical uncertainty is higher, such as in the case of the signal-modelling
 uncertainty.
 The statistical uncertainty in the total systematic uncertainty is calculated
 from the individual statistical uncertainties by the propagation of
 uncertainties.

\section{Results}
\label{sect:result}
 For the \mvabased\ selection, the likelihood fit to the data results in
%
\begin{linenomath*}
\begin{alignat*}{5}
  \mt &={}& &\EightLpJVal &{}\pm{}& \EightLpJSta\,\mathrm{(stat)}\,\GeV\,, \\
 \JSF &={}& &\EightLpJJSFVal &{}\pm{}& \EightLpJJSFSta\,\mathrm{(stat)}\,, 
              \textrm{ and} \\
\bJSF &={}& &\EightLpJbJSFVal &{}\pm{}& \EightLpJbJSFSta\,\mathrm{(stat)}\,. 
\end{alignat*}
\end{linenomath*}
%
 The statistical uncertainties are taken from the parabolic approximation of the
 likelihood profiles. The expected statistical uncertainties, calculated in
 \Sect{\ref{sect:templates}}, are compatible with those.
 The correlation matrices of the three variables with $i$ = 0, 1 and 2
 corresponding to \mt, \JSF\ and \bJSF\ are
%
\begin{linenomath*}
  \begin{align*}
    \rhoX{\mathrm{stat}} &= \left (
    \begin{array}{rrr}
      1 & \EightLpJRhomtJSFStat &  \EightLpJRhomtbJSFStat \\
      \EightLpJRhomtJSFStat  & 1 & \EightLpJRhoJSFbJSFStat  \\
      \EightLpJRhomtbJSFStat & \EightLpJRhoJSFbJSFStat & 1
    \end{array} 
    \right)\quad\quad 
    \mbox{and}\quad\quad
    \rhoX{\mathrm{tot}}  = \left (
    \begin{array}{rrr}
      1 & \EightLpJRhomtJSFTot & \EightLpJRhomtbJSFTot  \\
      \EightLpJRhomtJSFTot  & 1 & \EightLpJRhoJSFbJSFTot  \\
      \EightLpJRhomtbJSFTot & \EightLpJRhoJSFbJSFTot & 1
    \end{array} 
    \right )\,. 
  \end{align*}
\end{linenomath*}
%
 The left matrix corresponds to the correlations for statistical uncertainties
 only, while the right matrix is obtained by additionally taking into account
 all systematic uncertainties.

\begin{figure*}[tbp!]
\centering
\subfigure[Reconstructed top quark mass]
          {\includegraphics[width=0.48\textwidth]{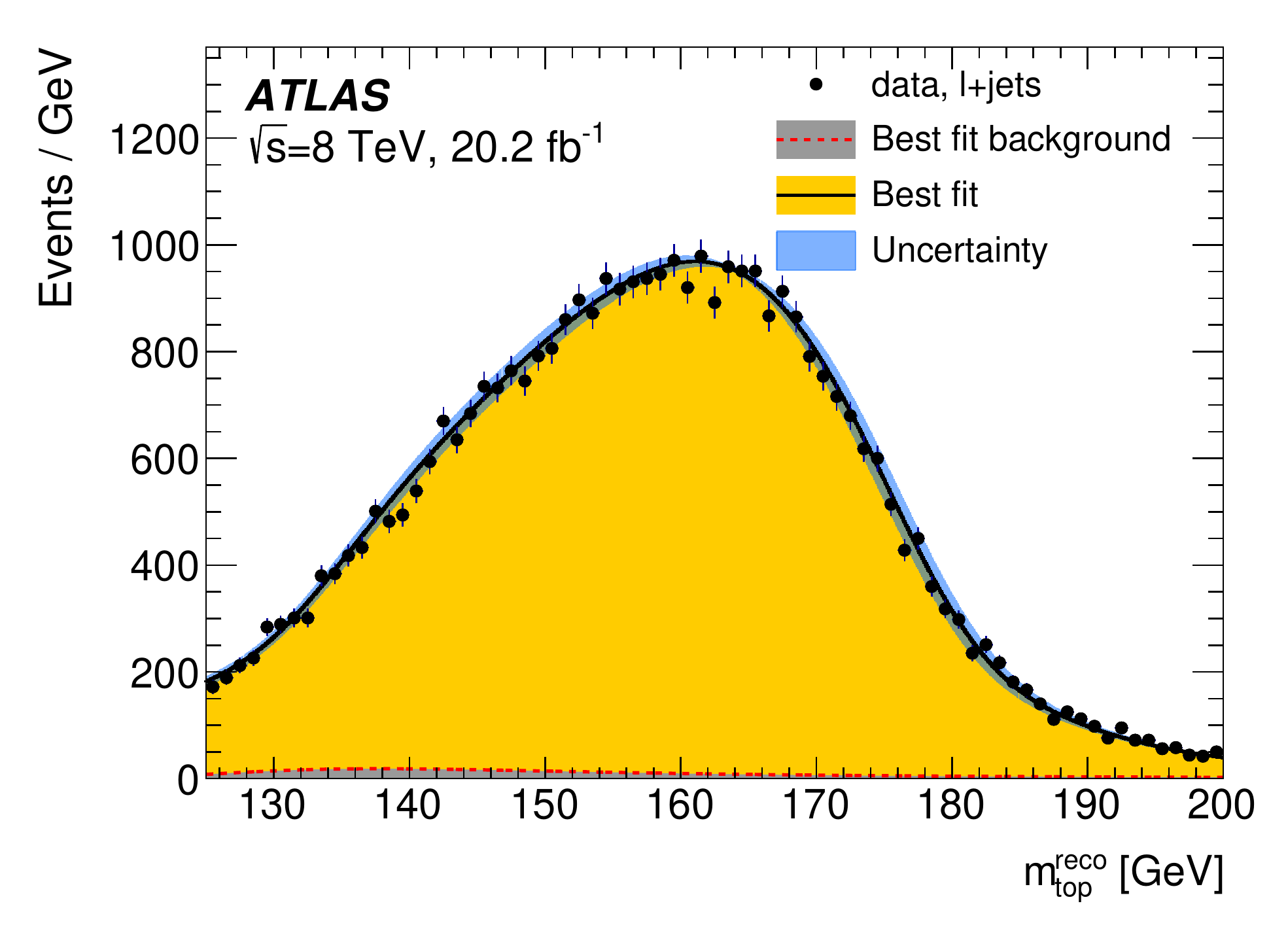}
            \label{fig:fig_05a}}
\subfigure[Reconstructed \Wboson\ boson mass]
          {\includegraphics[width=0.48\textwidth]{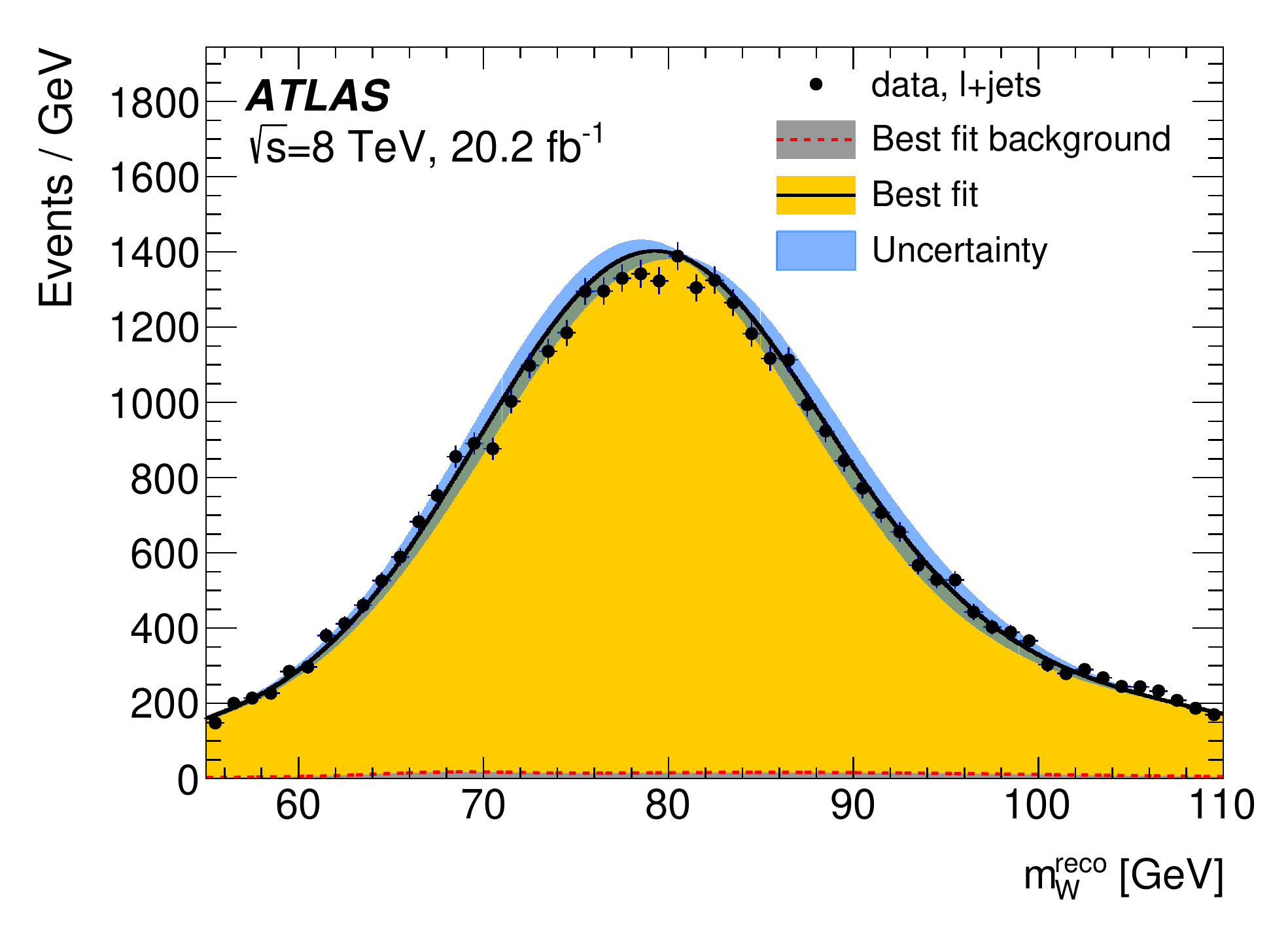}
            \label{fig:fig_05b}}
\hfill
\subfigure[Reconstructed ratio of jet transverse momenta]
          {\includegraphics[width=0.48\textwidth]{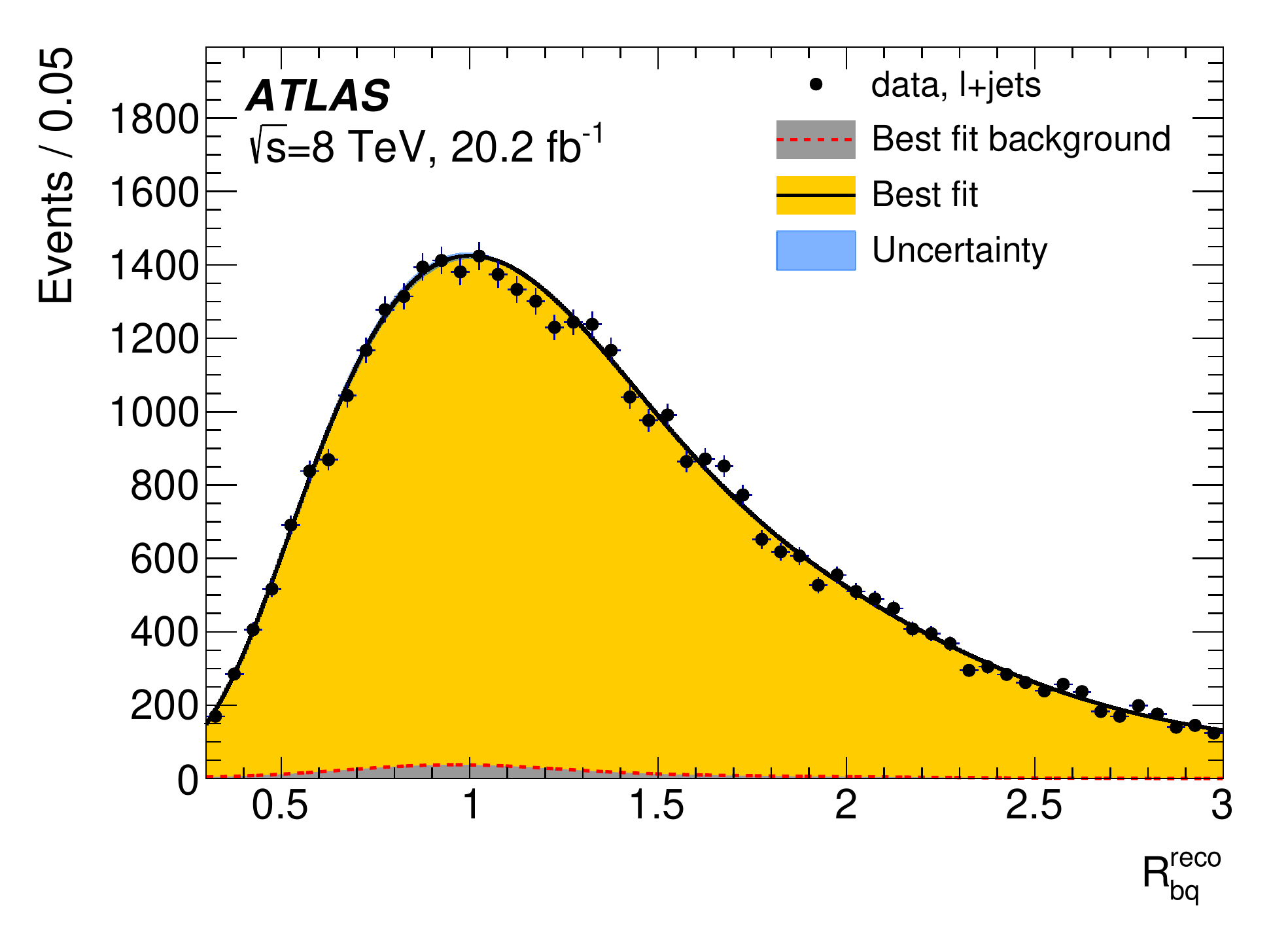}
            \label{fig:fig_05c}}
\caption{Results of the likelihood fit to the data.
 The figures show the data distributions of the three observables with
 statistical uncertainties together with the fitted probability density function
 for the background alone~(barely visible at the bottom of the figure) and for
 the sum of signal and background.
 The uncertainty band corresponds to the one standard deviation total
 uncertainty in the fit function. It is based on the total uncertainty in the
 three fitted parameters as explained in the text.
 Figure~(a) shows the distribution of the reconstructed top quark mass \mtr,
 figure~(b) shows the distribution of the reconstructed \Wboson\ boson mass \mWr
 and figure~(c) shows the reconstructed ratio of jet transverse momenta \rbqr.
\label{fig:fig_05}
}
\end{figure*}

 \Fig{\ref{fig:fig_05}} shows the \mtr, \mWr and \rbqr\ distributions in the
 data with statistical uncertainties together with the corresponding fitted
 probability density functions for the background alone and for the sum of
 signal and background.
%
%
 The uncertainty band attached to the fit to data is obtained in the following
 way.
 At each point in \mtr, \mWr and \rbqr, the band contains 68$\%$ of all fit
 function values obtained by randomly varying \mt, \JSF\ and \bJSF\ within their
 total uncertainties and taking into account their correlations.
 The waist in the uncertainty band is caused by the usage of normalized
 probability density functions.
 The band visualises the variations of the three template fit functions caused
 by all the uncertainties in \mt\ listed in \Tab{\ref{tab:LpJresults8TeV}}.
 The total uncertainty in all three fitted parameters is dominated by their
 systematic uncertainty. Therefore, the band shown is much wider than the band
 that would be obtained by fitting to the distributions with statistical
 uncertainties only.

 The measured value of \mt\ in the \ttbarlj\ channel at $\sqrts=8$~\TeV\ is
%
\begin{linenomath*}
  \begin{align*}
    \mt &=\XZ{\EightLpJVal}{\EightLpJSta}{\EightLpJSys}~\GeV
  \end{align*}
\end{linenomath*}
%
 with a total uncertainty of \EightLpJTot~\GeV. The statistical precision of the
 systematic uncertainty is \EightLpJSysUnc~\GeV.
 This result corresponds to a \EightLpJStandImp\ improvement on the result
 obtained using the standard selection on the same data.
 Compared with the result in the \ttbarlj\ channel at $\sqrts=7$~\TeV, the
 improvement is \EightLpJSevenImp.
 On top of the smaller statistical uncertainty, the increased precision is
 mainly driven by smaller theory modelling uncertainties achieved by the
 \mvabased\ selection.
 The larger number of events in the $\sqrts=8$~\TeV\ dataset is effectively
 traded for lower systematic uncertainties, resulting in a significant gain in
 total precision.
 The new ATLAS result in the \ttbarlj\ channel is more precise than the result
 from the CDF experiment, but less precise than the CMS and D0 results, measured
 in the same channel, as shown in \Fig{\ref{fig:TopMassLpJ}} in
 \App{\ref{sect:addcom}}.

\clearpage
\section{Combination with previous ATLAS results}
\label{sect:comb}
 This section presents the combination of the six \mt\ results of the ATLAS
 analyses in the \ttbarll, \ttbarlj\ and \ttbarjj\ channels at \cmes\ of
 $\sqrts=7$ and $8$~\TeV.
 The treatment of the results that are input to the combinations are described,
 followed by a detailed explanation of the evaluation of the estimator
 correlations for the various sources of systematic uncertainty.
 The compatibilities of the measured \mt\ values are investigated using a
 pairwise $\chiq$ for all pairs of measurements and by evaluating the
 compatibility of selected combinations.
 Finally, the six results are combined, displaying the effect of individual
 results on the combined result.
%
\subsection{Inputs to the combination and categorization of uncertainties}
\label{sect:combinp}
 The measured values of the individual analyses and their statistical and
 systematic uncertainties are given in \Tab{\ref{tab:results_indiv}}.
 For each result, the evaluated systematic uncertainties are shown together with
 their statistical uncertainties.
 The statistical uncertainties in the total systematic uncertainties and the
 total uncertainties are obtained from the propagation of
 uncertainties\footnote{For the previous results in the \ttbarll\ and
   \ttbarlj\ channels, the values quoted for the statistical uncertainties in
   the total systematic uncertainties differ from the ones in the original
   publications, where just the sum in quadrature of the statistical
   uncertainties in the individual systematic uncertainties was used.}.

 For the combinations to follow, the combined uncertainties for the previous
 results, namely $\ttbarll$ and $\ttbarlj$ at~$\sqrts=7$~\TeV\ from
 \Ref{\cite{TOPQ-2013-02}}, $\ttbarjj$ at~$\sqrts=7$~\TeV\ from
 \Ref{\cite{TOPQ-2013-03}}, $\ttbarll$ at~$\sqrts=8$~\TeV\ from
 \Ref{\cite{TOPQ-2016-03}} and $\ttbarjj$ at~$\sqrts=8$~\TeV\ from
 \Ref{\cite{TOPQ-2015-03}}, were all re-evaluated.
 In all cases, the numbers agree to within 0.01~\GeV\ with the original
 publications, which in any case is the rounding precision due to the precision
 of some of the inputs.
 On top of this, the results listed in \Tab{\ref{tab:results_indiv}} differ in
 some aspects from the original publications as explained below.
%
\begin{sidewaystable*}[tbp!]
\caption{The six measured values of \mt\ ($i=0, \ldots, 5$) and their
  statistical and systematic uncertainty sources $k$, numbered as given in the
  first column.
 For the individual measurements, the systematic uncertainty in \mt\ and its
 associated statistical uncertainty are given for each source of uncertainty.
 The last line refers to the sum in quadrature of the statistical and systematic
 uncertainties.
 The statistical uncertainties in the total systematic uncertainties and in the
 total uncertainties are calculated from the propagation of uncertainties.
 Systematic uncertainties listed as 0 are not evaluated, while an empty cell
 indicates an uncertainty not applicable to the correponding measurement.
 Statistical uncertainties quoted as 0.00 are smaller than 0.005.
\label{tab:results_indiv}}
\small
\begin{center}
\begin{tabular}{|c|l|r|r|r||r|r|r|} \cline{3-8}
\multicolumn{2}{c|}{} & \multicolumn{3}{c|}{$\sqrts=7$~\TeV} 
                      & \multicolumn{3}{c|}{$\sqrts=8$~\TeV}\\\cline{3-8}
\multicolumn{2}{c|}{}                      &   \mtdl\ [\GeV] &    \mtlj\ [\GeV] & \mtjj\ [\GeV]
                                           &   \mtdl\ [\GeV] &    \mtlj\ [\GeV] & \mtjj\ [\GeV] \\\hline
$k$ & Results~($i=0\ldots, 5$)             &          173.79 &     \SevenStaVal &          175.06       
                                           &          172.99 &     \EightLpJVal &          173.72\\\hline
 0 & Statistics                            &            0.54 &     \SevenStaSta &            1.35    
                                           &            0.41 &     \EightLpJSta &            0.55\\
   & {\it $~~$-- Stat.~comp.~(\mt)}          &                 &   \SevenStaStamt &                 
                                           &                 &   \EightLpJStamt &                \\
   & {\it $~~$-- Stat.~comp.~(\JSF)}         &                 &  \SevenStaStaJSF &                 
                                           &                 &  \EightLpJStaJSF &                \\
   & {\it $~~$-- Stat.~comp.~(\bJSF)}        &                 & \SevenStaStabJSF &                 
                                           &                 & \EightLpJStabJSF &                \\
 1 & Method                                & 0.09 $\pm$ 0.07 &  0.11 $\pm$ 0.10 & 0.42 $\pm$ 0.01
                                           & 0.05 $\pm$ 0.07 &  0.13 $\pm$ 0.11 &            0.11 \\ \hline
 2 & Signal Monte Carlo generator          & 0.26 $\pm$ 0.16 &  0.22 $\pm$ 0.21 & 0.30 $\pm$ 0.30
                                           & 0.09 $\pm$ 0.15 &  0.16 $\pm$ 0.17 & 0.18 $\pm$ 0.21\\
 3 & Hadronization                         & 0.53 $\pm$ 0.09 &  0.18 $\pm$ 0.12 & 0.50 $\pm$ 0.15
                                           & 0.22 $\pm$ 0.09 &  0.15 $\pm$ 0.10 & 0.64 $\pm$ 0.15\\
 4 & Initial- and final-state QCD radiation& 0.47 $\pm$ 0.05 &  0.32 $\pm$ 0.06 & 0.22 $\pm$ 0.11
                                           & 0.23 $\pm$ 0.07 &  0.08 $\pm$ 0.11 & 0.10 $\pm$ 0.28\\
 5 & Underlying event                      & 0.05 $\pm$ 0.05 &  0.15 $\pm$ 0.07 & 0.08 $\pm$ 0.10
                                           & 0.10 $\pm$ 0.14 &  0.08 $\pm$ 0.15 & 0.12 $\pm$ 0.16\\
 6 & Colour reconnection                   & 0.14 $\pm$ 0.05 &  0.11 $\pm$ 0.07 & 0.22 $\pm$ 0.10
                                           & 0.03 $\pm$ 0.14 &  0.19 $\pm$ 0.15 & 0.12 $\pm$ 0.16\\
 7 & Parton distribution function          & 0.10 $\pm$ 0.00 &  0.25 $\pm$ 0.00 & 0.09 $\pm$ 0.00
                                           & 0.05 $\pm$ 0.00 &  0.09 $\pm$ 0.00 & 0.09 $\pm$ 0.00\\\hline
 8 & Background normalization              & 0.04 $\pm$ 0.00 &  0.10 $\pm$ 0.00 &                
                                           & 0.03 $\pm$ 0.00 &  0.08 $\pm$ 0.00 &                \\
 9 & $W/Z$+jets shape                      & 0.00 $\pm$ 0.00 &  0.29 $\pm$ 0.00 &                
                                           &             $0$ &  0.11 $\pm$ 0.00 &                \\
10 & Fake leptons shape                    & 0.01 $\pm$ 0.00 &  0.05 $\pm$ 0.00 &                
                                           & 0.07 $\pm$ 0.00 &              $0$ &                \\
11 & Data-driven all-jets background       &                 &                  & 0.35 $\pm$ 0.21
                                           &                 &                  &            0.17 \\\hline
12 & Jet energy scale                      & 0.76 $\pm$ 0.09 &  0.58 $\pm$ 0.11 & 0.50 $\pm$ 0.05
                                           & 0.54 $\pm$ 0.04 &  0.54 $\pm$ 0.02 & 0.60 $\pm$ 0.03\\
13 & Relative $b$-to-light-jet energy scale& 0.68 $\pm$ 0.02 &  0.06 $\pm$ 0.03 & 0.62 $\pm$ 0.05
                                           & 0.30 $\pm$ 0.01 &  0.03 $\pm$ 0.01 & 0.34 $\pm$ 0.02\\
14 & Jet energy resolution                 & 0.19 $\pm$ 0.04 &  0.22 $\pm$ 0.11 & 0.01 $\pm$ 0.08
                                           & 0.09 $\pm$ 0.05 &  0.20 $\pm$ 0.04 & 0.10 $\pm$ 0.04\\
15 & Jet reconstruction efficiency         & 0.07 $\pm$ 0.00 &  0.12 $\pm$ 0.00 & 0.01 $\pm$ 0.01
                                           & 0.01 $\pm$ 0.00 &  0.02 $\pm$ 0.01 &             $0$\\
16 & Jet vertex fraction                   & 0.00 $\pm$ 0.00 &  0.01 $\pm$ 0.00 & 0.01 $\pm$ 0.01
                                           & 0.02 $\pm$ 0.00 &  0.09 $\pm$ 0.01 & 0.03 $\pm$ 0.01\\
17 & \btag                                 & 0.07 $\pm$ 0.00 &  0.50 $\pm$ 0.00 & 0.16 $\pm$ 0.00
                                           & 0.04 $\pm$ 0.02 &  0.38 $\pm$ 0.00 & 0.10 $\pm$ 0.00\\
18 & Leptons                               & 0.13 $\pm$ 0.00 &  0.04 $\pm$ 0.00 &             $0$
                                           & 0.14 $\pm$ 0.01 &  0.16 $\pm$ 0.01 & 0.01 $\pm$ 0.00\\
19 & Missing transverse momentum           & 0.04 $\pm$ 0.03 &  0.15 $\pm$ 0.04 & 0.02 $\pm$ 0.05
                                           & 0.01 $\pm$ 0.01 &  0.05 $\pm$ 0.01 & 0.01 $\pm$ 0.01\\
20 & Pile-up                               & 0.01 $\pm$ 0.00 &  0.02 $\pm$ 0.01 & 0.02 $\pm$ 0.00
                                           & 0.05 $\pm$ 0.01 &  0.15 $\pm$ 0.01 & 0.01 $\pm$ 0.00\\
21 & All-jets trigger                      &                 &                  & 0.01 $\pm$ 0.01
                                           &                 &                  & 0.08 $\pm$ 0.01\\
22 & Fast vs. full simulation              &                 &                  & 0.24 $\pm$ 0.18
                                           &                 &                  &                \\\hline
   & Total systematic uncertainty          & 1.31 $\pm$ 0.07
                                           & \SevenStaSys $\pm$ \SevenStaSysUnc
                                           & 1.21 $\pm$ 0.13
                                           & \EightDilSys $\pm$ \EightDilSysUnc
                                           & \EightLpJSys $\pm$ \EightLpJSysUnc
                                           & 1.02 $\pm$ 0.11 \\\hline
   & Total                                 & 1.42 $\pm$ 0.07
                                           & \SevenStaTot $\pm$ \SevenStaTotUnc
                                           & 1.82 $\pm$ 0.13
                                           & \EightDilTot $\pm$ \EightDilTotUnc
                                           & \EightLpJTot $\pm$ \EightLpJTotUnc
                                           & 1.16 $\pm$ 0.11 \\\hline
\end{tabular}
\end{center}
\end{sidewaystable*}

 The combination follows the approach developed for the combination of
 $\sqrts=7$~\TeV\ analyses in \Ref{\cite{TOPQ-2013-02}}, including the
 evaluation of the correlations given in \Sect{\ref{sect:combcor}} below.
 The treatment of uncertainty categories for the \ttbarll\ and
 \ttbarlj\ measurements at $\sqrts=7$~\TeV\ exactly follows
 \Ref{\cite{TOPQ-2013-02}}.
 The uncertainty categorizations for the \ttbarjj\ measurements at $\sqrts=7$
 and $8$~\TeV\ from \Refs{\cite{TOPQ-2013-03}}{\cite{TOPQ-2015-03}} closely
 follow this categorization but have some extra, analysis-specific sources of
 uncertainty, as shown in \Tab{\ref{tab:results_indiv}}.
 In addition, the \ttbarjj\ result at $\sqrts=8$~\TeV\ from
 \Ref{\cite{TOPQ-2015-03}} is based on a different treatment of the
 PDF-uncertainty-induced uncertainty in \mt.
 To allow the evaluation of the estimator correlations also for this uncertainty
 in \mt, for this combination, the respective uncertainty is newly evaluated
 according to the prescription given in \Sect{\ref{sect:unc}}.

 For the \ttbarjj\ result at~$\sqrts=7$~\TeV\ the statistical precisions in the
 systematic uncertainties were not evaluated in \Ref{\cite{TOPQ-2013-02}} but
 were calculated for this combination.
 For the \ttbarjj\ result at~$\sqrts=8$~\TeV\ in \Ref{\cite{TOPQ-2015-03}}, for
 some of the sources, the statistical uncertainty in the systematic uncertainty
 was not evaluated, such that the quoted statistical uncertainty in the total
 systematic uncertainty is a lower limit.

 For the mapping of uncertainty categories for data taken at different \cmes,
 the choice of \Ref{\cite{TOPQ-2016-03}} is employed.
 The most complex cases are the uncertainties involving eigenvector
 decompositions, such as the \JES\ and \btag\ scale factor uncertainties, and
 the uncertainty categories that do not apply to all input measurements.
 The JES-uncertainty-induced uncertainty in \mt\ is obtained from a number of
 \JES\ subcomponents. Some \JES\ subcomponents have an equivalent at the other
 \cme\ and others do not.
 As in \Ref{\cite{TOPQ-2016-03}}, the \JES\ subcomponents without an equivalent
 at the other \cme\ are treated as independent, resulting in vanishing estimator
 correlations for that part of the covariance matrix.
 For the remaining subcomponents, the estimator correlations are partly positive
 and partly negative.
 As an example, for the flavour part of the JES-uncertainty-induced uncertainty
 in \mt, the two most precise results, the \ttbarll\ and \ttbarlj\ measurements
 at $\sqrts=8$~\TeV, are negatively correlated.
 Consequently, for this pair, the resulting estimator correlation for the total
 JES-induced uncertainty in \mt\ is also negative.
 At the quoted precision, the two assumptions about the equivalence of the
 \JES\ subcomponents between the datasets at the two \cmes, i.e.~the weak and
 strong correlation scenarios described in \Tab{\ref{tab:jesresults8TeV}} in
 \App{\ref{sect:addlpj}}, leave the combined value and uncertainty unchanged.
 
 Following \Ref{\cite{TOPQ-2016-03}}, the $\sqrts=7$ and $8$~\TeV\ measurements
 are treated as uncorrelated for the nuisance parameters of the \btag,
 $c/\tau$-tagging, mistagging and JER uncertainties.
 In \Ref{\cite{TOPQ-2016-03}} it was shown that a correlated treatment of the
 flavour-tagging nuisance parameters results in an insignificant change in the
 combination.
 For the statistical, method calibration, MC-based background shape at
 $\sqrts=7$ and $8$~\TeV, and the pile-up uncertainties in \mt, the measurements
 are assumed to be uncorrelated.
 Details of the evaluation of the correlations for all remaining systematic
 uncertainties are discussed below.
%
\subsection{Mathematical framework and evaluation of estimator correlations}
\label{sect:combcor}
 All combinations are performed using the best linear unbiased estimate~(\BLUE)
 method~\cite{BLUE1,BLUERN} in a \Cpp\ implementation described in
 \Ref{\cite{BLUEcpp}}.
 The \BLUE\ method uses a linear combination of the inputs to combine
 measurements.
 The coefficients~(\BLUE\ weights) are determined via the minimization of the
 variance of the combined result. They can be used to construct measures for the
 importance of a given single measurement in the combination~\cite{BLUERN}.
 For any combination, the measured values \xX{i}, the list of uncertainties
 \sigmaX{ik}\ and the correlations \rhoX{ijk}\ of the estimators~($i, j$) for
 each source of uncertainty~($k$) have to be provided.
 For all uncertainties, a Gaussian probability distribution function is assumed.
 For the uncertainties in \mt\ for which the measurements are correlated, when
 using $\pm 1\sigma$ variations of a systematic effect, e.g.~when changing the
 \bJES\ by $\pm 1\sigma$, there are two possibilities.
 When simultaneously applying a variation for a systematic uncertainty,
 e.g.~$+1\sigma$ for the \bJES, to a pair~($i, j$) of measurements, e.g.~the
 \ttbarlj\ and \ttbarll\ measurements at $\sqrts=8$~\TeV, both analyses can
 result in a larger or smaller \mt\ value than the one obtained for the nominal
 case~(full correlation, $\rhoX{ijk}=+1$), or one analysis can result in a
 larger and the other in a smaller value~(full anti-correlation,
 $\rhoX{ijk}=-1$).
 Consequently, an uncertainty from a source only consisting of a single
 variation, such as the bJES-uncertainty-induced uncertainty or the uncertainty
 related to the choice of MC generator for signal events, results in a
 correlation of $\rhoX{ijk}=\pm 1$.
 The estimator correlations for composite uncertainties are evaluated by
 calculating the correlation from the subcomponents. As an example, for the
 \ttbarlj result at $\sqrts=8$~\TeV, the subcomponents of the \JES\ uncertainty
 are shown in \Tab{\ref{tab:jesresults8TeV}} in \App{\ref{sect:addlpj}}.
 For any pair of measurements~($i, j$), this evaluation is done by adding the
 covariance terms of the subcomponents $k$ with $\rhoX{ijk}=\pm 1$ and dividing
 by the total uncertainties for that source.
 The resulting estimator correlation is
%
\begin{linenomath*}
 \begin{align}
 \rhoX{ij} &= \frac{\sum_{k=1}^{N_\mathrm{comp}}\rhoX{ijk}\sigmaX{ik}\sigmaX{jk}}
              {\sigmaX{i}\sigmaX{j}}.
 \nonumber
 \end{align}
\end{linenomath*}
%
 The quantity $\sigmaXq{i}=\sum_{k=1}^{N_\mathrm{comp}}\sigmaXq{ik}$ is the sum
 of the single subcomponent variances in analysis $i$.
 This procedure is applied to all uncertainty sources that consist of more than
 one subcomponent to reduce the large list of uncertainty subcomponents per
 estimator of \order{100} to a suitable number of uncertainty sources, i.e.~to
 those given in \Tab{\ref{tab:results_indiv}}.
 Since the full covariance matrix is independent of how the subsets are chosen,
 this does not affect the combination.
%
\begin{figure*}[tbp!]
\centering
\subfigure[\ttbarll~($8$~\TeV) and \ttbarlj~($8$~\TeV)]
          {\includegraphics[width=0.47\textwidth]{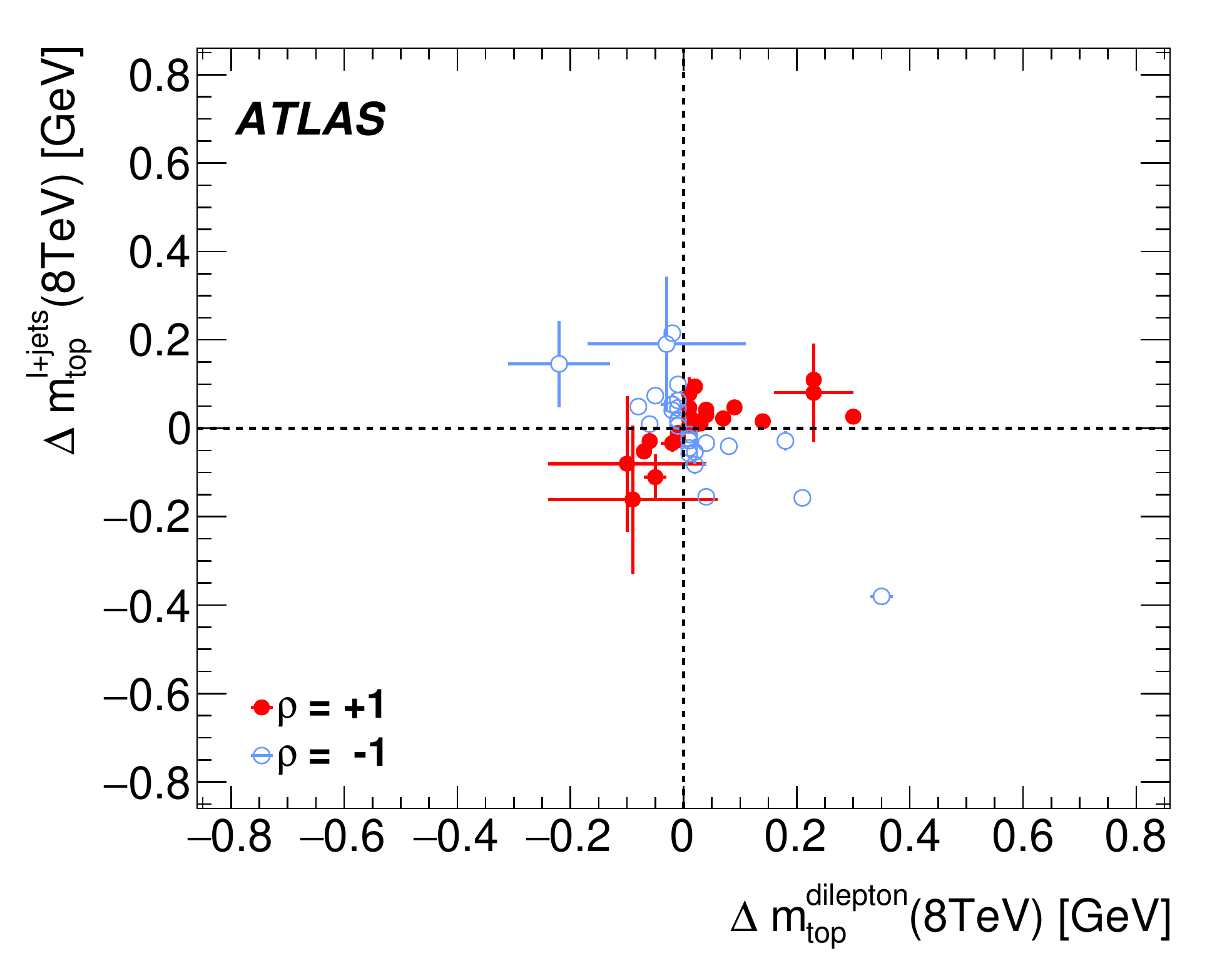}
            \label{fig:rho_dil8lpj8}
          }
\subfigure[\ttbarll~($8$~\TeV) and \ttbarjj~($8$~\TeV)]
          {\includegraphics[width=0.47\textwidth]{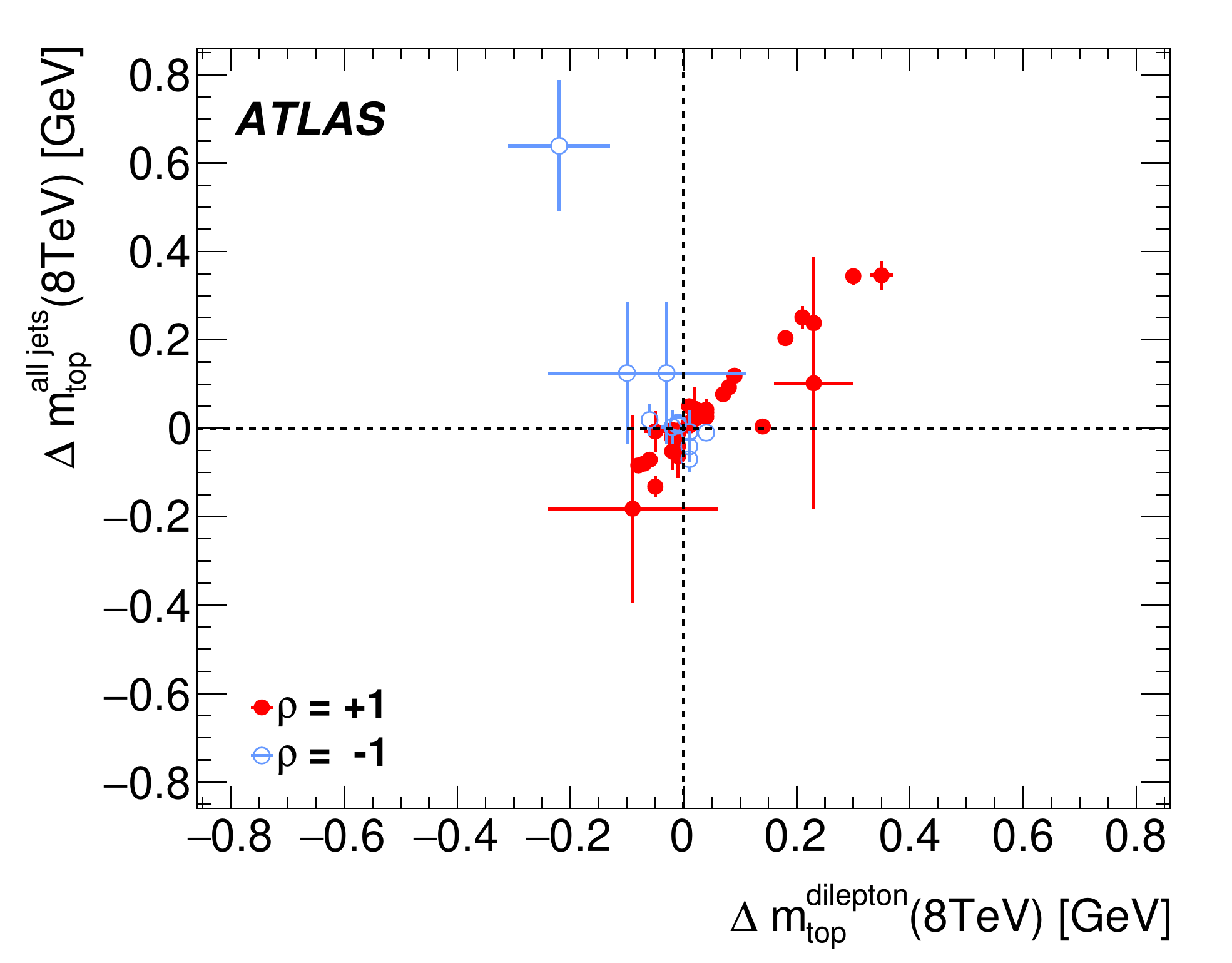}
            \label{fig:rho_dil8jet8}
          }
\hfill
\subfigure[\ttbarlj~($8$~\TeV) and \ttbarjj~($8$~\TeV)]
          {\includegraphics[width=0.47\textwidth]{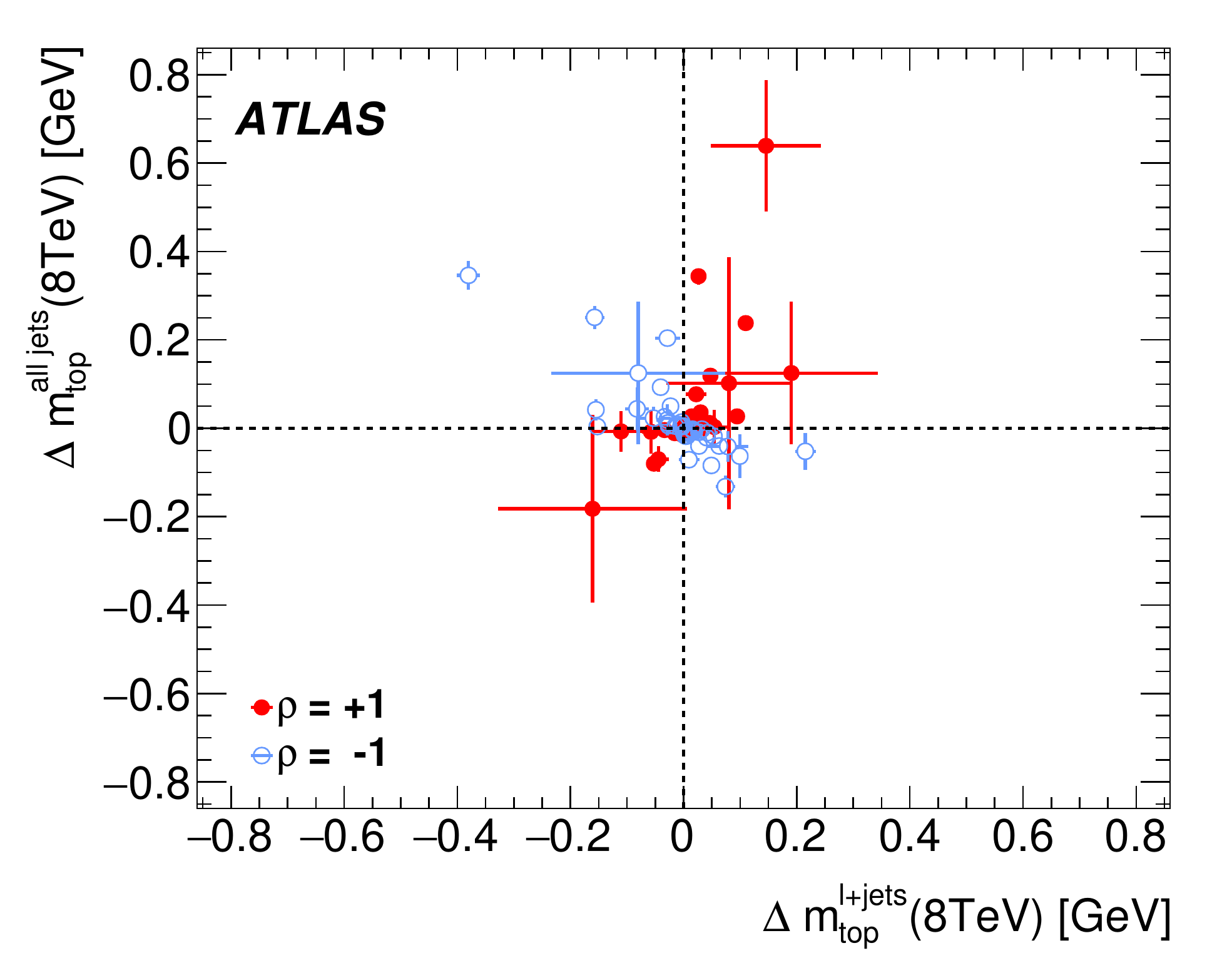}
            \label{fig:rho_lpj8jet8}
          }
\subfigure[\ttbarlj~($8$~\TeV) and \ttbarlj~($7$~\TeV)]
          {\includegraphics[width=0.47\textwidth]{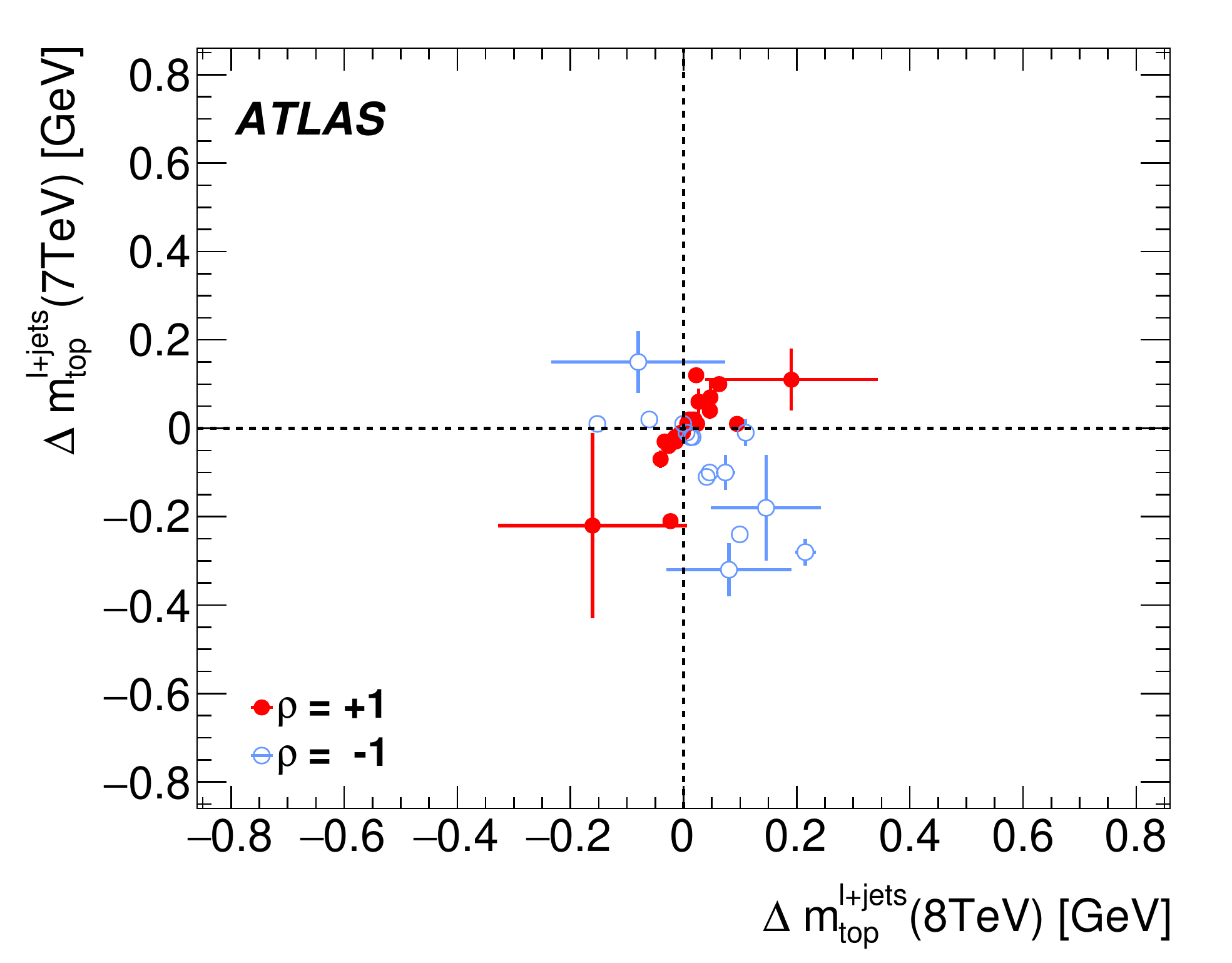}
            \label{fig:rho_lpj8lpj7}
          }
\caption{The pairwise shifts in \mt\ when simultaneously varying a pair of
  measurements for a systematic uncertainty or a subcomponent of a systematic
  uncertainty.
 Figures~(a) and (b) show the correlations of the \ttbarll\ measurement at
 $\sqrts=8$~\TeV\ with the two other measurements at the same \cme for all
 sources of uncertainty for which the estimators are correlated.
 Figure~(c) shows the correlations of the present measurement with the
 \ttbarjj\ measurement at $\sqrts=8$~\TeV, while figure~(d) shows the
 correlations of the present measurement with the \ttbarlj\ measurement at
 $\sqrts=7$~\TeV.
 The crosses indicate the statistical uncertainty in the systematic
 uncertainties.
 The solid points indicate the fully correlated cases, and the open points
 indicate the anti-correlated ones.
 \label{fig:rho}
}
\end{figure*}

 For the three analyses, the evaluated shifts in \mt\ per uncertainty
 subcomponent are referred to as \dmtdl, \dmtlj\ and \dmtjj.
 They are shown in \Fig{\ref{fig:rho}} for the various uncertainty
 subcomponents in selected pairs of analyses.
 The pairs using the results from $\sqrts=8$~\TeV\ data are shown in
 \Figrange{\ref{fig:rho_dil8lpj8}}{\ref{fig:rho_lpj8jet8}}, while
 \Fig{\ref{fig:rho_lpj8lpj7}} is for the two analyses in the \ttbarlj\ channel
 at the two \cmes.
 Each point represents the observed shifts for a systematic uncertainty or a
 subcomponent of a systematic uncertainty together with a cross, indicating the
 corresponding statistical precision in the systematic uncertainty in the two
 results.
 The solid points indicate the fully correlated cases, and the open points
 indicate the anti-correlated ones.\footnote{In the course of including more
   results into the combination of \Ref{\cite{TOPQ-2016-03}}, the definitions of
   the variations were homogenized while leaving the estimator correlations
   unchanged. As a consequence, for the corresponding figures some of the points
   now are located in the respective other quadrant, e.g.~for the
   \ttbarll\ result at $\sqrts=8$~\TeV.}

 For many significant sources of uncertainty in \Fig{\ref{fig:rho_dil8lpj8}},
 the \ttbarlj\ and \ttbarll\ measurements are anti-correlated.
 As shown in \Ref{\cite{TOPQ-2013-02}}, this is caused by the in situ
 determination of the \JSF\ and \bJSF\ in the three-dimensional
 \ttbarlj\ analysis.
 In contrast, for most sources of uncertainty, a positive estimator correlation
 is observed for the \ttbarll\ and \ttbarjj\ measurements at $\sqrts=8$~\TeV,
 shown in \Fig{\ref{fig:rho_dil8jet8}}.
 The prominent exception is the hadronization-uncertainty-induced uncertainty in
 \mt, i.e.~the single largest uncertainty in the \ttbarjj\ measurement at
 $\sqrts=8$~\TeV, for which the two measurements are anti-correlated.
 On the contrary, the \ttbarlj\ and \ttbarjj\ measurements at $\sqrts=8$~\TeV,
 shown in \Fig{\ref{fig:rho_lpj8jet8}}, are positively correlated for this
 uncertainty.
 Finally, the \ttbarlj\ measurements at the two \cmes\ in
 \Fig{\ref{fig:rho_lpj8lpj7}} show a rather low correlation.
 The correlations per source of uncertainty and the total estimator correlations
 are summarized in \Tab{\ref{tab:rho_indiv}}.
%
\begin{sidewaystable*}[tbp!]
\small
\caption{The pairwise correlations \rhoX{ijk}\ of the six measurements $i, j=0,
  \ldots, 5$ of \mt\ for each source of systematic uncertainty $k=0, \ldots,
  22$, along with the total estimator correlations and the compatibility of the
  measurements using \chiqij\ from \protect\Eqn{(\ref{Eq:compat})}.
 The indices $i$ and $j$ are 0 for $\ttbarll$ at~$\sqrts=7$~\TeV, 1 for
 $\ttbarlj$ at~$\sqrts=7$~\TeV, 2 for $\ttbarjj$ at~$\sqrts=7$~\TeV, 3 for
 $\ttbarll$ at~$\sqrts=8$~\TeV, 4 for $\ttbarlj$ at~$\sqrts=8$~\TeV, and 5 for
 $\ttbarjj$ at~$\sqrts=8$~\TeV.
 The correspondence of the indices $k=0, \ldots, 22$ and the sources of
 systematic uncertainty are given in \protect\Tab{\ref{tab:results_indiv}}.
 Correlations that are assigned, or cannot be evaluated because one uncertainty
 in the covariance term is zero at the quoted precision, are given as integer
 values, while evaluated correlations are shown as real values.
\label{tab:rho_indiv}}
\begin{center}
\begin{tabular}{|c||r|r|r|r|r||r|r|r|r||r|r|r||r|r||r|} \hline
       & \multicolumn{5}{c||}{\mtdl\ 7~\TeV}
       & \multicolumn{4}{c||}{\mtlj\ 7~\TeV}
       & \multicolumn{3}{c||}{\mtjj\ 7~\TeV}
       & \multicolumn{2}{c||}{\mtdl\ 8~\TeV}
       & \multicolumn{1}{c|}{\mtlj\ 8~\TeV}\\
       & \multicolumn{5}{c||}{$i=0$}
       & \multicolumn{4}{c||}{$i=1$}
       & \multicolumn{3}{c||}{$i=2$}
       & \multicolumn{2}{c||}{$i=3$}
       & \multicolumn{1}{c|}{$i=4$}\\\cline{2-16}
   $k$ &  $\rhoX{01}$ & $\rhoX{02}$ & $\rhoX{03}$ & $\rhoX{04}$ & $\rhoX{05}$  
       &  $\rhoX{12}$ & $\rhoX{13}$ & $\rhoX{14}$ & $\rhoX{15}$ 
       &  $\rhoX{23}$ & $\rhoX{24}$ & $\rhoX{25}$ 
       &  $\rhoX{34}$ & $\rhoX{35}$ 
       &  $\rhoX{45}$ \\\hline
     0 & $    0$ & $    0$ & $    0$ & $    0$ & $    0$
       & $    0$ & $    0$ & $    0$ & $    0$
       & $    0$ & $    0$ & $    0$
       & $    0$ & $    0$
       & $    0$ \\
     1 & $    0$ & $    0$ & $    0$ & $    0$ & $    0$
       & $    0$ & $    0$ & $    0$ & $    0$
       & $    0$ & $    0$ & $    0$
       & $    0$ & $    0$
       & $    0$ \\\hline
     2 & $ 1.00$ & $-1.00$ & $ 1.00$ & $ 1.00$ & $ 1.00$
       & $-1.00$ & $ 1.00$ & $ 1.00$ & $ 1.00$
       & $-1.00$ & $-1.00$ & $-1.00$
       & $ 1.00$ & $ 1.00$
       & $ 1.00$ \\
     3 & $ 1.00$ & $ 1.00$ & $ 1.00$ & $-1.00$ & $-1.00$
       & $ 1.00$ & $ 1.00$ & $-1.00$ & $-1.00$
       & $ 1.00$ & $-1.00$ & $-1.00$
       & $-1.00$ & $-1.00$
       & $ 1.00$ \\
     4 & $-1.00$ & $ 1.00$ & $ 1.00$ & $ 1.00$ & $ 1.00$
       & $-1.00$ & $-1.00$ & $-1.00$ & $-1.00$
       & $ 1.00$ & $ 1.00$ & $ 1.00$
       & $ 1.00$ & $ 1.00$
       & $ 1.00$ \\
     5 & $-1.00$ & $ 1.00$ & $ 1.00$ & $ 1.00$ & $-1.00$
       & $-1.00$ & $-1.00$ & $-1.00$ & $ 1.00$
       & $ 1.00$ & $ 1.00$ & $-1.00$
       & $ 1.00$ & $-1.00$
       & $-1.00$ \\
     6 & $-1.00$ & $ 1.00$ & $ 1.00$ & $-1.00$ & $-1.00$
       & $-1.00$ & $-1.00$ & $ 1.00$ & $ 1.00$
       & $ 1.00$ & $-1.00$ & $-1.00$
       & $-1.00$ & $-1.00$
       & $ 1.00$ \\
     7 & $ 0.53$ & $ 0.22$ & $-0.02$ & $ 0.72$ & $-0.61$
       & $-0.36$ & $-0.32$ & $ 0.72$ & $-0.81$
       & $ 0.41$ & $-0.05$ & $ 0.27$
       & $-0.48$ & $ 0.40$
       & $-0.76$ \\\hline
     8 & $ 1.00$ & $    0$ & $ 0.31$ & $-0.77$ & $    0$
       & $    0$ & $ 0.31$ & $-0.74$ & $    0$
       & $    0$ & $    0$ & $    0$
       & $-0.06$ & $    0$
       & $    0$ \\
     9 & $    0$ & $    0$ & $    0$ & $    0$ & $    0$
       & $    0$ & $    0$ & $    0$ & $    0$
       & $    0$ & $    0$ & $    0$
       & $    0$ & $    0$
       & $    0$ \\
    10 & $ 0.20$ & $    0$ & $    0$ & $    0$ & $    0$
       & $    0$ & $ 0.27$ & $    0$ & $    0$
       & $    0$ & $    0$ & $    0$
       & $    0$ & $    0$
       & $    0$ \\
    11 & $    0$ & $    0$ & $    0$ & $    0$ & $    0$
       & $    0$ & $    0$ & $    0$ & $    0$
       & $    0$ & $    0$ & $    1$
       & $    0$ & $    0$
       & $    0$ \\\hline
    12 & $-0.24$ & $ 0.86$ & $ 0.36$ & $ 0.18$ & $ 0.36$
       & $ 0.10$ & $ 0.04$ & $-0.29$ & $ 0.13$
       & $ 0.41$ & $ 0.09$ & $ 0.42$
       & $-0.54$ & $ 0.98$
       & $-0.57$ \\
    13 & $ 1.00$ & $ 1.00$ & $ 1.00$ & $ 1.00$ & $ 1.00$
       & $ 1.00$ & $ 1.00$ & $ 1.00$ & $ 1.00$
       & $ 1.00$ & $ 1.00$ & $ 1.00$
       & $ 1.00$ & $ 1.00$
       & $ 1.00$ \\
    14 & $-1.00$ & $ 1.00$ & $    0$ & $    0$ & $    0$
       & $-1.00$ & $    0$ & $    0$ & $    0$
       & $    0$ & $    0$ & $    0$
       & $ 0.22$ & $-0.07$
       & $-0.17$ \\
    15 & $ 1.00$ & $ 1.00$ & $ 1.00$ & $ 1.00$ & $    0$
       & $ 1.00$ & $ 1.00$ & $ 1.00$ & $    0$
       & $ 1.00$ & $ 1.00$ & $    0$
       & $ 1.00$ & $    0$
       & $    0$ \\
    16 & $-1.00$ & $ 1.00$ & $-1.00$ & $-1.00$ & $-1.00$
       & $-1.00$ & $ 1.00$ & $ 1.00$ & $ 1.00$
       & $-1.00$ & $-1.00$ & $-1.00$
       & $ 1.00$ & $ 1.00$
       & $ 1.00$ \\
    17 & $-0.80$ & $-0.03$ & $    0$ & $    0$ & $    0$
       & $    0$ & $    0$ & $    0$ & $    0$
       & $    0$ & $    0$ & $    0$
       & $-0.23$ & $    1$
       & $    1$ \\
    18 & $-0.35$ & $    0$ & $ 0.93$ & $-0.08$ & $ 0.42$
       & $    0$ & $-0.51$ & $-0.17$ & $ 0.02$
       & $    0$ & $    0$ & $    0$
       & $ 0.11$ & $ 0.28$
       & $-0.36$ \\
    19 & $ 0.00$ & $-0.26$ & $-0.26$ & $-0.12$ & $ 0.04$
       & $ 0.84$ & $ 0.26$ & $ 0.22$ & $ 0.16$
       & $    0$ & $    0$ & $    0$
       & $ 0.97$ & $ 0.86$
       & $ 0.96$ \\
    20 & $    0$ & $    0$ & $    0$ & $    0$ & $    0$
       & $    0$ & $    0$ & $    0$ & $    0$
       & $    0$ & $    0$ & $    0$
       & $    0$ & $    0$
       & $    0$ \\
    21 & $    0$ & $    0$ & $    0$ & $    0$ & $    0$
       & $    0$ & $    0$ & $    0$ & $    0$
       & $    0$ & $    0$ & $ 1.00$
       & $    0$ & $    0$
       & $    0$\\
    22 & $    0$ & $    0$ & $    0$ & $    0$ & $    0$
       & $    0$ & $    0$ & $    0$ & $    0$
       & $    0$ & $    0$ & $    0$
       & $    0$ & $    0$
       & $    0$\\\hline
 Total & \Rdlsljs & \Rdlsjjs & \Rdlsdle & \Rdlslje & \Rdlsjje
       & \Rljsjjs & \Rljsdle & \Rljslje & \Rljsjje
       & \Rjjsdle & \Rjjslje & \Rjjsjje
       & \Rdlelje & \Rdlejje
       & \Rljejje \\\hline
 \chiqij & \Cdlsljs & \Cdlsjjs & \Cdlsdle & \Cdlslje & \Cdlsjje
       & \Cljsjjs & \Cljsdle & \Cljslje & \Cljsjje
       & \Cjjsdle & \Cjjslje & \Cjjsjje
       & \Cdlelje & \Cdlejje
       & \Cljejje \\\hline
\end{tabular}
\end{center}
\end{sidewaystable*}

 The improvement in the combination obtained by the use of evaluated
 correlations compared with using estimator correlations assigned solely by
 physics assessments~(here referred to as assigned correlations) is quantified
 using an example.
 Using the choices of assigned correlations from \Ref{\cite{ATLAS:2014wva}} for
 the ATLAS results in the \ttbarll\ and \ttbarlj\ channels at
 $\sqrts=7$~\TeV\ listed in \Tab{\ref{tab:results_indiv}} gives a combined value
 of
 $\mt=\XZ{\SevendillpjwcVal}{\SevendillpjwcSta}{\SevendillpjwcSys}$~\GeV\ 
 compared
 with $\mt=\XZ{\SevendillpjVal}{\SevendillpjSta}{\SevendillpjSys}$~\GeV.
 The significant improvement in the precision of the combination demonstrates
 the particular importance of evaluating the correlations.

 For the combinations presented in this paper, most estimator correlations could
 be evaluated.
 The most prominent exception is for the \btag\ uncertainty, where the
 $\ttbarjj$ measurement at $\sqrts=8$~\TeV\ is based on a different
 \btag\ algorithm and calibration than the \ttbarll\ and \ttbarlj\ measurements
 at $\sqrts=8$~\TeV.
 It was verified that assignments of the estimator correlations of
 $\rhoX{i5k}\in[-1, 1]$, with $i=3, 4$ and $k=17$, yield insignificant
 differences in the full combination.
 Estimator correlations of $\rhoX{i5} = 1$ are assigned for this case, as this
 choice gives the largest uncertainty in the combination.
 A similar situation arises for the data-driven all-jets background uncertainty
 in the two \ttbarjj\ measurements, where the method used for the background
 estimate is similar but not identical for the two measurements.
 Consequently, the conservative ad hoc assignment of $\rhoX{25k}=1$ was also
 made for this source $k=11$.
%
\subsection{Compatibility of the inputs and selected combinations}
\label{sect:combcon}
 Before any combination is performed, the compatibility of the input results is
 verified.
 For each pair of results, their compatibility is expressed by the ratio of the
 squared difference between the pair of measured values and the uncertainty in
 this difference~\cite{BLUERN} as
%
\begin{linenomath*}
\begin{align}
 \chiqij &= \frac{(x_i-x_j)^2}{\sigma_i^2+\sigma_j^2-2\rhoX{ij}\sigma_i\sigma_j}.
\label{Eq:compat}
\end{align}
\end{linenomath*}
%
 The corresponding values are given in \Tab{\ref{tab:rho_indiv}}.
 Analysing the \chiqij\ values reveals good \chiq\ probabilities, with the
 smallest \chiq\ probability being $\chipropone=\ProCompatMin$.
 The largest sum of \chiqij\ values by far is observed for the \ttbarjj\ result
 at $\sqrts=7$~\TeV.

\begin{figure*}[tbp!]
\centering
\subfigure[\ttbarll~($8$~\TeV) and \ttbarlj~($8$~\TeV)]
          {\includegraphics[width=0.45\textwidth]{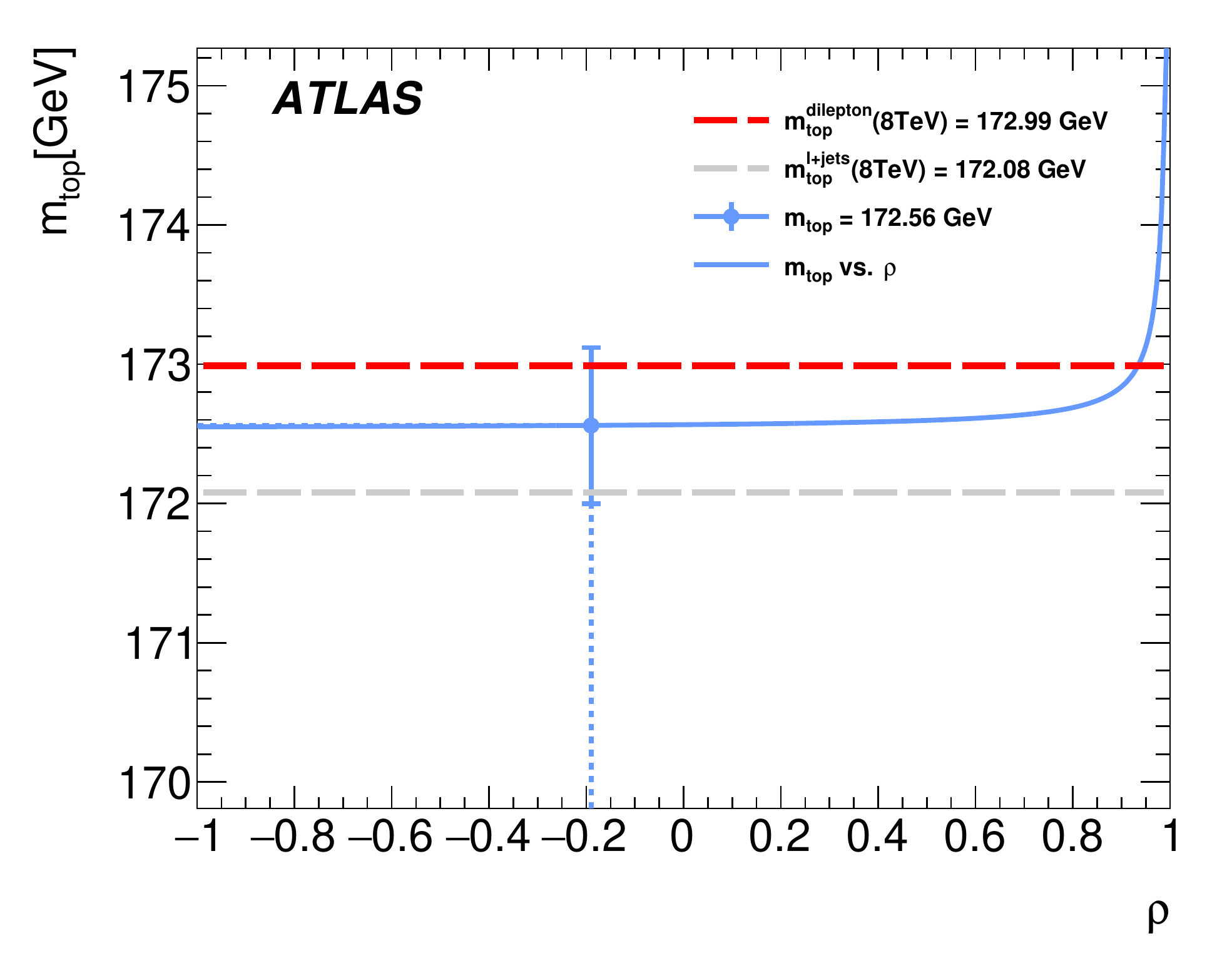}
            \label{fig:comb_val_dil8lpj8}
          }
\subfigure[\ttbarll~($8$~\TeV) and \ttbarlj~($8$~\TeV)]
          {\includegraphics[width=0.45\textwidth]{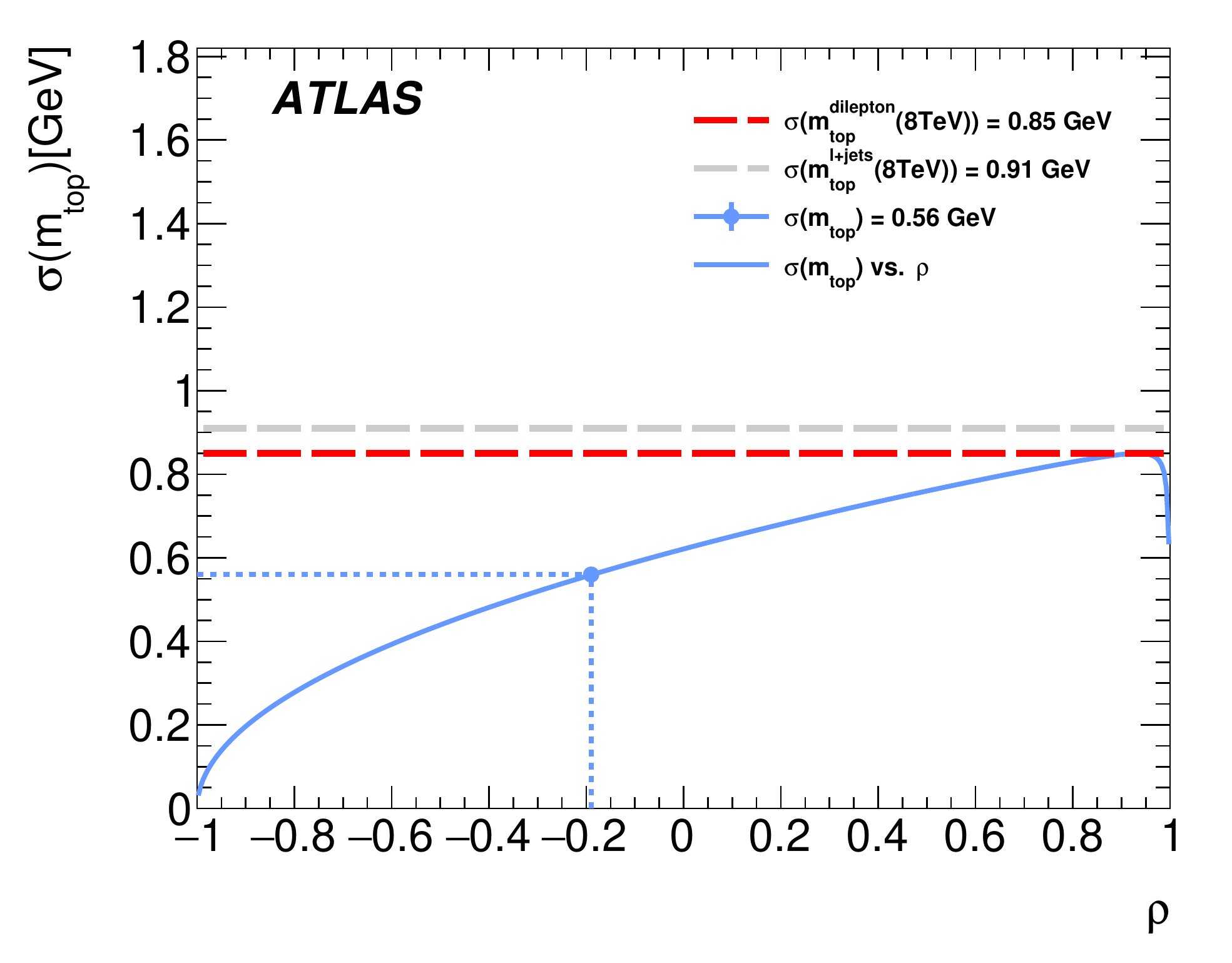}
            \label{fig:comb_unc_dil8lpj8}
          }
\hfill
\subfigure[\ttbarll~($8$~\TeV) and \ttbarjj~($8$~\TeV)]
          {\includegraphics[width=0.45\textwidth]{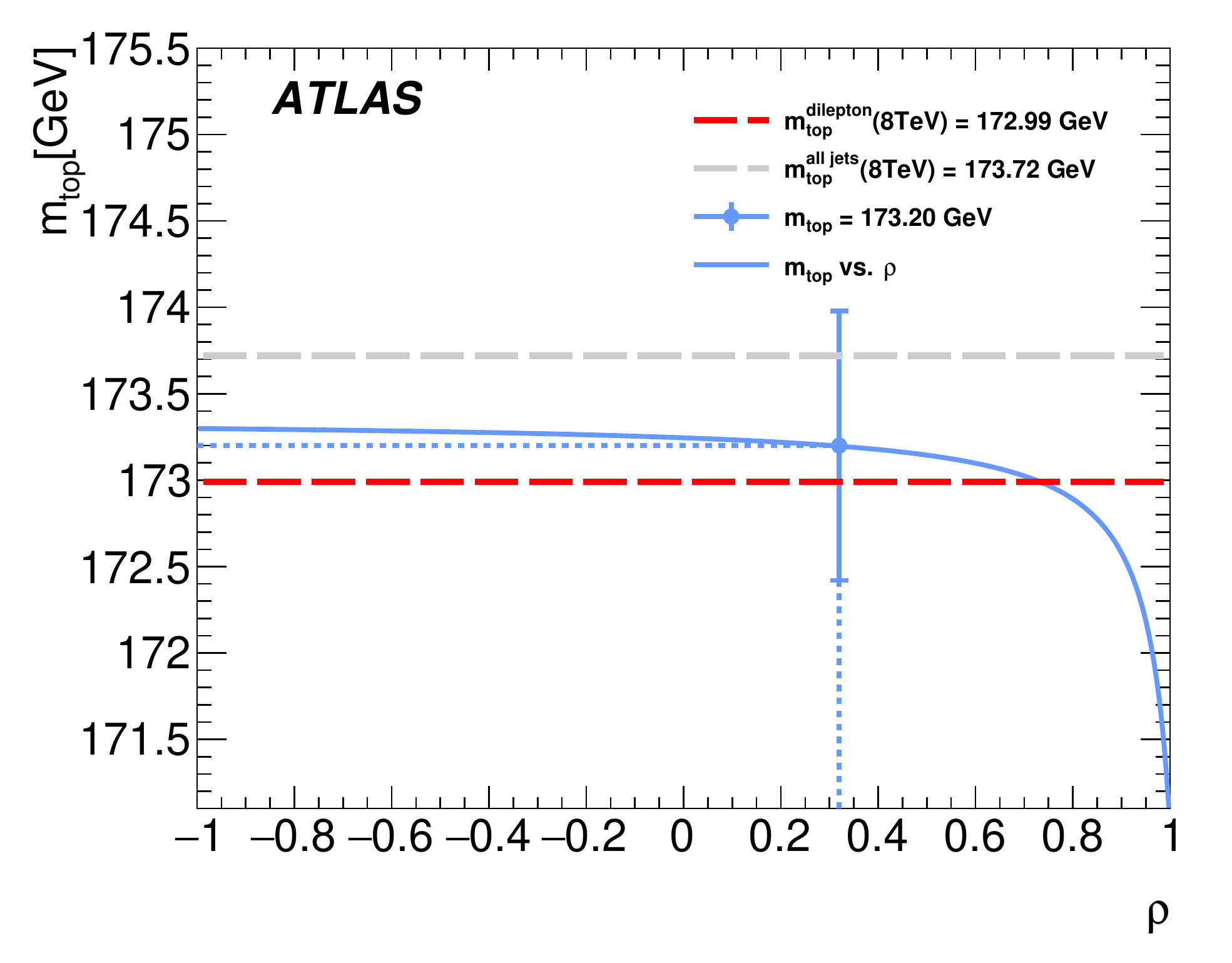}
            \label{fig:comb_val_dil8jet8}
          }
\subfigure[\ttbarll~($8$~\TeV) and \ttbarjj~($8$~\TeV)]
          {\includegraphics[width=0.45\textwidth]{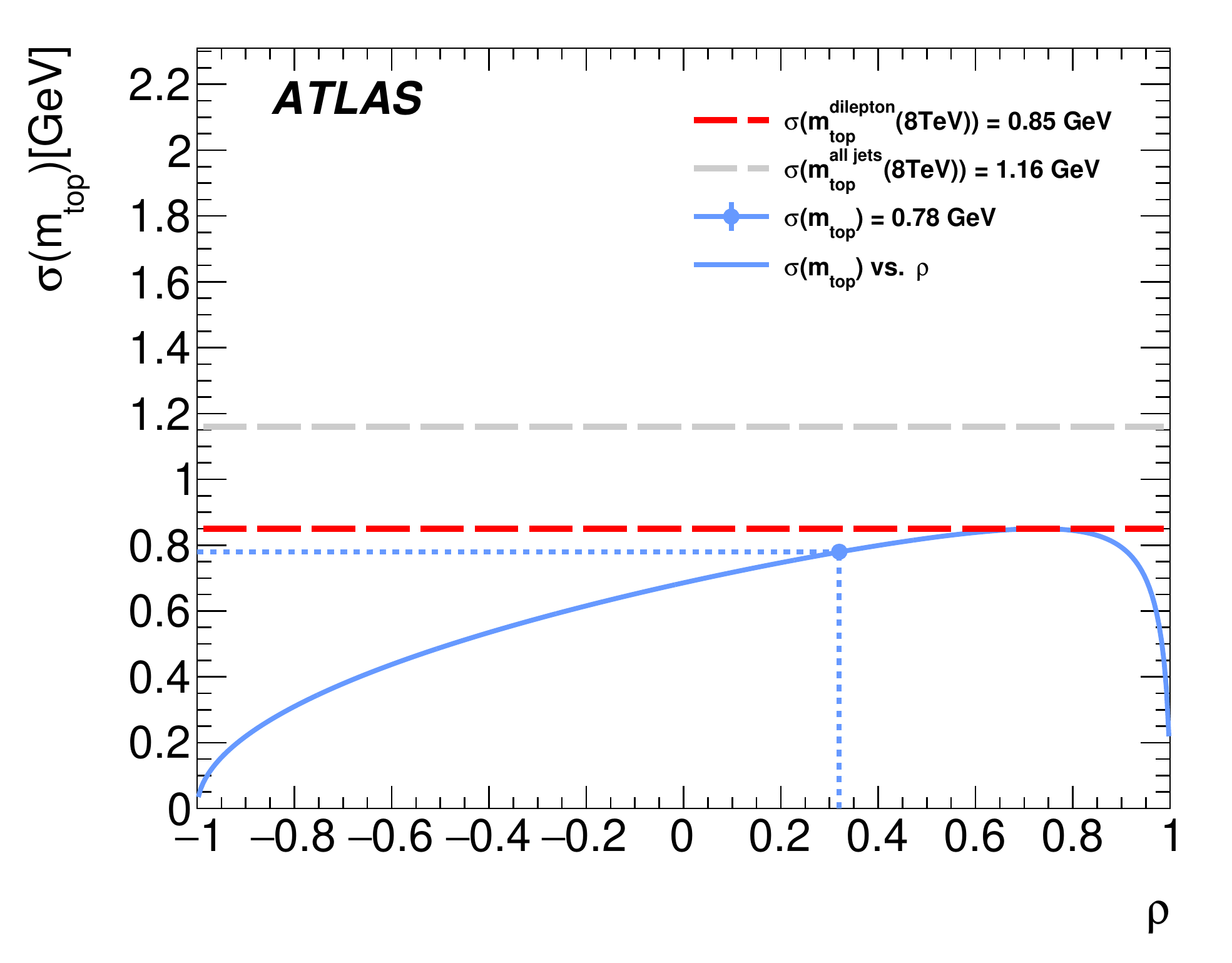}
            \label{fig:comb_unc_dil8jet8}
          }
\hfill
\subfigure[\ttbarlj~($8$~\TeV) and \ttbarjj~($8$~\TeV)]
          {\includegraphics[width=0.45\textwidth]{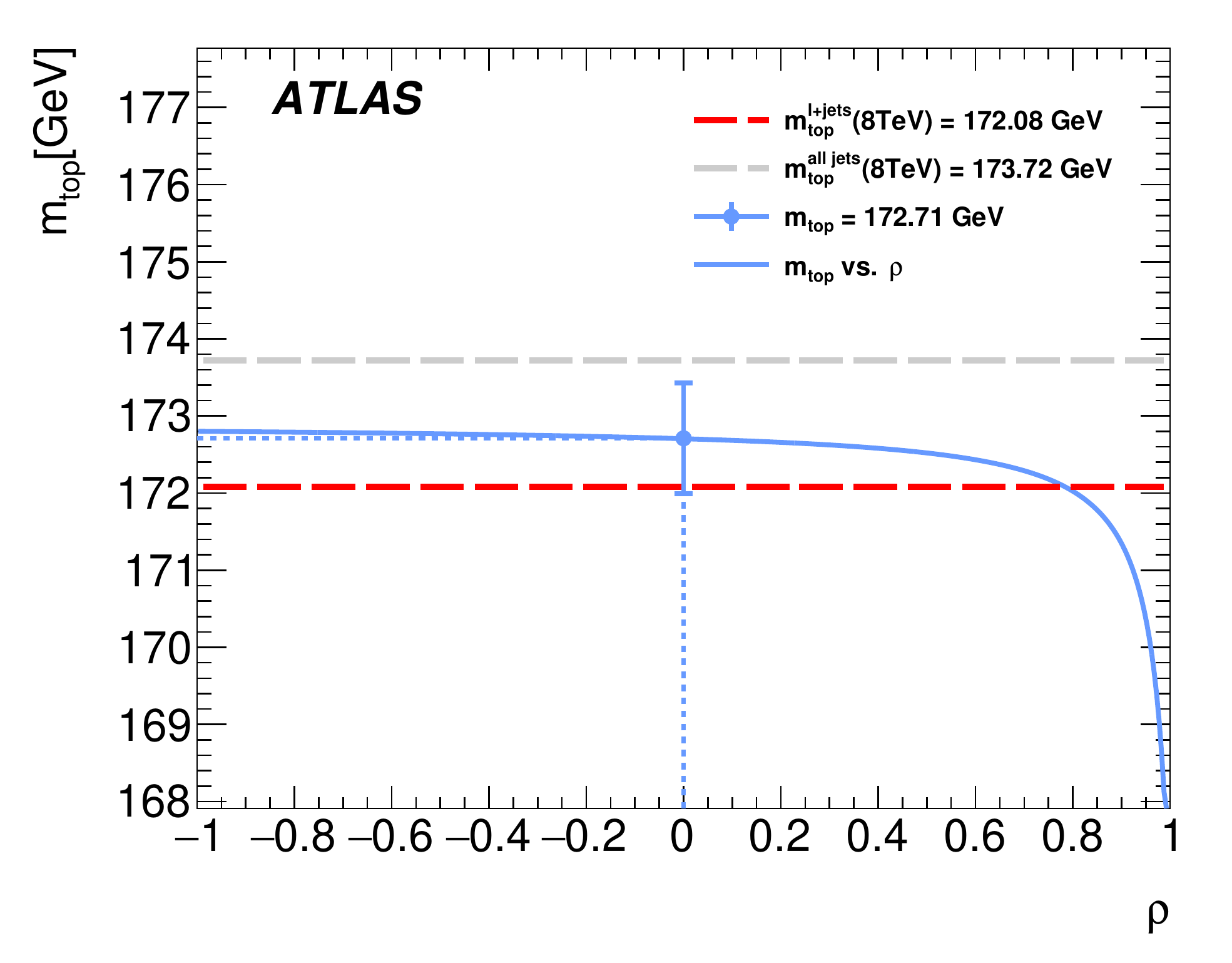}
            \label{fig:comb_val_lpj8jet8}
          }
\subfigure[\ttbarlj~($8$~\TeV) and \ttbarjj~($8$~\TeV)]
          {\includegraphics[width=0.45\textwidth]{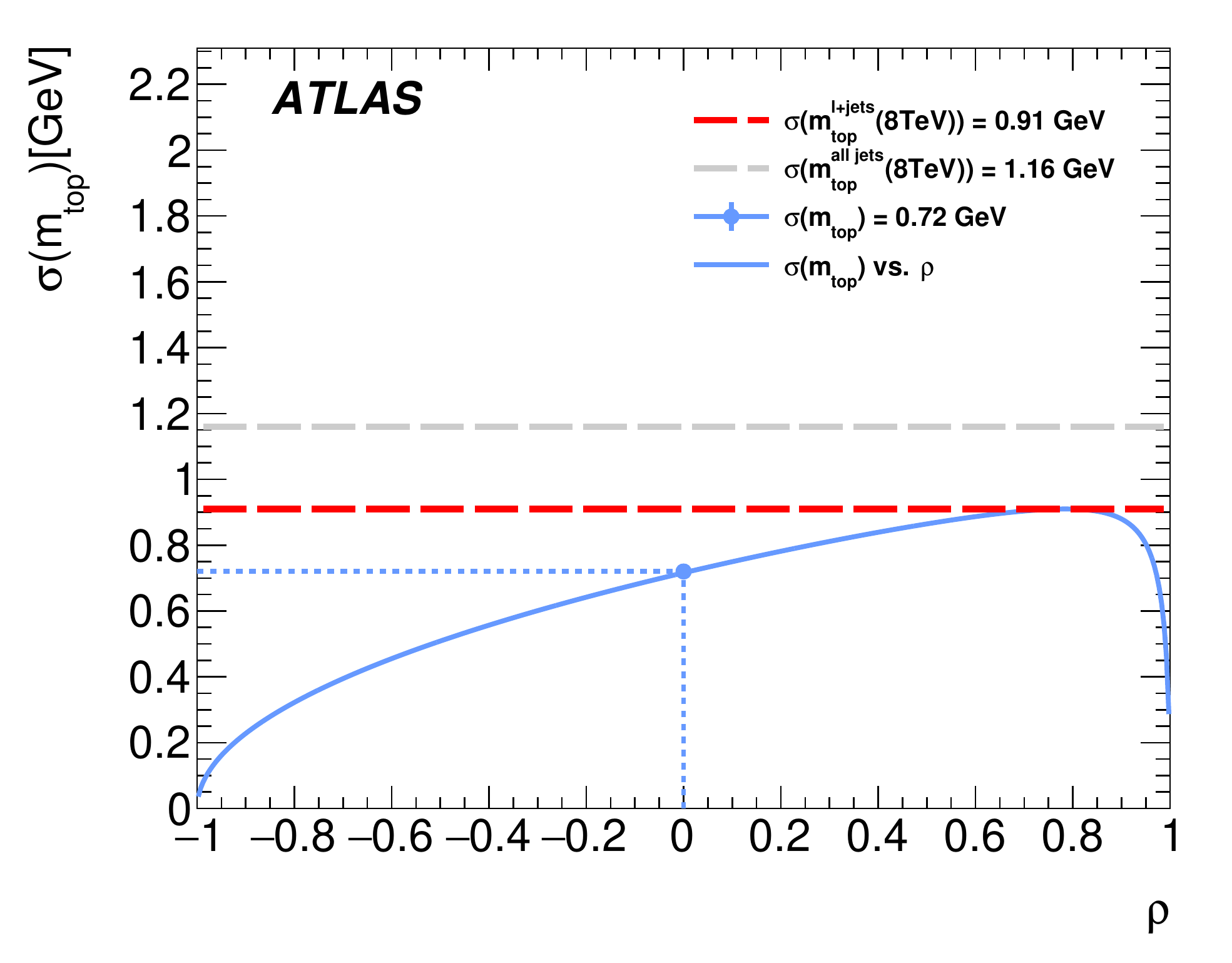}
            \label{fig:comb_unc_lpj8jet8}
          }
\caption{The combined values~(left) and uncertainties~(right) of the combination
  of pairs of individual results at $\sqrts=8$~\TeV, shown as functions of the
  total correlation $\rho$~(solid lines).
 The combination of the \ttbarll\ and \ttbarlj\ results is shown in the top row.
 The middle row is for the combination of the \ttbarll\ and \ttbarjj\ results.
 Finally, the combination of the \ttbarlj\ and \ttbarjj\ results is shown in the
 bottom row.
 For comparison, the corresponding values for the input results are also
 shown~(dashed lines).
  \label{fig:comb}
}
\end{figure*}

 The dependences of the combined values and their uncertainties on the total
 correlation for pairwise combinations of results are analysed.
 The dependences for pairs of the three results from $\sqrts=8$~\TeV\ data are
 shown in \Fig{\ref{fig:comb}}. The largest information gain is achieved by
 combining the \ttbarll\ and \ttbarlj\ results at $\sqrts=8$~\TeV, shown in
 \Figs{\ref{fig:comb_val_dil8lpj8}}{\ref{fig:comb_unc_dil8lpj8}}, which are 
 anti-correlated, i.e.~$\rho=\Rdlelje$.
%

\begin{figure*}[tbp!]
\centering
\includegraphics[width=0.95\textwidth]{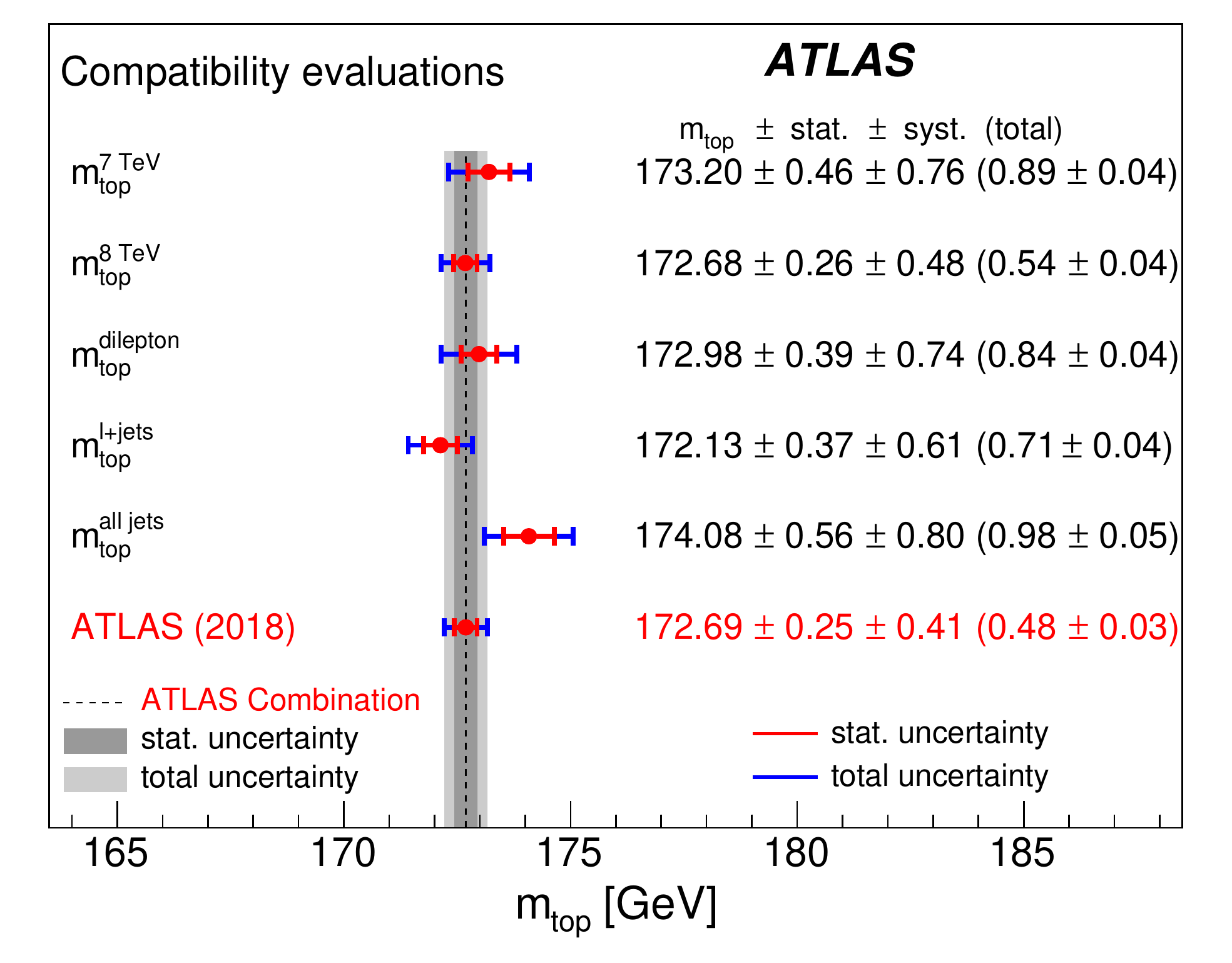}
\caption{The compatibility combinations performed for the six ATLAS results of
  \mt.
 The figure shows the combined results listed in
 \protect\Tab{\protect\ref{tab:results_comb}}, which are obtained from data
 taken at different \cmes\ and for the three decay channels, in comparison to
 the new ATLAS result.
 The values quoted are the combined value, the statistical uncertainty, the
 systematic uncertainty, the total uncertainty and the uncertainty in the total
 uncertainty.
 The results are compared with the new ATLAS combination listed in the last line
 and shown as the grey vertical bands.
 \label{fig:combchkall}}
\end{figure*}

 Based on \Tabs{\ref{tab:results_indiv}}{\ref{tab:rho_indiv}}, selected
 combinations are analysed, yielding the results given in
 \Tab{\ref{tab:results_comb}} and shown in \Fig{\ref{fig:combchkall}}.
 The \BLUE\ weights and the pulls\footnote{Using the individual results
   $\xX{i}\pm\sigmaX{i}$ and the combined result $x\pm\sigmaX{x}$, the pull of
   result $i$ is calculated as $(\xX{i}-x)/\sqrt{\sigmaXq{i}-\sigmaXq{x}}$ and
   should be Gaussian distributed with mean zero and width unity. The pull is a
   measure of the likeliness of the \xX{i}\ measured in data.} of the results
 are given in \Tab{\ref{tab:BlueCorAll}}.
%
\begin{sidewaystable*}[tbp!]
\small
\caption{The results for selected combinations based on the six results at
  $\sqrts=7$ and $8$~\TeV.
 The left two columns of results show the combination of the three results
 at~$\sqrts=7$~\TeV~(\mtseven) or at~$\sqrts=8$~\TeV~(\mteight), both
 combinations neglecting the results at the respective other \cme.
 The middle three columns show the combination of the six results as if pairs of
 measurements would determine a decay-specific top quark mass, namely \mtdl,
 \mtlj\ and \mtjj.
 Finally, shown on the right is the combination of the three most important
 results in the combination denoted by \mtsig\ and the combination of all
 results, i.e.~the ATLAS result for \mt\ shown in
 \protect\Fig{\ref{fig:combimp}}.
 For each combination, the uncertainty is given for each source of uncertainty.
 Uncertainties quoted as 0.00 are smaller than 0.005, while empty cells
 indicate uncertainties that do not apply to the respective combination.
 Finally, the total systematic uncertainty and the sum in quadrature of the
 statistical and systematic uncertainties are given.
 Both are quoted including the precision at which the respective uncertainty is
 known.
\label{tab:results_comb}}
\begin{center}
\begin{tabular}{|l|r|r||r|r|r||r|r|}\cline{2-8}
\multicolumn{1}{c|}{} & \mtseven\ [\GeV] & \mteight\ [\GeV]
                      &    \mtdl\ [\GeV] &    \mtlj\ [\GeV] & \mtjj\ [\GeV] 
                      &   \mtsig\ [\GeV] & \mt\ [\GeV] \\ \hline
Results                               &\SevenVal&\EightVal& 172.98 & 172.13 & 174.08 & \CombSigVal &\CombVal\\ \hline
Statistics                            &\SevenSta&\EightSta&   0.39 &   0.37 &   0.56 & \CombSigSta &\CombSta\\
Method                                &    0.08 &    0.06 &   0.05 &   0.10 &   0.14 &        0.06 &   0.06 \\\hline
Signal Monte Carlo generator          &    0.17 &    0.13 &   0.12 &   0.15 &   0.03 &        0.14 &   0.12 \\
Hadronization                         &    0.35 &    0.04 &   0.18 &   0.03 &   0.32 &        0.07 &   0.00 \\
Initial- and final-state QCD radiation&    0.03 &    0.15 &   0.26 &   0.07 &   0.13 &        0.07 &   0.07 \\
Underlying event                      &    0.05 &    0.07 &   0.10 &   0.01 &   0.06 &        0.05 &   0.03 \\
Colour reconnection                   &    0.02 &    0.08 &   0.02 &   0.16 &   0.02 &        0.08 &   0.08 \\
Parton distribution function          &    0.15 &    0.03 &   0.05 &   0.13 &   0.08 &        0.07 &   0.05 \\\hline
Background normalization              &    0.07 &    0.04 &   0.03 &   0.04 &   0.00 &        0.03 &   0.02 \\
$W/Z$+jets shape                      &    0.15 &    0.05 &   0.01 &   0.12 &   0.00 &        0.07 &   0.06 \\
Fake leptons shape                    &    0.03 &    0.03 &   0.07 &   0.02 &   0.00 &        0.03 &   0.03 \\
Data-driven all-jets background       &    0.04 &    0.02 &   0.01 &   0.01 &   0.22 &             &   0.03 \\\hline
Jet energy scale                      &    0.40 &    0.27 &   0.53 &   0.34 &   0.51 &        0.21 &   0.22 \\
Relative $b$-to-light-jet energy scale&    0.35 &    0.19 &   0.32 &   0.01 &   0.41 &        0.15 &   0.17 \\
Jet energy resolution                 &    0.04 &    0.10 &   0.09 &   0.16 &   0.07 &        0.10 &   0.09 \\
Jet reconstruction efficiency         &    0.09 &    0.01 &   0.01 &   0.05 &   0.00 &        0.03 &   0.03 \\
Jet vertex fraction                   &    0.00 &    0.05 &   0.02 &   0.07 &   0.02 &        0.05 &   0.05 \\
\btag                                 &    0.24 &    0.18 &   0.05 &   0.30 &   0.09 &        0.17 &   0.17 \\
Leptons                               &    0.04 &    0.10 &   0.14 &   0.11 &   0.01 &        0.09 &   0.08 \\
Missing transverse momentum           &    0.08 &    0.03 &   0.01 &   0.07 &   0.01 &        0.04 &   0.04 \\
Pile-up                               &    0.01 &    0.07 &   0.05 &   0.10 &   0.01 &        0.06 &   0.06 \\
All-jets trigger                      &    0.00 &    0.01 &   0.00 &   0.00 &   0.06 &             &   0.01 \\ 
Fast vs. full simulation              &    0.03 &         &   0.01 &   0.01 &   0.07 &             &   0.01 \\\hline
Total systematic uncertainty          & \SevenSys $\pm$ 0.04 & \EightSys $\pm$ 0.04 &        0.74 $\pm$ 0.04
                                      &      0.61 $\pm$ 0.04 &      0.80 $\pm$ 0.05 & \CombSigSys $\pm$ \CombSigUncStab 
                                      &  \CombSys $\pm$ \CombUncStab \\\hline
Total                                 & \SevenUnc $\pm$ 0.04 & \EightUnc $\pm$ 0.04 &        0.84 $\pm$ 0.04
                                      &      0.71 $\pm$ 0.04 &      0.98 $\pm$ 0.05 & \CombSigUnc $\pm$ \CombSigUncStab
                                      &  \CombUnc $\pm$ \CombUncStab \\\hline
\end{tabular}
\end{center}
\end{sidewaystable*}
%
\begin{table}[tbp!]
\begin{center}
\caption{The \BLUE\ weights and the pulls of the results for the combinations
  reported in \protect\Tab{\protect\ref{tab:results_comb}}.
 The upper part refers to the independent combinations of the three results per
 centre-of-mass energy resulting in uncorrelated results \mtseven\ and \mteight.
 The middle part is for the combination of the three observables from pairs of
 results per \ttbar\ decay channel, resulting in correlated results \mtdl,
 \mtlj\ and \mtjj.
 The lower part refers to the combination of the three most important results
 \mtsig\ and of all results \mt.
 \label{tab:BlueCorAll}}
\renewcommand{\arraystretch}{1.2}
\begin{tabular}{|r|r|r|r|r|r|r|r|}\cline{3-8}
\multicolumn{2}{c|}{} & \multicolumn{3}{c|}{$\sqrts=7$~\TeV}
                      & \multicolumn{3}{c|}{$\sqrts=8$~\TeV} \\\cline{3-8}
\multicolumn{2}{c|}{} & \mtdl & \mtlj & \mtjj & \mtdl & \mtlj & \mtjj \\\hline
 \mtseven & weight & $ 0.36$& $ 0.51$& $ 0.13$&       &       &        \\
          &   pull & $ 0.54$& $-0.95$& $ 1.18$&       &       &        \\
 \mteight & weight &        &        &        & $0.45$& $ 0.44$& $0.12$\\
          &   pull &        &        &        & $0.48$& $-0.82$& $1.02$\\\hline
    \mtdl & weight & $ 0.08$& $-0.04$& $-0.04$& $0.92$& $ 0.04$& $ 0.04$\\
    \mtlj & weight & $-0.09$& $ 0.33$& $ 0.03$& $0.09$& $ 0.67$& $-0.03$\\
    \mtjj & weight & $-0.04$& $-0.01$& $ 0.30$& $0.04$& $ 0.01$& $ 0.71$\\
          &   pull & $ 0.71$& $ 0.18$& $ 0.64$& $0.05$& $-0.09$& $-0.58$\\\hline
   \mtsig & weight &        & $ 0.17$&        & $0.43$& $ 0.40$&        \\
          &   pull &        & $-0.16$&        & $0.70$& $-0.57$&        \\
      \mt & weight & $-0.03$& $ 0.16$&  $0.04$& $0.35$& $ 0.37$& $ 0.10$\\
          &   pull & $ 0.83$& $-0.30$&  $1.36$& $0.43$& $-0.79$& $ 0.98$\\\hline
\end{tabular}
\renewcommand{\arraystretch}{1.0}
\end{center}
\end{table}

 To investigate the difference in precision of combined results obtained from
 $\sqrts=7$ and $8$~\TeV\ results, two independent combinations of the three
 results per \cme\ are performed.
 For each decay channel, the results at $\sqrts=8$~\TeV\ are significantly more
 precise than those at $\sqrts=7$~\TeV.
 In addition, the two most precise results per \cme\ are significantly less
 correlated at $\sqrts=8$~\TeV\ than at $\sqrts=7$~\TeV.
 Consequently, the size of the uncertainty of the combined result at
 $\sqrts=8$~\TeV~(\mteight) is \EightSevenImp\ smaller than the one obtained
 from the results at $\sqrts=7$~\TeV~(\mtseven).
 As shown in \Figs{\ref{fig:TopMassAccImpSev}}{\ref{fig:TopMassAccImpEig}} in
 \App{\ref{sect:addcom}}, for both \cmes, the combination is dominated by the
 results in the \ttbarll\ and \ttbarlj\ channels.

 To investigate whether the measured \mt\ depends on the \ttbar\ decay mode, a
 combination of the six results is performed in which the results in the three
 \ttbar\ decay channels are treated as determining potentially different masses,
 namely \mtdl, \mtlj\ and \mtjj.
 In such a combination, results obtained in one decay channel influence the
 combined result in another decay channel by means of their estimator
 correlation.
 Therefore, for each observable, e.g.~\mtdl, by construction the sum of weights
 of the results in the corresponding decay channel equals unity, while for each
 of the other decay channels the sum of weights of the results equals
 zero~\cite{BLUE2}.
 The combination yields compatible results for the three masses listed in
 \Tab{\ref{tab:results_comb}}. Consequently, the data do not show any sign of a
 decay-channel-dependent \mt.
 The correlation matrix of the three observables 0, 1 and 2 corresponding to
 \mtdl, \mtlj\ and \mtjj\ is
%
\begin{linenomath*}
 \begin{align}
  \rhoX{\mt} &= \left ( \begin{array}{rrr}
                         1 & \Rmdlmlj & \Rmdlmjj   \\
                  \Rmdlmlj &        1 & \Rmljmjj  \\
                  \Rmdlmjj & \Rmljmjj & 1
                 \end{array} \right) ,
 \nonumber
 \end{align}
\end{linenomath*}
%
 and the smallest \chiq\ probability of any pair of combined results for
 determining the same \mt\ is $\chipropone = \ProCompatThree$.
 As shown in \Figrange{\ref{fig:TopMassAccImpDiL}}{\ref{fig:TopMassAccImpHad}}
 in \App{\ref{sect:addcom}}, for the combination of the three observables, the
 results based on $\sqrts=7$~\TeV\ data lead to significant improvements on
 their more precise counterparts obtained from $\sqrts=8$~\TeV\ data, apart from
 the \ttbarll\ channel.
 
 Given that no dependence of \mt\ on the \cme\ or the \ttbar\ decay channel is
 expected, the above examples of combinations are merely additional
 investigations of the compatibility of the input results.
 The compatibility combinations are summarized in \Fig{\ref{fig:combchkall}} and
 listed in \Tab{\ref{tab:results_comb}}.
 For all combinations, the values quoted in \Fig{\ref{fig:combchkall}} are the
 combined value, the statistical uncertainty, the systematic uncertainty, the
 total uncertainty and the statistical uncertainty in the total uncertainty.
%
%
\subsection{The combined result of \mt}
\label{sect:combres}
\begin{figure*}[tbp!]
\centering
\centering
\subfigure[Inputs to the combination]{
  \includegraphics[width=0.48\textwidth]{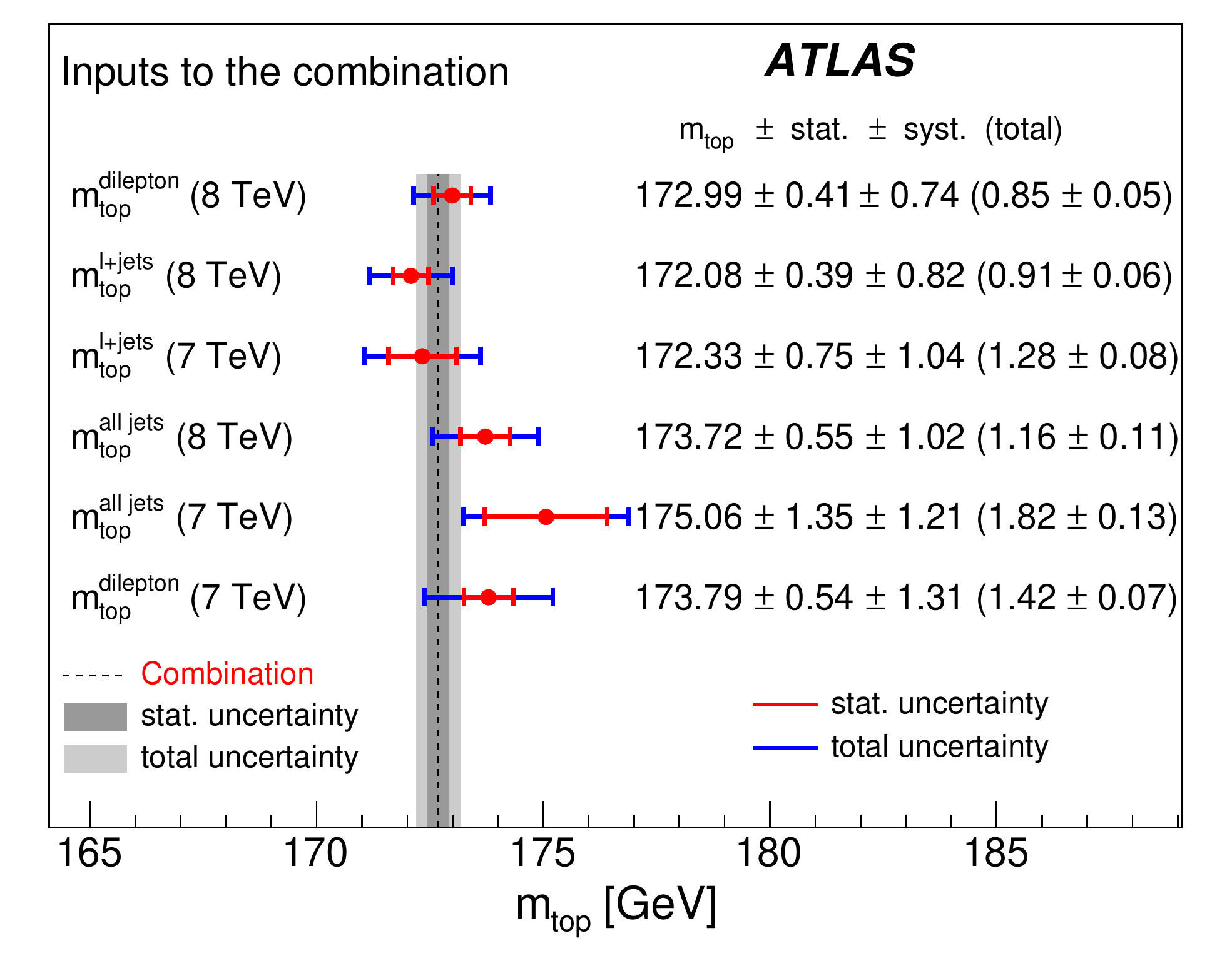}
  \label{fig:combimpinp}
}
\subfigure[Combination according to importance]{
  \includegraphics[width=0.48\textwidth]{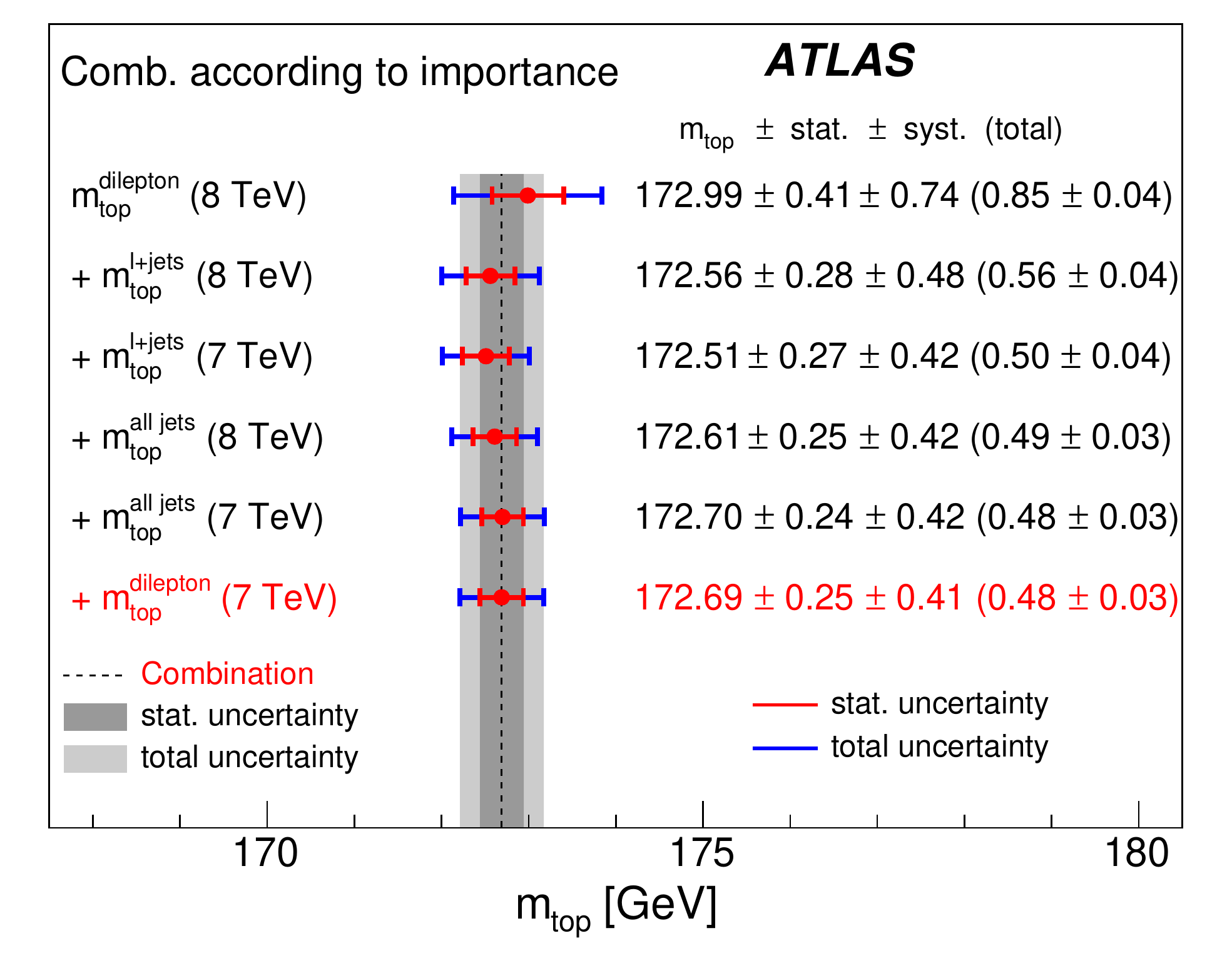}
  \label{fig:combimpcom}
}
\caption{The combination of the six ATLAS results of \mt\ according to
  importance~\protect\Ref{\cite{BLUERN}}.
 Figure~(a) shows the inputs to the combination.
 Figure~(b) shows results of the combination when successively adding results to
 the most precise one.
 The values quoted are the combined value, the statistical uncertainty, the
 systematic uncertainty, the total uncertainty and the uncertainty in the total
 uncertainty.
 In this figure, each line shows the combined result when adding the result
 listed to the combination indicated by a `+'.
 The new ATLAS combination is given in the last line, and shown in both figures
 as the vertical grey bands.
 \label{fig:combimp}
}
\end{figure*}

 The use of the statistical uncertainties in the systematic uncertainties has
 two main advantages.
 Firstly, it allows a determination of the uncertainties in the evaluation of
 the total correlations of the estimators, avoiding the need to perform ad hoc
 variations.
 Secondly, it enables the monitoring of the evolution of the combined result in
 relation to the precision in its uncertainty while including results, thereby
 evaluating their influence on the combination.
 The significance of the individual results in the combination is shown in
 \Fig{\ref{fig:combimp}}.
 The individual results are shown in \Fig{\ref{fig:combimpinp}}.
 Their combination is displayed in \Fig{\ref{fig:combimpcom}} where, following
 \Ref{\cite{BLUERN}}, starting from the most precise result, i.e.~the
 \ttbarll\ measurement at $\sqrts=8$~\TeV, results are added to the combination
 one at a time according to their importance, and the combined result is
 reported.
%
%
 Each following line of this figure shows the combined result when adding the
 result listed to the input of the combination, indicated by the `+' in front of
 the name of the added estimate.
 The last line in \Fig{\ref{fig:combimpcom}} shows the new ATLAS combined value
 of \mt.

 The inclusion of the \ttbarlj\ result at $\sqrts=8$~\TeV\ leads to the result
 quoted in the second line, which improves the combined uncertainty by much more
 than the statistical precision in the uncertainty of the most precise result.
 The same is found when adding the \ttbarlj\ result at $\sqrts=7$~\TeV\ and
 comparing with the statistical uncertainty in the previous combination, albeit
 at a much reduced significance.
 The corresponding result obtained from these three results, denoted by \mtsig,
 is also listed in \Tab{\ref{tab:results_comb}}.

 The improvement in the combination by applying the \mvabased\ selection to the
 \ttbarlj\ analysis at $\sqrts=8$~\TeV\ is sizeable.
 This is seen from repeating the combination of \mtsig but using the result from
 the standard selection from \Tab{\ref{tab:LpJresults8TeV}}.
 With this, the correlation of the $\sqrts=8$~\TeV\ \ttbarlj\ result with the
 $\sqrts=8$~\TeV\ \ttbarll\ result changes from \Rdlelje\ to \CombSigUncStdCor.
 The resulting uncertainty in the combination is
 $\XZtot{\CombSigUncStd}{\CombSigUncStdStab}$~\GeV, i.e.~the combination is
 \CombSigUncStdImp\ less precise than \mtsig\ obtained using the result from the
 \mvabased\ selection.
 Adding the remaining results reduces the quoted combined uncertainty by
 \CombSixImp~\GeV, which is smaller than the statistical precision in the
 uncertainty of the previously achieved result of \mtsig.

 The changes in statistical uncertainties in the combined value and its
 uncertainty due to variations of the input systematic uncertainties within
 their uncertainties are evaluated for two cases, namely for \mtsig\ and for the
 combination of all results.
 Following \Ref{\cite{TOPQ-2016-03}}, the distributions of the combined values
 and uncertainties are calculated from 500 combinations, where for each
 combination, the sizes of the uncertainties as well as the correlations are
 newly evaluated.
 Due to the re-evaluation of the correlation, the resulting distributions are
 not Gaussian and are also not exactly centred around the combined value and the
 combined uncertainty.
 For \mtsig, the root mean square of the distribution of the combined value is
 \CombSigValStab~\GeV, and that of the distribution of its uncertainty is
 \CombSigUncStab~\GeV.
 The corresponding values for the new ATLAS combination are
 \CombValStab~\GeV\ and \CombUncStab~\GeV, respectively.

 The full breakdown of uncertainties for the new combined ATLAS result for
 \mt\ is reported in the last column of \Tab{\ref{tab:results_comb}}.
 The combined result is
%
\begin{linenomath*}
  \begin{align*}
    \mt &=\XZ{\CombVal}{\CombSta}{\CombSys}~\GeV
  \nonumber
  \end{align*}
\end{linenomath*}
%
 with a total uncertainty of $\CombUnc\pm\CombUncStab$~\GeV, where the quoted
 uncertainty in this uncertainty is statistical.
 This means that the uncertainty in this combined result is only known to this
 precision, which, given its size, is fully adequate.

 The \chiq\ probability of \mtsig\ is \CombSigPro.
 Driven by the larger pulls of the remaining three results listed in
 \Tab{\ref{tab:BlueCorAll}}, the \chiq\ probability of \CombPro\ for the new
 ATLAS combination of \mt\ is lower but still good.
 The new ATLAS combined result of \mt\ provides a \CombImp\ improvement relative
 to the most precise single input result, which is the \ttbarll\ analysis at
 $\sqrts=8$~\TeV.
 With a relative precision of \CombPre, it improves on the previous combination
 in \Ref{\cite{TOPQ-2016-03}} by \CombImpPre\ and supersedes it.
 As shown in \App{\ref{sect:addcom}}, the new ATLAS combined result of \mt\ is
 more precise than the results from the CDF and D0 experiments, and has a
 precision similar to the CMS combined result.
%
\begin{figure*}[tbp!]
\centering
\includegraphics[width=0.95\textwidth]{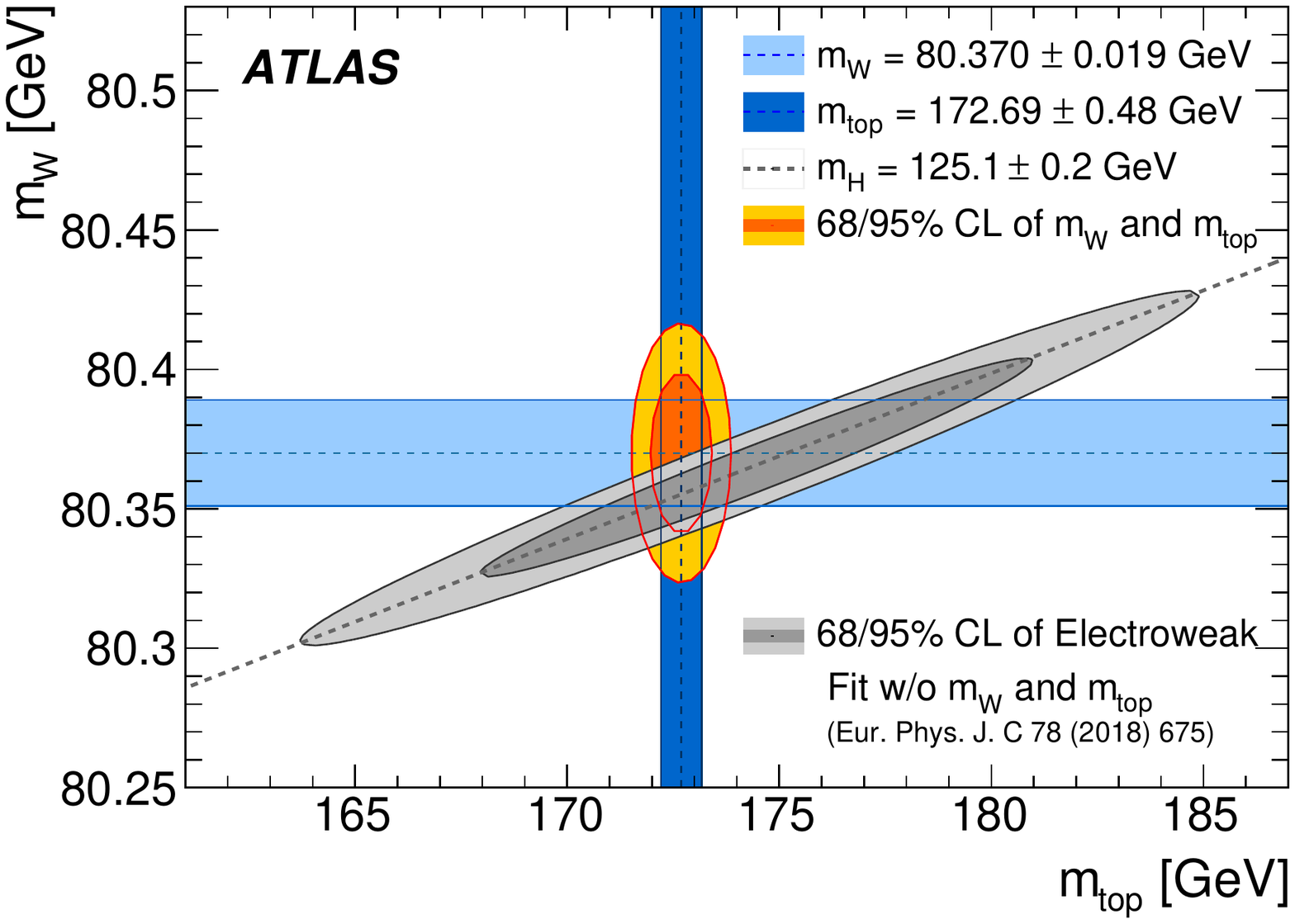}
\caption{Comparison of indirect determinations and direct measurements of the
  top quark and \Wboson\ boson masses.
 The direct ATLAS measurements of \mW\ and \mt\ are shown as the horizontal and
 vertical bands, respectively.
 Their 68$\%$ and 95$\%$ confidence-level (CL) contours are compared with the
 corresponding results from the electroweak fit.
 \label{fig:fig_EWfit}
}
\end{figure*}

 In \Fig{\ref{fig:fig_EWfit}}, the 68$\%$ and 95$\%$ confidence-level contours
 of the indirect determination of \mW\ and \mt\ from the global electroweak fit
 in \Ref{\cite{Baak2014}} are compared with the corresponding confidence-level
 contours of the direct ATLAS measurements of the two masses.
 The top quark mass used in this figure was obtained above, while the
 \Wboson\ boson mass is taken from \Ref{\cite{STDM-2014-18}}.
 The electroweak fit uses as input the LHC combined result of the Higgs boson
 mass of $\mH = 125.09 \pm 0.24$~\GeV\ from \Ref{\cite{HIGG-2014-14}}.
 There is good agreement between the direct ATLAS mass measurements and their
 indirect determinations by the electroweak fit.

\clearpage
\FloatBarrier
\section{Conclusion}
\label{sect:conclusion}
 The top quark mass is measured via a three-dimensional template method in the
 \ttbarlj\ channel and combined with previous ATLAS \mt\ measurements at the
 LHC.

 For the \ttbarlj\ analysis from $\sqrts=8$~\TeV\ proton--proton collision data
 with an integrated luminosity of about $\atlumo$~\invfb, the event selection of
 the corresponding $\sqrts=7$~\TeV\ analysis is refined.
 An optimization employing a \mvabased\ selection to efficiently suppress
 less-well-reconstructed events results in a significant reduction in total
 uncertainty, driven by a significant decrease in theory-modelling-induced
 uncertainties.
 With this approach, the measured value of \mt\ is
%
\begin{linenomath*}
  \begin{align*}
    \mt &=\XZ{\EightLpJVal}{\EightLpJSta}{\EightLpJSys}~\GeV
  \end{align*}
\end{linenomath*}
%
 with a total uncertainty of $\EightLpJTot\pm\EightLpJTotUnc$~\GeV, where the
 quoted uncertainty in the total uncertainty is statistical.
 The precision is limited by systematic uncertainties, mostly by uncertainties
 in the calibration of the jet energy scale, \btag\ and the Monte Carlo
 modelling of signal events.
 This result is more precise than the result from the CDF experiment, but less
 precise than the CMS and D0 results, measured in the same channel.

 The correlations of six measurements of \mt, performed in the three
 \ttbar\ decay channels from $\sqrts=7$ and 8~\TeV\ ATLAS data, are evaluated
 for all sources of the systematic uncertainty.
 Using a dedicated mapping of uncertainty categories, combinations are
 performed, where measurements are added one at a time according to their
 importance.
 Treating the pairs of measurements in the three \ttbar\ decay channels as
 determining potentially different masses, namely \mtdl, \mtlj\ and \mtjj,
 yields consistent values within uncertainties, i.e.~the data do not show any
 sign of a decay-channel-dependent \mt.

 The combined result of \mt\ from the six measurements is
%
\begin{linenomath*}
\begin{align*}
 \mt &= \XZ{\CombVal}{\CombSta}{\CombSys}~\GeV
\end{align*}
\end{linenomath*}
%
 with a total uncertainty of $\CombUnc\pm\CombUncStab$~\GeV, where the quoted
 uncertainty in the total uncertainty is statistical.
 This combination is dominated by three input measurements: the measurement in
 the \ttbarll\ channel from $\sqrts=8$~\TeV\ data and the two measurements in
 the \ttbarlj\ channel from $\sqrts=8$ and 7~\TeV\ data.
 With a relative precision of \CombPre, this new ATLAS combination of \mt\ is
 more precise than the result from the CDF and D0 experiments and has a
 precision similar to the CMS combined result.
 This result supersedes the previous combined ATLAS result.

 With this precision in \mt\ achieved, precise knowledge of the relation between
 the mass definition of the experimental analysis and the pole mass is becoming
 relevant.
 The combined result is mostly limited by the uncertainties in the calibration
 of the jet energy scales, \btag\ and in the Monte Carlo modelling of signal
 events.

\section*{Acknowledgements}

We thank CERN for the very successful operation of the LHC, as well as the
support staff from our institutions without whom ATLAS could not be
operated efficiently.

We acknowledge the support of ANPCyT, Argentina; YerPhI, Armenia; ARC, Australia; BMWFW and FWF, Austria; ANAS, Azerbaijan; SSTC, Belarus; CNPq and FAPESP, Brazil; NSERC, NRC and CFI, Canada; CERN; CONICYT, Chile; CAS, MOST and NSFC, China; COLCIENCIAS, Colombia; MSMT CR, MPO CR and VSC CR, Czech Republic; DNRF and DNSRC, Denmark; IN2P3-CNRS, CEA-DRF/IRFU, France; SRNSFG, Georgia; BMBF, HGF, and MPG, Germany; GSRT, Greece; RGC, Hong Kong SAR, China; ISF and Benoziyo Center, Israel; INFN, Italy; MEXT and JSPS, Japan; CNRST, Morocco; NWO, Netherlands; RCN, Norway; MNiSW and NCN, Poland; FCT, Portugal; MNE/IFA, Romania; MES of Russia and NRC KI, Russian Federation; JINR; MESTD, Serbia; MSSR, Slovakia; ARRS and MIZ\v{S}, Slovenia; DST/NRF, South Africa; MINECO, Spain; SRC and Wallenberg Foundation, Sweden; SERI, SNSF and Cantons of Bern and Geneva, Switzerland; MOST, Taiwan; TAEK, Turkey; STFC, United Kingdom; DOE and NSF, United States of America. In addition, individual groups and members have received support from BCKDF, CANARIE, CRC and Compute Canada, Canada; COST, ERC, ERDF, Horizon 2020, and Marie Sk{\l}odowska-Curie Actions, European Union; Investissements d' Avenir Labex and Idex, ANR, France; DFG and AvH Foundation, Germany; Herakleitos, Thales and Aristeia programmes co-financed by EU-ESF and the Greek NSRF, Greece; BSF-NSF and GIF, Israel; CERCA Programme Generalitat de Catalunya, Spain; The Royal Society and Leverhulme Trust, United Kingdom. 

The crucial computing support from all WLCG partners is acknowledged gratefully, in particular from CERN, the ATLAS Tier-1 facilities at TRIUMF (Canada), NDGF (Denmark, Norway, Sweden), CC-IN2P3 (France), KIT/GridKA (Germany), INFN-CNAF (Italy), NL-T1 (Netherlands), PIC (Spain), ASGC (Taiwan), RAL (UK) and BNL (USA), the Tier-2 facilities worldwide and large non-WLCG resource providers. Major contributors of computing resources are listed in Ref.~\cite{ATL-GEN-PUB-2016-002}.

\clearpage
\appendix
\part*{Appendices}
\addcontentsline{toc}{part}{Appendices}
\section{Results from the BDT optimization and individual
  sources of systematic uncertainty}
\label{sect:addlpj}
 This appendix has additional details of the measurement of \mt\ in the
 \ttbarlj\ channel from $\sqrts=8$~\TeV\ data discussed in the main text.
%
\begin{figure*}[tbp!]
\centering
\subfigure[Reconstructed top quark mass]
          {\includegraphics[width=0.49\textwidth]{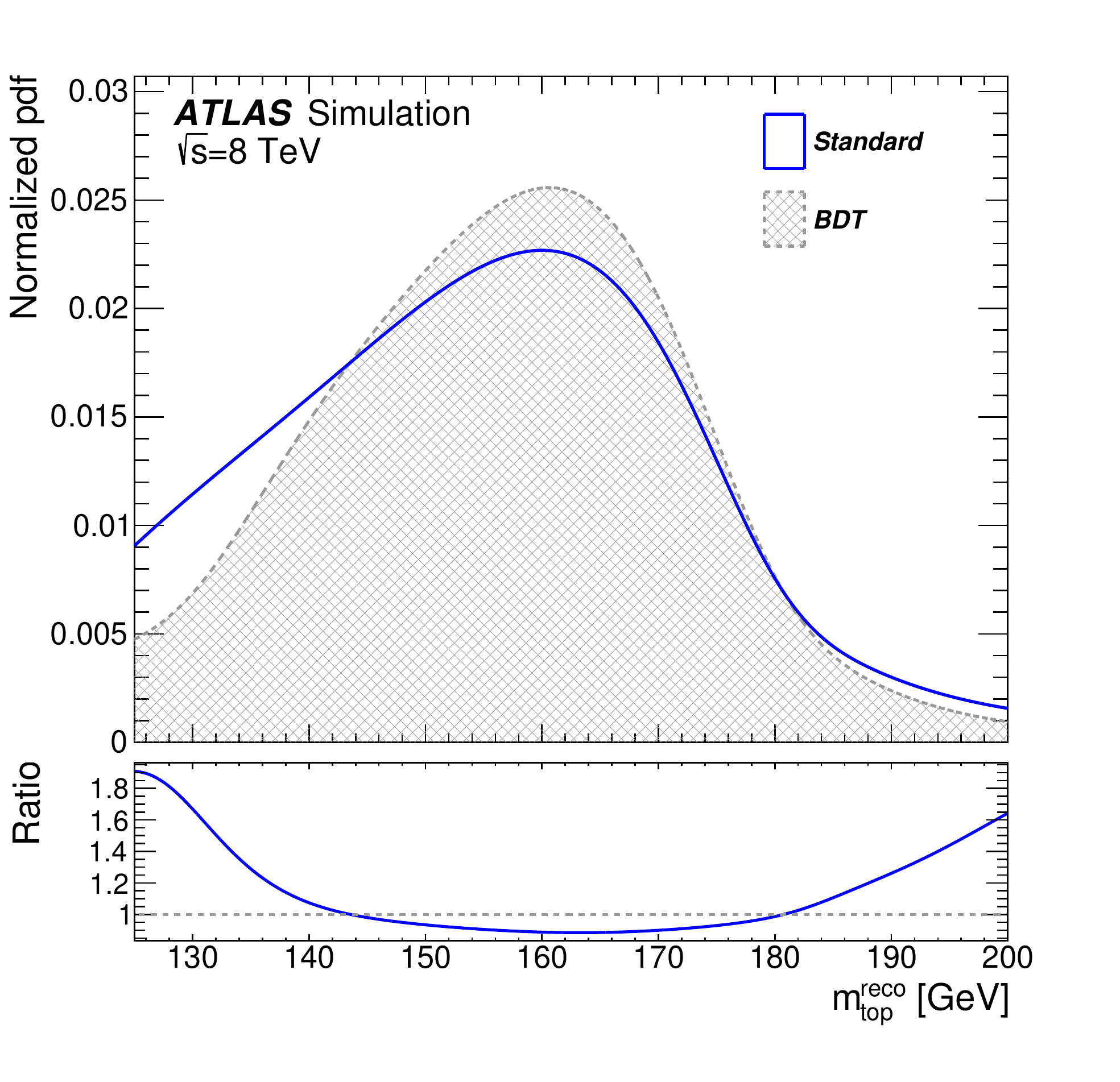}}
\subfigure[Reconstructed \Wboson\ boson mass]
          {\includegraphics[width=0.49\textwidth]{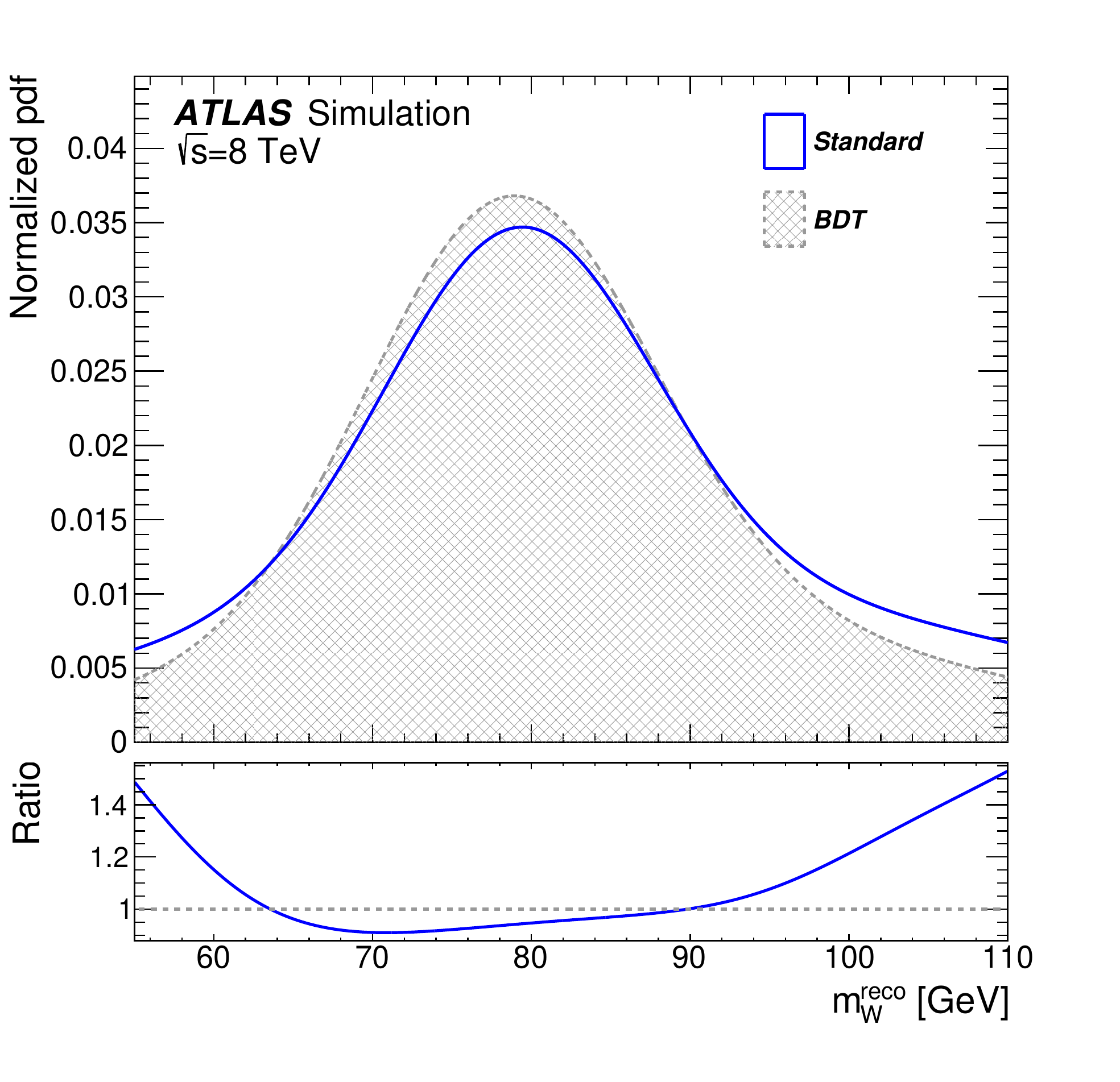}}
\hfill
\subfigure[Reconstructed ratio of jet transverse momenta]
          {\includegraphics[width=0.49\textwidth]{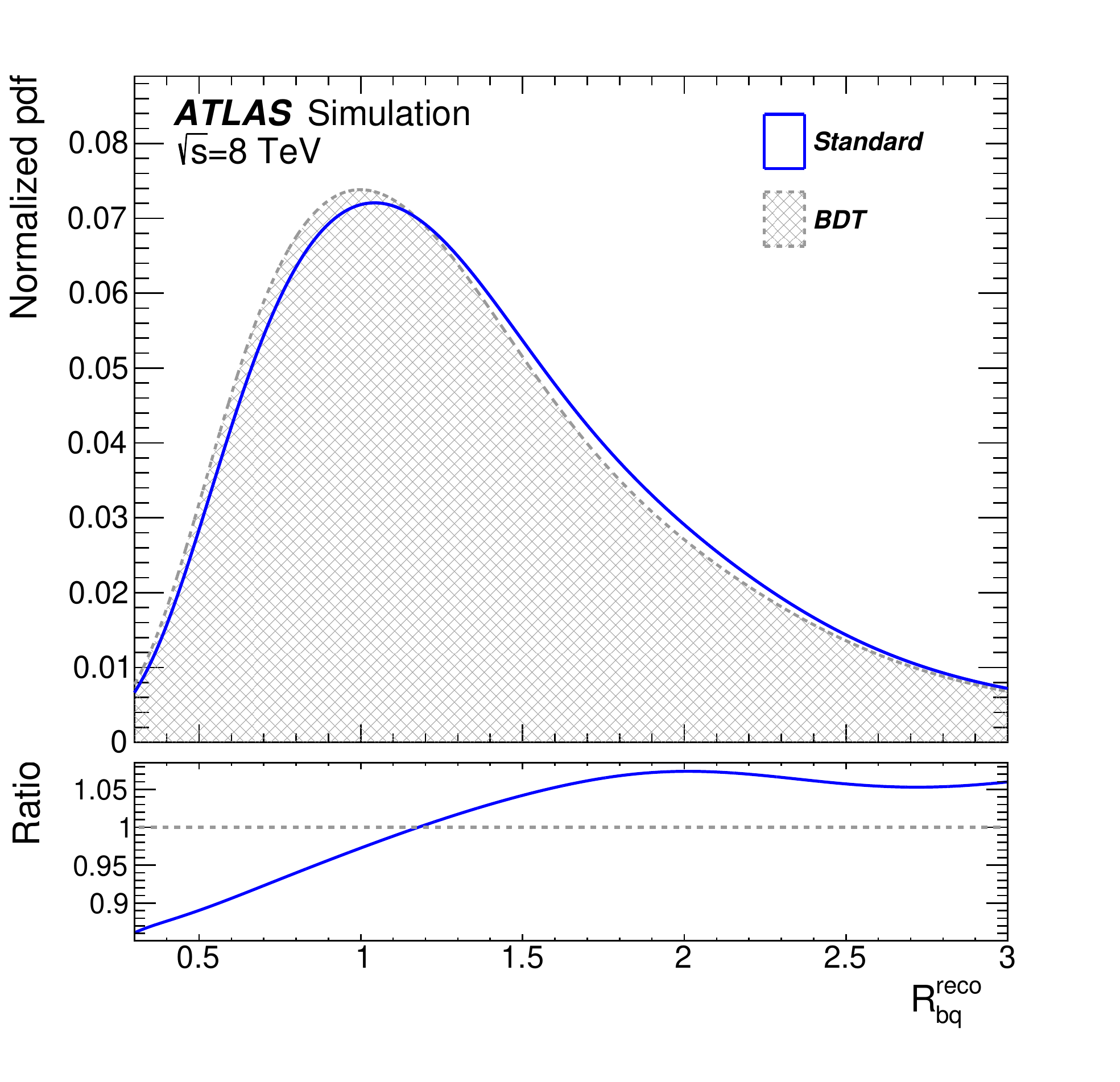}}
\caption{Comparison of the template fit functions of the three observables for
  the standard event selection~(solid line) and the \mvabased\ event
  selection~(dashed line).
 The ratios (standard / \mvabased) of the pairs of functions are also shown.
 Figure~(a) shows the reconstructed top quark mass \mtr,
 figure~(b) shows the reconstructed \Wboson\ boson mass \mWr and
 figure~(c) shows the reconstructed jet-\pT\ ratio \rbqr.
 \label{fig:Resolution}
}
\end{figure*}
 
 In \Fig{\ref{fig:Resolution}}, the template fit functions of the three
 observables are compared for the standard and the \mvabased\ event selection.
 The distributions of \mtr\ and \mWr\ are narrower for the \mvabased\ event
 selection, which means the resolution in the two masses is improved compared 
 with what is observed for the standard selection.
 The \rbqr\ distribution is slightly shifted to lower values for the
 \mvabased\ event selection, but the difference is small.
 
 For the \mvabased\ selection, a number of systematic uncertainties listed in
 \Tab{\ref{tab:LpJresults8TeV}} are calculated by performing pseudo-experiments
 for more than one systematic variation.
 The individual components are given in
 \Tabrange{\ref{tab:LpJresults8TeVbreak}}{\ref{tab:btagresults8TeV}} below.
 Whenever the uncertainty is obtained from just a pair of samples, the shift
 \dmt\ is listed together with the definition of the difference evaluated
 indicating the direction of the shift.
 For the other cases, the shift in \mt\ is quoted relative to the value measured
 in the central sample for the upward variation~(\dmtup), the downward
 variation~(\dmtdw) and the final shift assigned to this uncertainty
 component~(\dmt).
 For most of the cases the signs of \dmtup\ and \dmtdw\ are different,
 indicating that \mt\ from the central sample is surrounded by the values from
 the two variations. In this case $\vert\dmt\vert$ is
 $0.5\cdot\vert\dmtup-\dmtdw\vert$, otherwise it is the maximum of
 $\vert\dmtup\vert$ and $\vert\dmtdw\vert$. 
 In both cases, the sign of \dmt\ is the one from \dmtup.
%
\begin{table*}[tbph!]
\caption{The individual components of the uncertainty sources considered for the
  \ttbarlj\ analysis at $\sqrts=8$~\TeV\ for the sources of uncertainty not
  documented in
  \protect\Tabrange{\ref{tab:pdfresults8TeV}}{\ref{tab:btagresults8TeV}}.
 The uncertainties together with their statistical precisions are listed in
 boldface and given with 0.01~\GeV\ precision. The uncertainty per source is
 calculated as the sum in quadrature of the subcomponents.
 Uncertainties quoted as 0.00~(0.000) are smaller than 0.005~(0.0005).
 \label{tab:LpJresults8TeVbreak}
}
\begin{center}
\begin{tabular}{|l|r|r|c|}\hline
 Uncertainty & \dmtup~[\GeV] & \dmtdw~[\GeV] &  \dmt~[\GeV] \\
\hline
\textbf{Signal Monte Carlo generator} & 
\multicolumn{3}{c|}{\boldmath $0.16 \pm 0.17$} \\ \hline
$~~~$ \PowhegBox\ - \Mcatnlo~(\HERWIG) &      &          & $-0.161 \pm 0.168$ \\
\hline
\textbf{Hadronization} & 
\multicolumn{3}{c|}{\boldmath $0.15 \pm 0.10$} \\ \hline
$~~~$ \PowhegPythia\ - \PowhegHerwig\  &      &          & $+0.146 \pm 0.098$ \\
\hline
\textbf{Initial- and final-state QCD radiation} & 
\multicolumn{3}{c|}{\boldmath $0.08 \pm 0.11$} \\ \hline
$~~~$ less I/FSR - more I/FSR      & $+$0.086 & $-$0.075 & $+0.080 \pm 0.111$ \\
\hline
\textbf{Underlying event} & 
\multicolumn{3}{c|}{\boldmath $0.08 \pm 0.15$} \\ \hline
$~~~$  P2012 - P2012 \mpiHi\       &          &          & $-0.080 \pm 0.153$ \\
\hline
\textbf{Colour reconnection} & 
\multicolumn{3}{c|}{\boldmath $0.19 \pm 0.15$} \\ \hline
$~~~$ P2012 - P2012 \loCR\         &          &          & $+0.191 \pm 0.154$ \\
\hline
\textbf{Background normalization} & 
\multicolumn{3}{c|}{\boldmath $0.08 \pm 0.00$} \\ \hline
$~~~$ $Z$+jets norm.               & $+$0.007 & $-$0.015 & $+0.011 \pm 0.000$ \\
$~~~$ $W$+jets norm.               & $-$0.017 & $-$0.061 & $-0.061 \pm 0.000$ \\
$~~~$ Fake lepton norm.            &          &          & $+0.046 \pm 0.000$ \\
\hline
\textbf{ $W/Z$+jets shape} & 
\multicolumn{3}{c|}{\boldmath $0.11 \pm 0.00$} \\ \hline
$~~~$ $W$+jets HF0                 & $-$0.001 & $-$0.070 & $-0.070 \pm 0.000 $ \\
$~~~$ $W$+jets HF1                 & $-$0.005 & $-$0.087 & $-0.087 \pm 0.000 $ \\
\hline
\textbf{Jet reconstruction efficiency} & 
\multicolumn{3}{c|}{\boldmath $0.02 \pm 0.01$} \\ \hline
$~~~$ nominal - 0.23$\%$ drop      &          &          & $+0.022 \pm 0.013$ \\
\hline
\textbf{Jet vertex fraction} & 
\multicolumn{3}{c|}{\boldmath $0.09 \pm 0.01$} \\ \hline
$~~~$                              & $+$0.077 & $-$0.112 & $0.095 \pm 0.009$ \\
\hline
\textbf{Leptons} & 
\multicolumn{3}{c|}{\boldmath $0.16 \pm 0.01$} \\ \hline
$~~~$ Electron energy scale        & $+$0.025 & $-$0.006 & $+0.016 \pm 0.006$ \\
$~~~$ Electron energy resolution   & $-$0.152 & $-$0.145 & $-0.152 \pm 0.013$ \\
$~~~$ Muon resolution~(muon spectrometer)
                        &          &          & $+0.027 \pm 0.000$ \\
$~~~$ Muon resolution~(inner detector) 
             &          &          & $+0.023 \pm 0.000$ \\
$~~~$ Muon scale                   & $-$0.013 & $+$0.015 & $-0.014 \pm 0.000$ \\
$~~~$ Lepton trigger SF            & $-$0.005 & $-$0.003 & $-0.005 \pm 0.001$ \\
$~~~$ Lepton identification SF     & $+$0.005 & $-$0.011 & $+0.008 \pm 0.001$ \\
$~~~$ Lepton reconstruction SF     & $+$0.003 & $-$0.008 & $+0.005 \pm 0.000$ \\
\hline
\textbf{Missing transverse momentum}(\met) & 
\multicolumn{3}{c|}{\boldmath $0.05 \pm 0.01$} \\ \hline
$~~~$ \met~(resolution soft term)  & $+$0.003 & $+$0.012 & $+0.012 \pm 0.018$ \\
$~~~$ \met~(scale soft term)       & $+$0.054 & $-$0.039 & $+0.047 \pm 0.009$ \\
\hline
\end{tabular}
\end{center}
\end{table*}
%
\begin{table*}[tbp!]
\caption{The individual components of the PDF uncertainty considered for the
  \ttbarlj\ analysis at $\sqrts=8$~\TeV, the resulting PDF-uncertainty-induced
  shifts in \mt and the final uncertainty in \mt.
 The components~\cite{Pumplin2002,MAR-0901,Ball:2012cx} together with their
 statistical precisions are listed in boldface. The total uncertainty in the
 CT10 variations is calculated as the sum in quadrature of the CT10
 subcomponents.
 The total uncertainty is given with 0.01~\GeV\ precision.
 Uncertainties quoted as 0.00~(0.000) are smaller than 0.005~(0.0005).
 The term {\it nuisance parameter} is denoted by {\it NuP}.
 The last line refers to the sum in quadrature of the PDF subcomponents.
\label{tab:pdfresults8TeV}
}
\begin{center}
\begin{tabular}{|l|r|}\hline
PDF uncertainty components  &           \dmt~[\GeV] \\\hline
\textbf{CT10 variations}    & \boldmath$ 0.09  \pm 0.00 $ \\
$~~~$ CT10 NuP2 - NuP1      &          $-0.001 \pm 0.000$ \\
$~~~$ CT10 NuP4 - NuP3      &          $+0.000 \pm 0.000$ \\
$~~~$ CT10 NuP6 - NuP5      &          $+0.015 \pm 0.000$ \\
$~~~$ CT10 NuP8 - NuP7      &          $+0.005 \pm 0.003$ \\
$~~~$ CT10 NuP10 - NuP9     &          $+0.004 \pm 0.000$ \\
$~~~$ CT10 NuP12 - NuP11    &          $+0.002 \pm 0.000$ \\
$~~~$ CT10 NuP14 - NuP13    &          $-0.026 \pm 0.001$ \\
$~~~$ CT10 NuP16 - NuP15    &          $-0.004 \pm 0.000$ \\
$~~~$ CT10 NuP18 - NuP17    &          $-0.015 \pm 0.001$ \\
$~~~$ CT10 NuP20 - NuP19    &          $+0.013 \pm 0.001$ \\
$~~~$ CT10 NuP22 - NuP21    &          $+0.006 \pm 0.001$ \\
$~~~$ CT10 NuP24 - NuP23    &          $+0.063 \pm 0.001$ \\
$~~~$ CT10 NuP26 - NuP25    &          $+0.000 \pm 0.001$ \\
$~~~$ CT10 NuP28 - NuP27    &          $+0.009 \pm 0.000$ \\
$~~~$ CT10 NuP30 - NuP29    &          $-0.004 \pm 0.000$ \\
$~~~$ CT10 NuP32 - NuP31    &          $+0.007 \pm 0.001$ \\
$~~~$ CT10 NuP34 - NuP33    &          $+0.019 \pm 0.002$ \\
$~~~$ CT10 NuP36 - NuP35    &          $-0.011 \pm 0.001$ \\
$~~~$ CT10 NuP38 - NuP37    &          $-0.001 \pm 0.000$ \\
$~~~$ CT10 NuP40 - NuP39    &          $-0.001 \pm 0.001$ \\
$~~~$ CT10 NuP42 - NuP41    &          $-0.005 \pm 0.001$ \\
$~~~$ CT10 NuP44 - NuP43    &          $-0.003 \pm 0.000$ \\
$~~~$ CT10 NuP46 - NuP45    &          $-0.002 \pm 0.000$ \\
$~~~$ CT10 NuP48 - NuP47    &          $+0.027 \pm 0.000$ \\
$~~~$ CT10 NuP50 - NuP49    &          $+0.002 \pm 0.000$ \\
$~~~$ CT10 NuP52 - NuP51    &          $-0.003 \pm 0.001$ \\
\textbf{NNPDF - CT10}       & \boldmath$-0.034 \pm 0.001$ \\
\textbf{MSTW - CT10}        & \boldmath$-0.024 \pm 0.002$ \\\hline
Total                       &          $ 0.09  \pm 0.00 $ \\\hline
\end{tabular}
\end{center}
\end{table*}
%
\begin{table*}[tbp!]
\caption{The individual components of the \JES\ uncertainty considered for the
  \ttbarlj\ analysis at $\sqrts=8$~\TeV, the resulting \JES- and
  bJES-uncertainty-induced shifts in \mt\ and the final uncertainty in \mt.
 The components~\cite{PERF-2012-01} together with their statistical precisions
 are listed in boldface and, wherever applicable, calculated as the sum in
 quadrature of the respective subcomponents.
 A shift listed as `0' means that the corresponding variation resulted in an
 unchanged event sample.
 Uncertainties quoted as 0.00~(0.000) are smaller than 0.005~(0.0005).
 In the rightmost column, the mapping to the uncertainty components used for
 $\sqrts=7$~\TeV\ data is given for the weak and the strong correlation
 scenarios.
 The `+' sign indicates corresponding components at the two \cmes\ for the weak
 and strong scenario, while the~`(+)' sign indicates components that only
 correspond for the strong scenario. Finally, mentioning a name indicates that
 the mapped sources carry different names at $\sqrts=7$ and $8$~\TeV.
 The uncertainty components and the total uncertainty are given with
 0.01~\GeV\ precision.
 The term {\it nuisance parameter} is denoted by {\it NuP}.
 The last line refers to the sum in quadrature of the JES components.
\label{tab:jesresults8TeV}
}
\begin{center}
\begin{tabular}{|l|r|r|r|l|}\hline
 JES uncertainty components & \dmtup~[\GeV]                    & \dmtdw~[\GeV]
                            & \multicolumn{1}{c|}{\dmt~[\GeV]} & Map to $\sqrts=7$~\TeV \\
\hline
\textbf{Statistical~(total)}                 & \multicolumn{3}{c|}{\boldmath$ 0.17  \pm 0.01 $} & \\
\hline
$~~~$ Statistical NuP1                       & $-$0.159 & $+$0.151 &         $-0.155 \pm 0.013$ & \\
$~~~$ Statistical NuP2                       & $-$0.008 & $+$0.048 &         $-0.028 \pm 0.006$ & \\
$~~~$ Statistical NuP3                       & $+$0.063 & $-$0.035 &         $+0.049 \pm 0.009$ & \\
$~~~$ Statistical NuP4                       & $+$0.020 & $+$0.022 &         $+0.022 \pm 0.018$ & \\
$~~~$ $\eta$ inter-calibration~(stat.)       & $-$0.062 & $+$0.020 &         $-0.041 \pm 0.011$ & \\
\hline
\textbf{Modelling~(total)}                   & \multicolumn{3}{c|}{\boldmath$ 0.38  \pm 0.02 $} & \\
\hline
$~~~$ Modelling NuP1                         & $-$0.372 & $+$0.389 &         $-0.380 \pm 0.020$ & (+) \\
$~~~$ Modelling NuP2                         & $+$0.029 & $-$0.000 &         $+0.014 \pm 0.006$ & (+) \\
$~~~$ Modelling NuP3                         & $+$0.010 & $+$0.005 &         $+0.010 \pm 0.018$ & (+) \\
$~~~$ Modelling NuP4                         & $+$0.034 & $-$0.026 &         $+0.030 \pm 0.006$ & (+) \\
$~~~$ $\eta$ inter-calibration~(model)       & $+$0.056 & $-$0.038 &         $+0.047 \pm 0.013$ & + \\  
\hline
\textbf{Detector~(total)}                    & \multicolumn{3}{c|}{\boldmath$ 0.11  \pm 0.01 $} & \\
\hline
$~~~$ Detector NuP1                          & $+$0.116 & $-$0.103 &         $+0.110 \pm 0.011$ & + \\
$~~~$ Detector NuP2                          & $-$0.015 & $+$0.017 &         $-0.016 \pm 0.009$ &  \\
$~~~$ Detector NuP3                          & $+$0.015 & $-$0.014 &         $+0.015 \pm 0.006$ & (+) Detector NuP2 \\
\hline
\textbf{Mixed~(total)}                       & \multicolumn{3}{c|}{\boldmath$ 0.09  \pm 0.01 $} & \\ 
\hline
$~~~$ Mixed NuP1                             & $-$0.004 & $-$0.029 &         $-0.029 \pm 0.022$ & (+) \\
$~~~$ Mixed NuP2                             & $+$0.053 & $-$0.054 &         $+0.054 \pm 0.009$ & \\
$~~~$ Mixed NuP3                             & $-$0.044 & $+$0.061 &         $-0.052 \pm 0.009$ & \\
$~~~$ Mixed NuP4                             & $+$0.039 & $-$0.016 &         $+0.028 \pm 0.006$ & (+) Mixed NuP2 \\
\hline
\textbf{Single particle high-\boldmath$\pt$} & \multicolumn{3}{c|}{\boldmath$0.01 \pm 0.00$} & \\
\hline
$~~~$ Single particle high-\boldmath$\pt$    &        0 & $+$0.005 &         $+0.005 \pm 0.000$ & \\
\hline
\textbf{Pile-up~(total)}                     & \multicolumn{3}{c|}{\boldmath$ 0.18  \pm 0.02 $} & \\
\hline
$~~~$ Pile-up: offset~($\mu$)                & $+$0.041 & $-$0.040 &         $+0.041 \pm 0.011$ & + \\         
$~~~$ Pile-up: offset~(\nvtx)                & $+$0.065 & $-$0.083 &         $+0.074 \pm 0.017$ & + \\ 
$~~~$ Pile-up: \pt                           & $+$0.042 & $+$0.040 &         $+0.042 \pm 0.018$ & \\ 
$~~~$ Pile-up: $\rho$                        & $-$0.173 & $+$0.141 &         $-0.157 \pm 0.017$ & \\
\hline
\textbf{Punch-through}                       & \multicolumn{3}{c|}{\boldmath$0.02 \pm 0.01$} & \\
\hline
Punch-through                                & $+$0.013 & $+$0.017 &         $+0.017 \pm 0.013$ & \\
\hline
\textbf{Flavour~(total)}                     & \multicolumn{3}{c|}{\boldmath$ 0.24  \pm 0.02 $} & \\
\hline
$~~~$ Flavour composition                    & $+$0.079 & $-$0.119 &         $+0.099 \pm 0.000$ & + \\         
$~~~$ Flavour response                       & $+$0.220 & $-$0.211 &         $+0.215 \pm 0.018$ & + \\
\hline
\textbf{bJES}                                & \multicolumn{3}{c|}{\boldmath$0.03 \pm 0.01$} & + \\
\hline
bJES                                         & $+$0.006 & $-$0.047 &         $+0.026 \pm 0.013$ & + \\
\hline 
   Total~(without \bJES)                     & \multicolumn{3}{c|}{$ 0.54  \pm 0.02 $} & \\ \hline
\end{tabular}
\end{center}
\end{table*}
%
\begin{table*}[tbp!]
\caption{The individual components of the JER uncertainty considered for the
  \ttbarlj\ analysis at $\sqrts=8$~\TeV, the resulting JER-uncertainty-induced
  shifts in \mt and the final uncertainty in \mt.
 The {\it data versus simulation difference} and {\it noise forward region}
 components are quoted as the difference from the nominal sample.
 The total uncertainty is calculated as the sum in quadrature of the
 subcomponents and given with 0.01~\GeV\ precision.
 The term {\it nuisance parameter} is denoted by {\it NuP}.
\label{tab:jerresults8TeV}
}
\begin{center}
\begin{tabular}{|l|r|r|c|}\hline
JER uncertainty components    & \dmtup~[\GeV] & \dmtdw~[\GeV] & \dmt~[\GeV] \\ \hline
JER data versus simulation difference &       &            & $-0.034 \pm 0.018$ \\
JER noise forward region      &        &            & $+0.032 \pm 0.013$ \\
JER NuP1~(only down var.)     &        &            & $-0.111 \pm 0.052$ \\
JER NuP2                      & $-$0.034 & $+$0.055 & $-0.044 \pm 0.018$ \\
JER NuP3                      & $+$0.025 & $-$0.084 & $+0.054 \pm 0.024$ \\
JER NuP4                      & $-$0.074 & $+$0.090 & $-0.082 \pm 0.021$ \\
JER NuP5                      & $+$0.078 & $+$0.016 & $+0.078 \pm 0.037$ \\
JER NuP6                      & $-$0.041 & $+$0.017 & $-0.029 \pm 0.019$ \\
JER NuP7                      & $-$0.039 & $+$0.076 & $-0.057 \pm 0.016$ \\
JER NuP8                      & $-$0.053 & $-$0.013 & $-0.053 \pm 0.029$ \\
JER NuP9~(only up var.)       &          &          & $+0.036 \pm 0.018$ \\
\hline
Total                         &          &          & $ 0.20  \pm 0.04 $ \\
\hline
\end{tabular}
\end{center}
\end{table*}
%
\begin{table*}[tbp!]
\caption{The individual components of the flavour-tagging uncertainty considered
  for the \ttbarlj\ analysis at $\sqrts=8$~\TeV, the resulting shifts in
  \mt\ and the final \btag-uncertainty-induced uncertainty in \mt.
 The uncertainty components together with their statistical precisions are
 listed in boldface and calculated as the sum in quadrature of the respective
 subcomponents.
 The uncertainty components and the total uncertainty are given with
 0.01~\GeV\ precision.
 Uncertainties quoted as 0.00~(0.000) are smaller than 0.005~(0.0005).
 The term {\it nuisance parameter} is denoted by {\it NuP}.
 The last line refers to the sum in quadrature of the components.
\label{tab:btagresults8TeV}
}
\begin{center}
\begin{tabular}{|l|r|r|c|}\hline
Flavour-tagging uncertainty components     & \dmtup~[\GeV]& \dmtdw~[\GeV]& \dmt~[\GeV] \\ \hline
\textbf{\btag\ scale factor variations}    & \multicolumn{3}{c|}{\boldmath$ 0.31 \pm 0.00 $} \\
\hline
$~~~$ \btag\ NuP1                          & $+$0.051 & $-$0.062 & $+0.057 \pm 0.000$ \\
$~~~$ \btag\ NuP2                          & $-$0.230 & $+$0.143 & $-0.187 \pm 0.000$ \\
$~~~$ \btag\ NuP3                          & $-$0.090 & $+$0.005 & $-0.047 \pm 0.001$ \\
$~~~$ \btag\ NuP4                          & $+$0.090 & $-$0.175 & $+0.132 \pm 0.000$ \\
$~~~$ \btag\ NuP5                          & $-$0.230 & $+$0.148 & $-0.189 \pm 0.000$ \\
$~~~$ \btag\ NuP6                          & $+$0.023 & $-$0.105 & $+0.064 \pm 0.001$ \\
\hline
\textbf{\ctautag\ scale factor variations} & \multicolumn{3}{c|}{\boldmath$ 0.15  \pm 0.00 $} \\
\hline
$~~~$ \ctautag\ NuP1                       & $+$0.098 & $-$0.102 & $+0.100 \pm 0.000$ \\
$~~~$ \ctautag\ NuP2                       & $-$0.057 & $-$0.026 & $-0.057 \pm 0.000$ \\
$~~~$ \ctautag\ NuP3                       & $+$0.046 & $-$0.125 & $+0.085 \pm 0.000$ \\
$~~~$ \ctautag\ NuP4                       & $-$0.057 & $-$0.023 & $-0.057 \pm 0.000$ \\
\hline
\textbf{\Mistag\ scale factor variations}  & \multicolumn{3}{c|}{\boldmath$ 0.16  \pm 0.00 $} \\
\hline
$~~~$ \Mistag\ NuP1                        & $-$0.005 & $+$0.003 & $-0.004 \pm 0.000$ \\
$~~~$ \Mistag\ NuP2                        & $-$0.042 & $-$0.039 & $-0.042 \pm 0.000$ \\
$~~~$ \Mistag\ NuP3                        & $-$0.038 & $-$0.036 & $-0.038 \pm 0.000$ \\
$~~~$ \Mistag\ NuP4                        & $-$0.032 & $-$0.040 & $-0.040 \pm 0.000$ \\
$~~~$ \Mistag\ NuP5                        & $-$0.037 & $-$0.044 & $-0.044 \pm 0.000$ \\
$~~~$ \Mistag\ NuP6                        & $-$0.036 & $-$0.045 & $-0.045 \pm 0.000$ \\
$~~~$ \Mistag\ NuP7                        & $-$0.034 & $-$0.040 & $-0.040 \pm 0.000$ \\
$~~~$ \Mistag\ NuP8                        & $-$0.041 & $-$0.040 & $-0.041 \pm 0.000$ \\
$~~~$ \Mistag\ NuP9                        & $-$0.029 & $-$0.045 & $-0.045 \pm 0.000$ \\
$~~~$ \Mistag\ NuP10                       & $-$0.073 & $-$0.001 & $-0.073 \pm 0.000$ \\
$~~~$ \Mistag\ NuP11                       & $-$0.026 & $-$0.055 & $-0.055 \pm 0.000$ \\
$~~~$ \Mistag\ NuP12                       & $+$0.007 & $-$0.095 & $+0.051 \pm 0.000$ \\\hline
Total                                      &  \multicolumn{3}{c|}{$ 0.38  \pm 0.00 $} \\\hline
\end{tabular}
\end{center}
\end{table*}
%

\clearpage
\section{Additional information about the various combinations}
\label{sect:addcom}
 This appendix gives additional information about the various combinations
 discussed in the main text.
 For all combinations the values quoted are the combined value, the statistical
 uncertainty, the systematic uncertainty, the total uncertainty and the
 uncertainty in the total uncertainty, which is statistical.
%
\begin{figure*}[tbp!]
\centering
\subfigure[$\sqrts=7$~\TeV]{
  \includegraphics[width=0.45\textwidth]{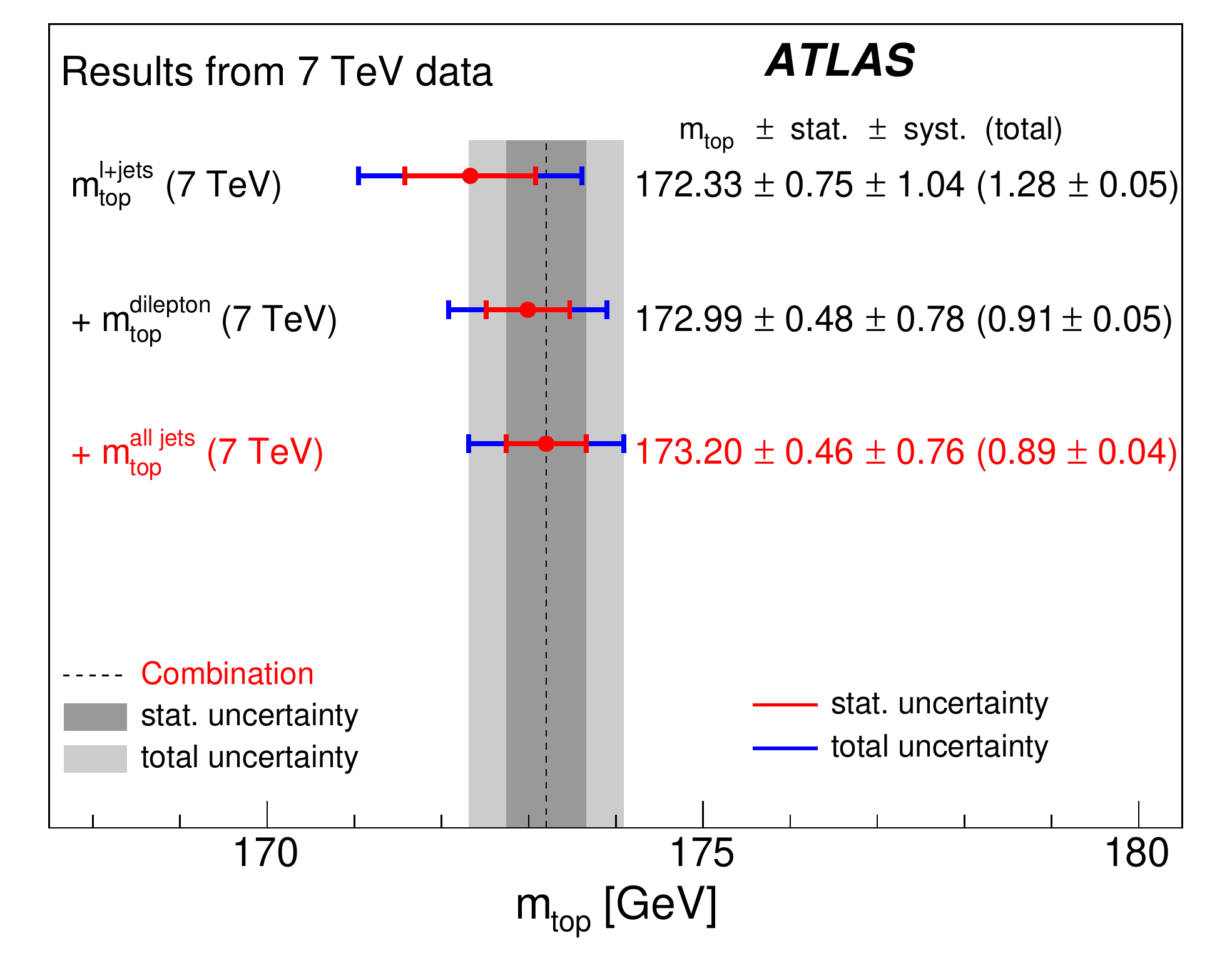}
  \label{fig:TopMassAccImpSev}}
\subfigure[$\sqrts=8$~\TeV]{
  \includegraphics[width=0.45\textwidth]{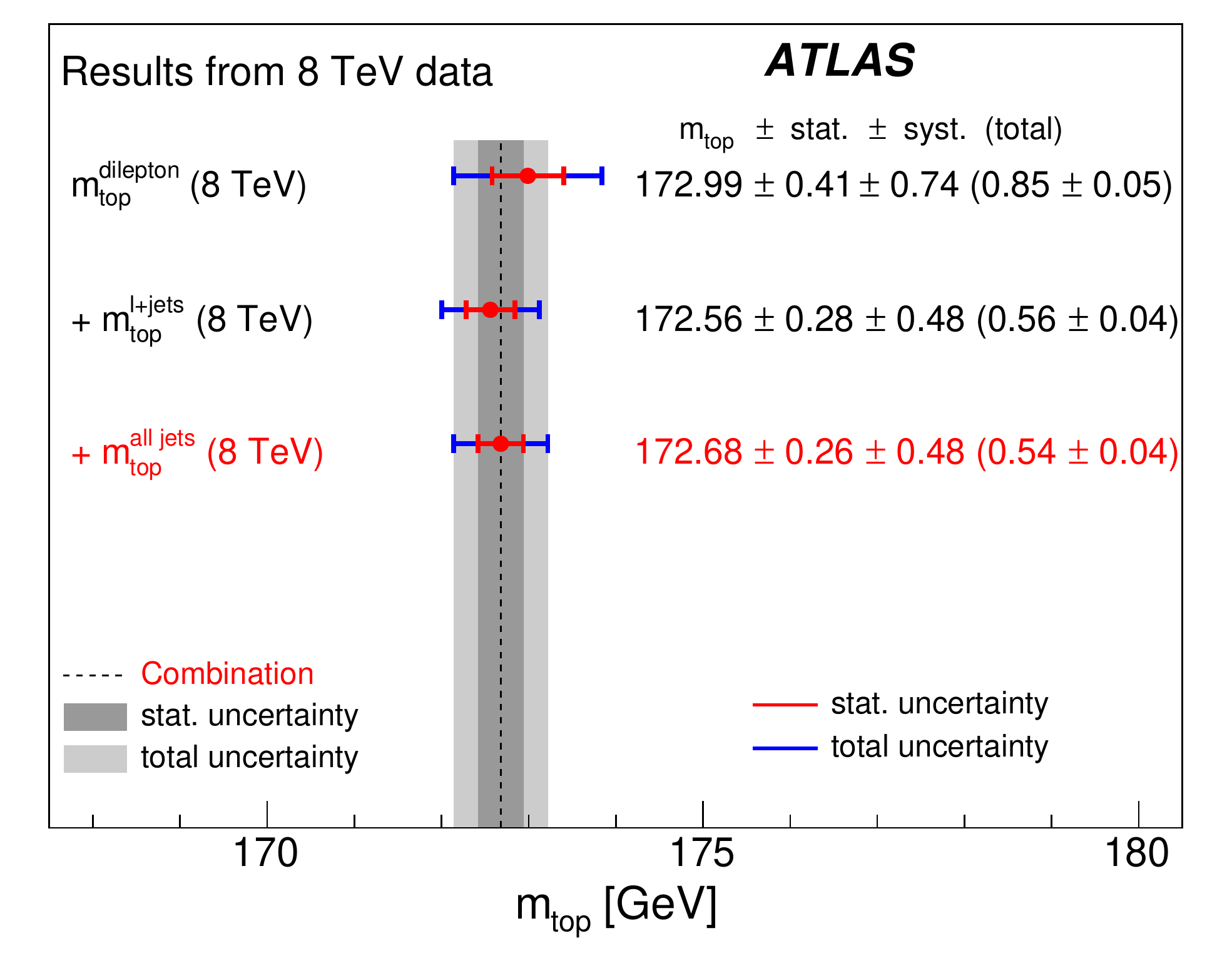}
  \label{fig:TopMassAccImpEig}}
\hfill
\subfigure[\ttbarll\ channel]{
  \includegraphics[width=0.45\textwidth]{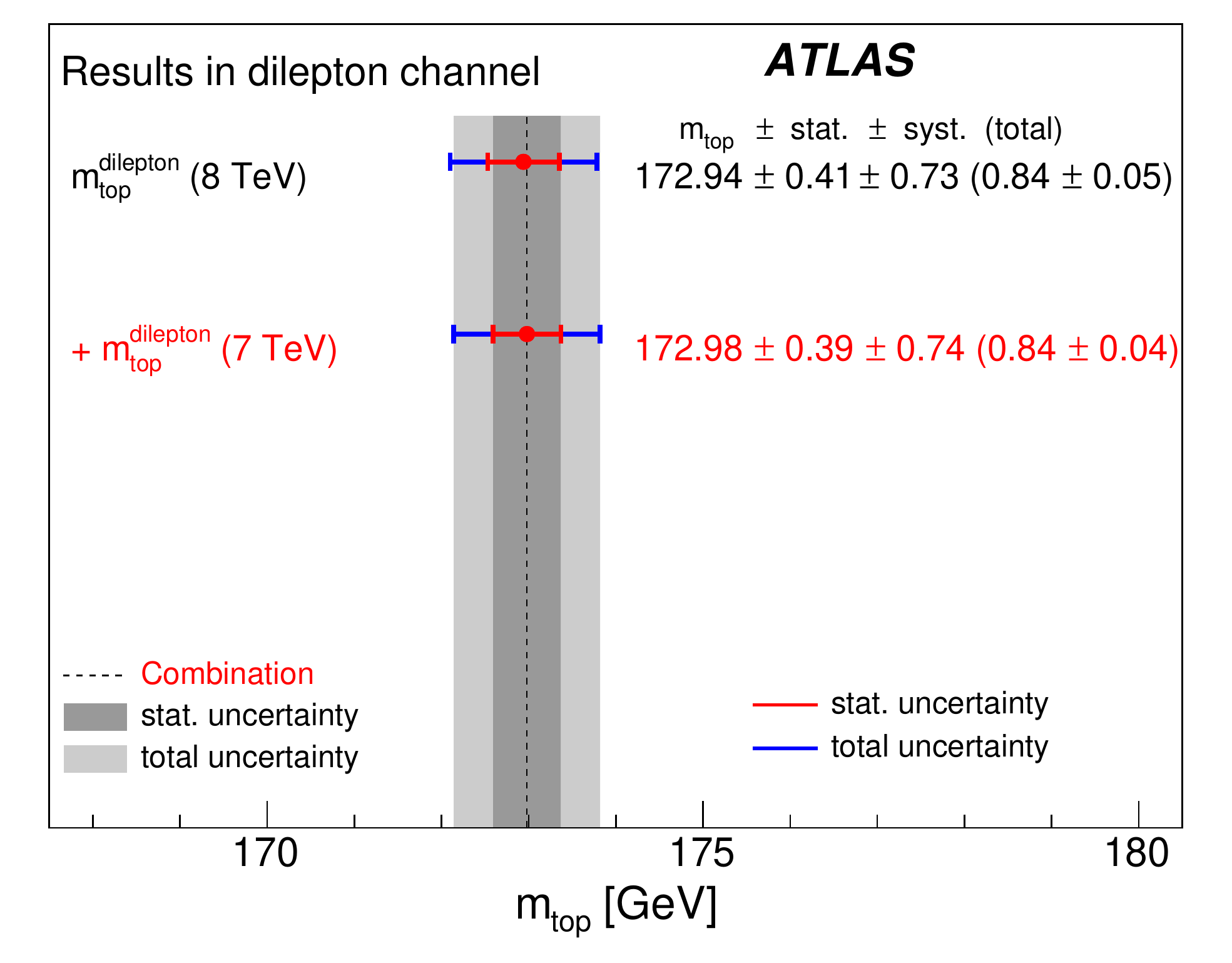}
  \label{fig:TopMassAccImpDiL}}
\subfigure[\ttbarlj\ channel]{
  \includegraphics[width=0.45\textwidth]{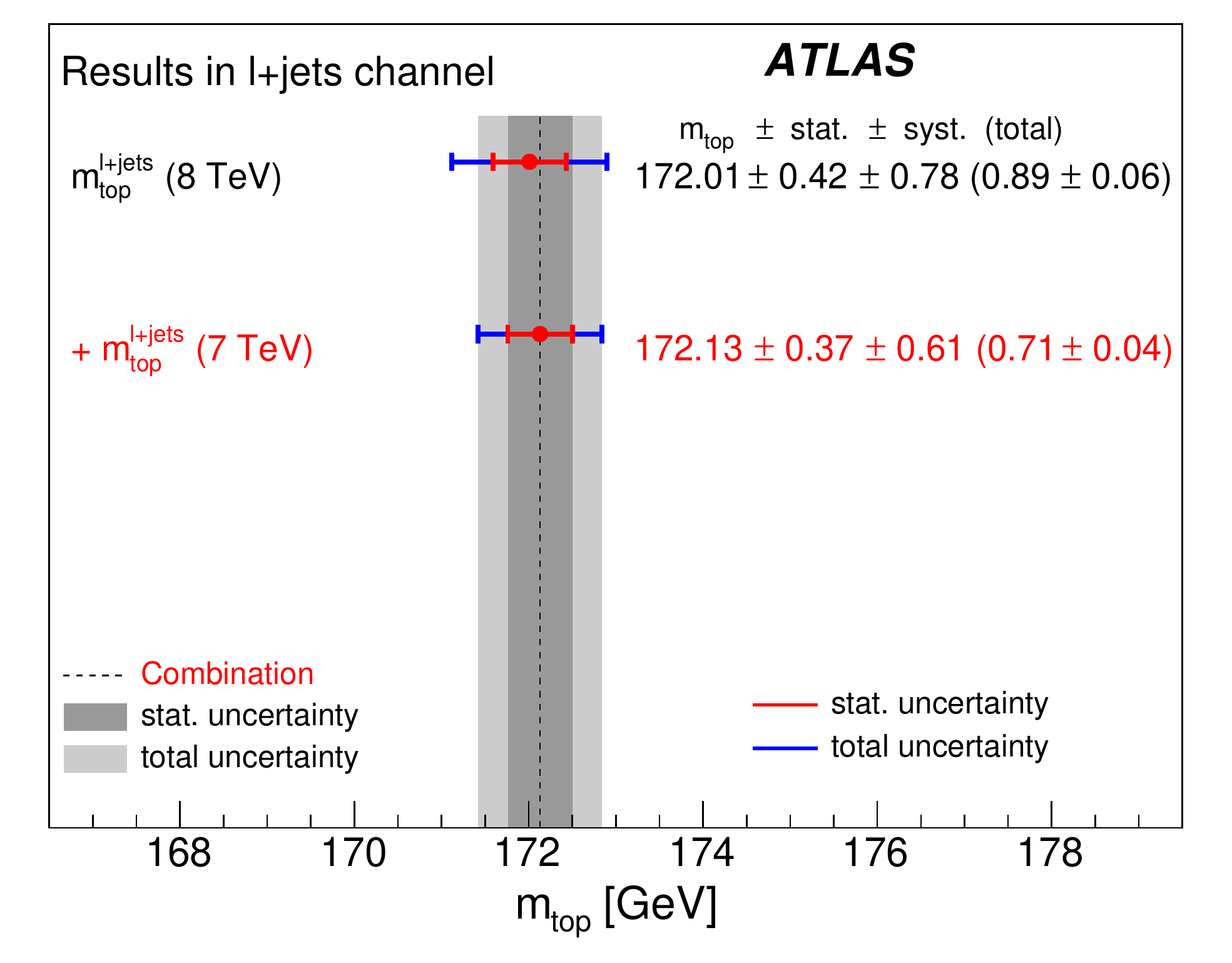}
  \label{fig:TopMassAccImpLpJ}}
\hfill
\subfigure[\ttbarjj\ channel]{
  \includegraphics[width=0.45\textwidth]{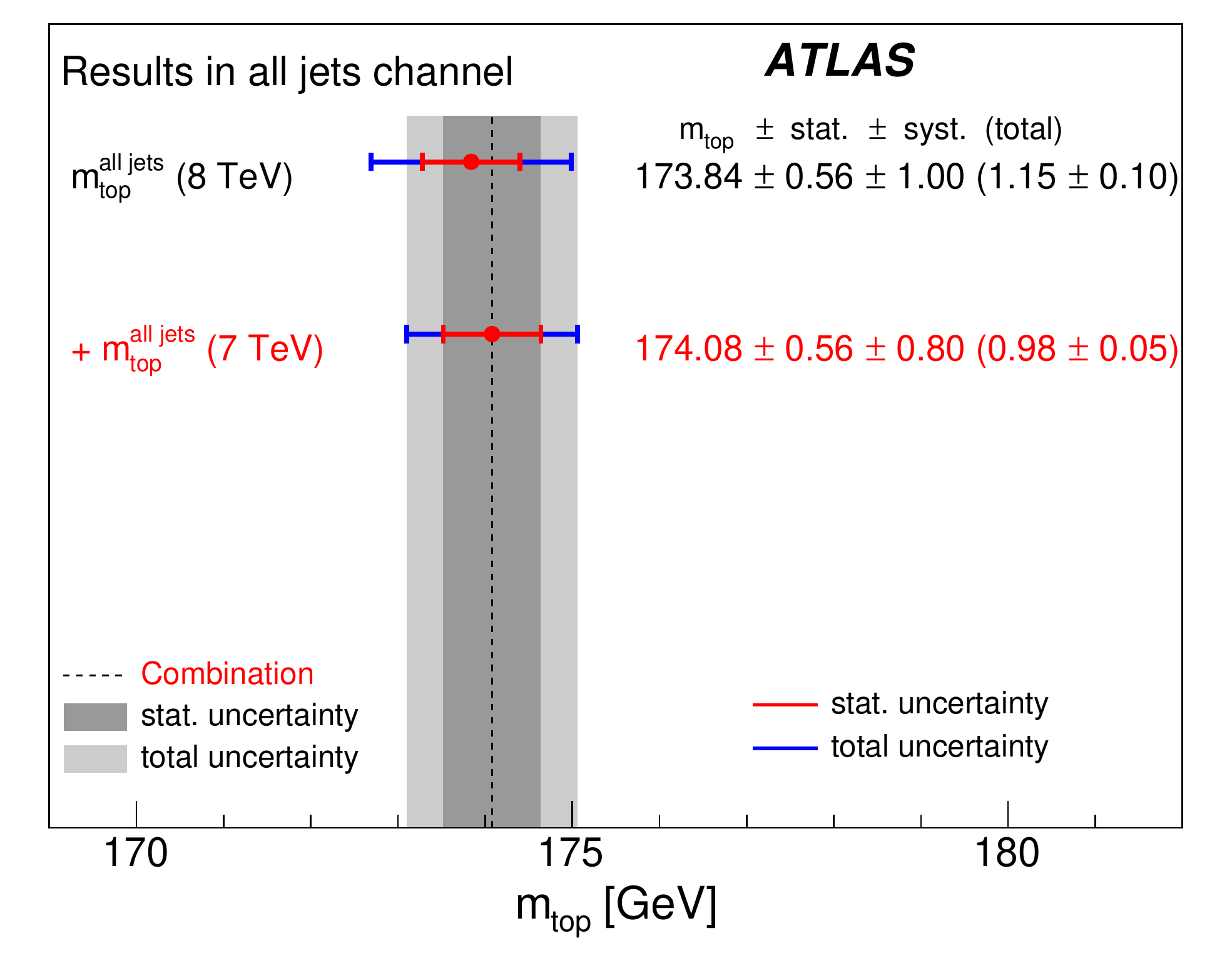}
  \label{fig:TopMassAccImpHad}}
\caption{Selected combinations of the six ATLAS measurements of \mt\ according
  to importance.
 The figures show the combined result when successively adding results to the
 most precise one of a given category.
 Each line of this figure shows the combined result when adding the result
 listed to the combination, indicated by a `+'.
 The values quoted are the combined value, the statistical uncertainty, the
 systematic uncertainty, the total uncertainty and the uncertainty in the total
 uncertainty, which is statistical.
 Figures~(a) and (b) refer to the independent combinations of the three
 measurements per centre-of-mass energy resulting in uncorrelated results
 \mtseven\ and \mteight.
 Figures~(c)--(e) refer to the combination of the three correlated observables
 from pairs of measurements per \ttbar\ decay channel, resulting in \mtdl,
 \mtlj\ and \mtjj.
 \label{fig:TopMassAccImp}
}
\end{figure*}

 \Figs{\ref{fig:TopMassAccImpSev}}{\ref{fig:TopMassAccImpEig}} show the
 independent combinations per \cme.
 For both \cmes, the combination is dominated by the results in the
 \ttbarll\ and \ttbarlj\ channels.
 Using only those in the combinations yields combined results of
 $\mtseven=\XZ{\SevendillpjVal}{\SevendillpjSta}{\SevendillpjSys}$~\GeV\ with an
 uncertainty of $\SevendillpjUnc\pm\SevendillpjUncSta$~\GeV\
 at $\sqrts=7$~\TeV\ and
 $\mteight=\XZ{\EightdillpjVal}{\EightdillpjSta}{\EightdillpjSys}$~\GeV\ with an
 uncertainty of $\EightdillpjUnc\pm\EightdillpjUncSta$~\GeV\
 at $\sqrts=8$~\TeV.
 At both \cmes, the difference between the combined uncertainties of the partial
 and full combination is much smaller than the respective statistical precision
 in the total systematic uncertainties.
 This statistical precision is obtained from varying each systematic uncertainty
 within its statistical precision and repeating the combination, as explained in
 the main text.

 \Figrange{\ref{fig:TopMassAccImpDiL}}{\ref{fig:TopMassAccImpHad}} show the
 dependent combinations per \ttbar\ decay channel.
 The combined result of \mtdl based only on the \ttbarll\ measurement from
 $\sqrts=8$~\TeV\ data and the measurements in the other decay channels is
 $\mtdl=\XZ{\DilSigVal}{\DilSigSta}{\DilSigSys}$~\GeV\ with an uncertainty of
 $\DilSigUnc\pm\DilSigUncSta~\GeV$.
 As a consequence of the influence of the measurements in the other decay
 channels discussed in the main text, this result does not coincide with the
 \ttbarll\ result at $\sqrts=8$~\TeV.
%
\begin{figure*}[tbp!]
\centering
\subfigure[\ttbarll\ channel]{
  \includegraphics[width=0.49\textwidth]{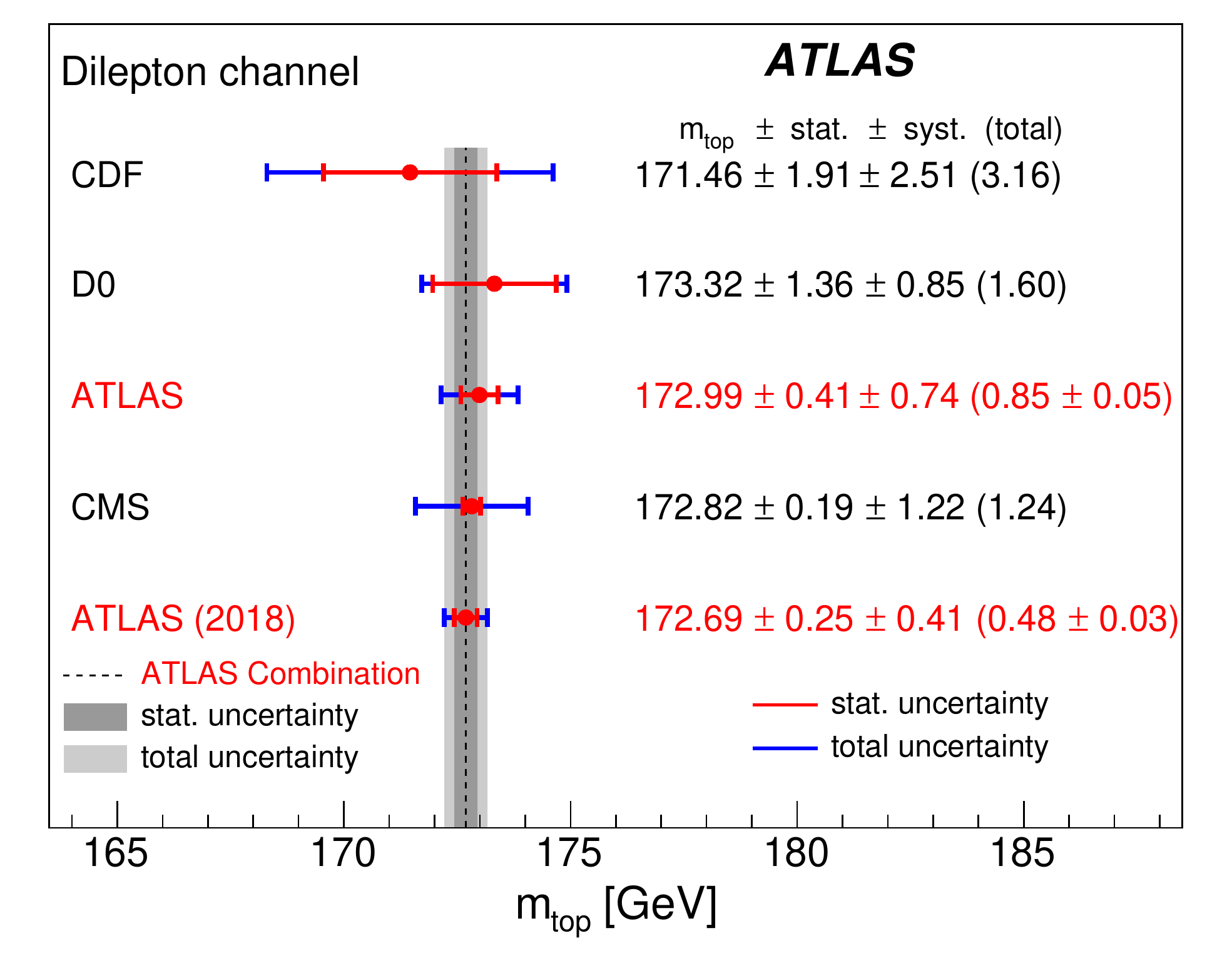}\label{fig:TopMassDiL}}
\subfigure[\ttbarlj\ channel]{
  \includegraphics[width=0.49\textwidth]{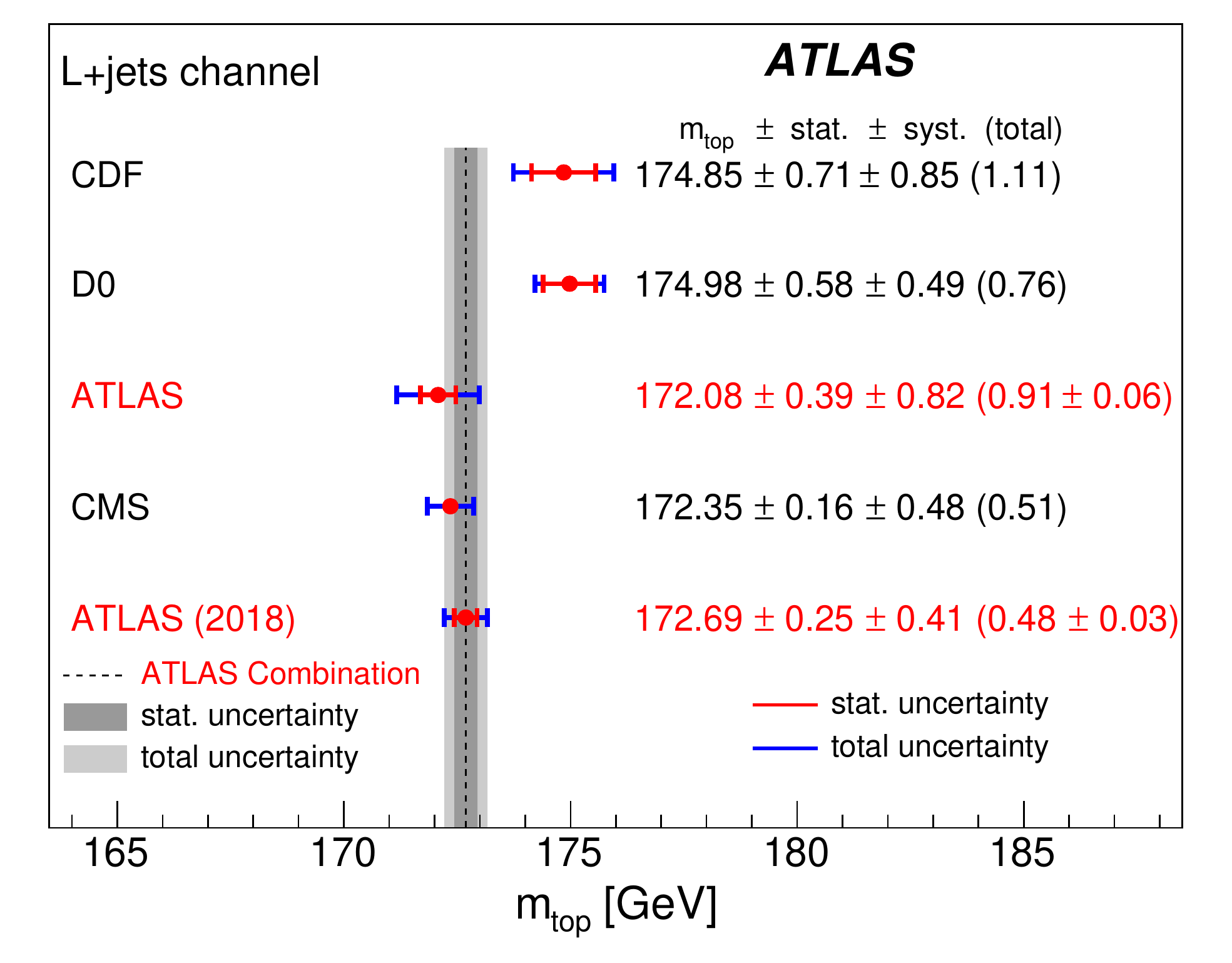}\label{fig:TopMassLpJ}}
\hfill
\subfigure[\ttbarjj\ channel]{
  \includegraphics[width=0.49\textwidth]{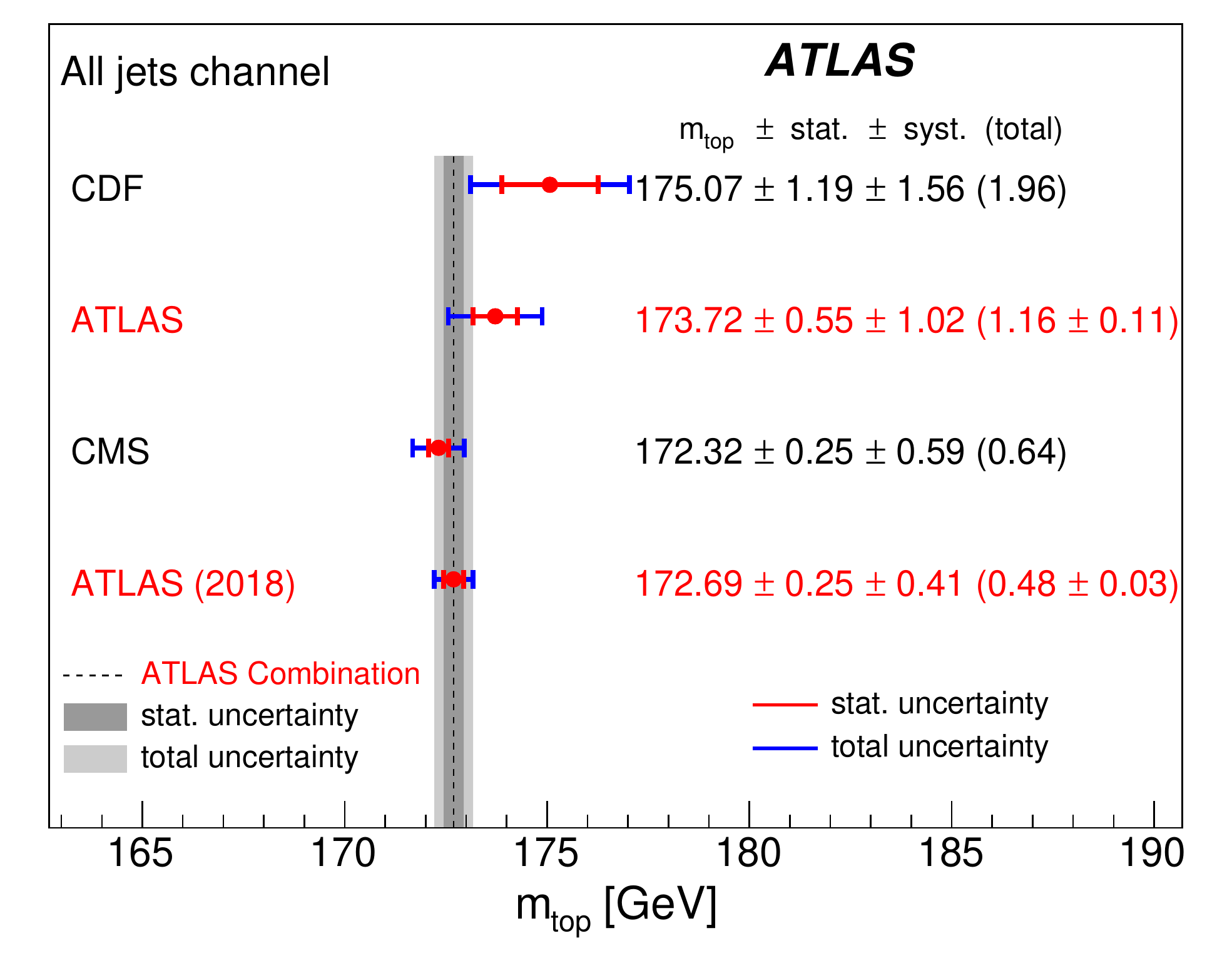}\label{fig:TopMassHad}}
\subfigure[Combinations]{
  \includegraphics[width=0.49\textwidth]{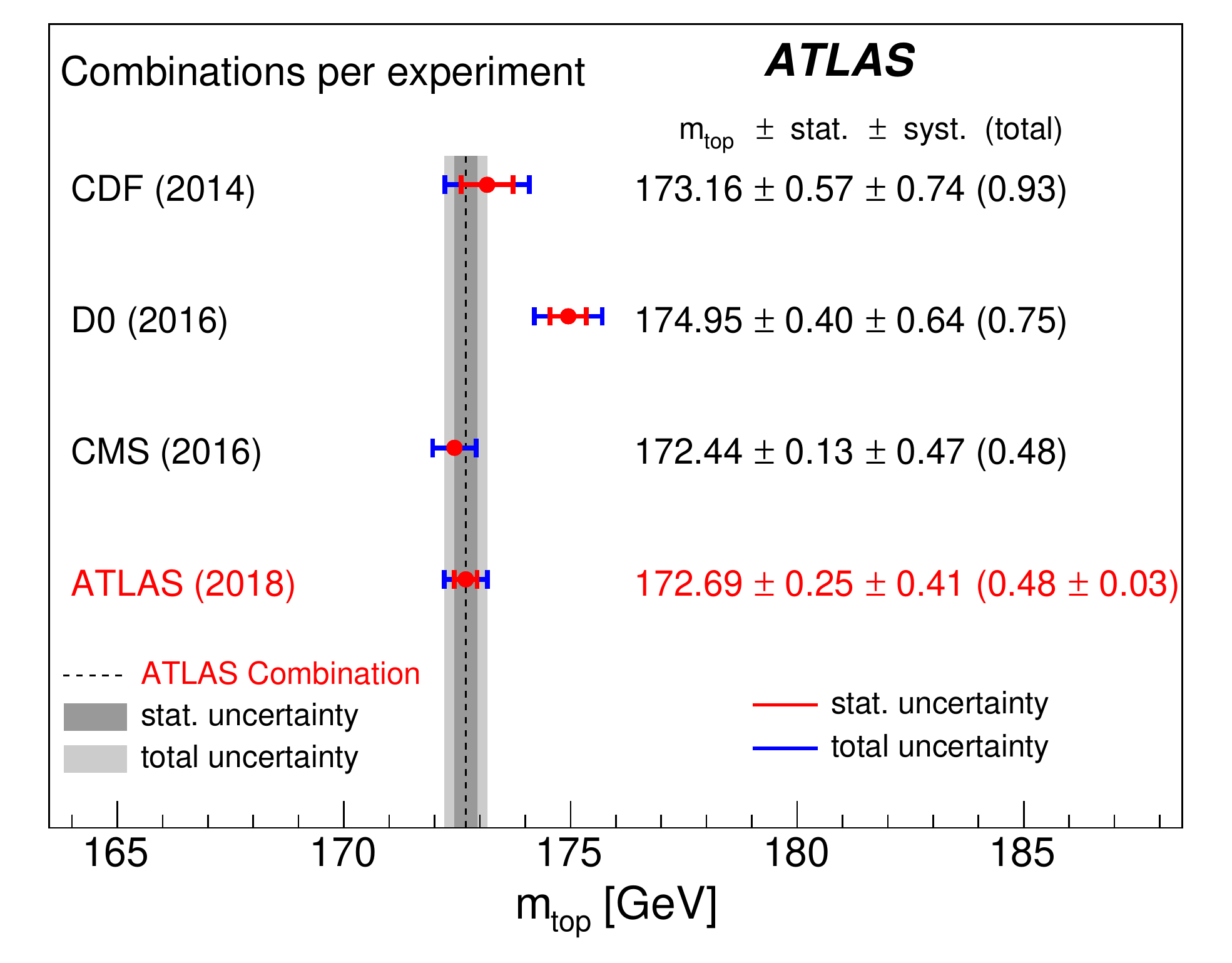}\label{fig:TopMassCom}}
\caption{The most precise result of \mt\ per experiment in the different
  \ttbar\ decay channels and \mt\ from the latest combinations performed by the
  individual experiments.
 Figure~(a) shows the measurements in the \ttbarll\ channel,
 figure~(b) shows those in the \ttbarlj\ and
 figure~(c) shows the ones in the \ttbarjj\ channel.
 Figure~(d) shows the results of \mt\ from the latest combination per
 experiment.
 In all figures the vertical band corresponds to the new ATLAS combined value of
 \mt.
 The values quoted are the combined value, the statistical uncertainty, the
 systematic uncertainty, the total uncertainty and for ATLAS results also the
 uncertainty in the total uncertainty, which is statistical.
 For CDF, the separation into statistical and systematic uncertainties is
 different for the result in the \ttbarlj\ channel and the combination.
 For the former, the statistical component caused by the in situ determination
 of the jet scale factor is included in the statistical uncertainty, while for
 the latter, this uncertainty is part of the systematic uncertainty.
\label{fig:TopMassComp}
}
\end{figure*}

 In \Fig{\ref{fig:TopMassComp}}, the most precise ATLAS results of \mt\ per
 decay channel and the new ATLAS combined value of \mt\ are compared with the
 respective results from the CDF, D0 and CMS experiments.

\FloatBarrier
\printbibliography
\clearpage 
 
\begin{flushleft}
{\Large The ATLAS Collaboration}

\bigskip

M.~Aaboud$^\textrm{\scriptsize 34d}$,    
G.~Aad$^\textrm{\scriptsize 99}$,    
B.~Abbott$^\textrm{\scriptsize 124}$,    
O.~Abdinov$^\textrm{\scriptsize 13,*}$,    
B.~Abeloos$^\textrm{\scriptsize 128}$,    
D.K.~Abhayasinghe$^\textrm{\scriptsize 91}$,    
S.H.~Abidi$^\textrm{\scriptsize 164}$,    
O.S.~AbouZeid$^\textrm{\scriptsize 39}$,    
N.L.~Abraham$^\textrm{\scriptsize 153}$,    
H.~Abramowicz$^\textrm{\scriptsize 158}$
% ...
\bigskip
\\

% ...
$^{13}$Institute of Physics, Azerbaijan Academy of Sciences, Baku; Azerbaijan.\\
% ...
$^{34}$$^{(a)}$Facult\'e des Sciences Ain Chock, R\'eseau Universitaire de Physique des Hautes Energies - Universit\'e Hassan II, Casablanca;$^{(b)}$Centre National de l'Energie des Sciences Techniques Nucleaires (CNESTEN), Rabat;$^{(c)}$Facult\'e des Sciences Semlalia, Universit\'e Cadi Ayyad, LPHEA-Marrakech;$^{(d)}$Facult\'e des Sciences, Universit\'e Mohamed Premier and LPTPM, Oujda;$^{(e)}$Facult\'e des sciences, Universit\'e Mohammed V, Rabat; Morocco.\\
% ...
$^{39}$Niels Bohr Institute, University of Copenhagen, Copenhagen; Denmark.\\
% ...
$^{91}$Department of Physics, Royal Holloway University of London, Egham; United Kingdom.\\
% ...
$^{99}$CPPM, Aix-Marseille Universit\'e, CNRS/IN2P3, Marseille; France.\\
% ...
$^{124}$Homer L. Dodge Department of Physics and Astronomy, University of Oklahoma, Norman OK; United States of America.\\
% ...
$^{128}$LAL, Universit\'e Paris-Sud, CNRS/IN2P3, Universit\'e Paris-Saclay, Orsay; France.\\
% ...
$^{153}$Department of Physics and Astronomy, University of Sussex, Brighton; United Kingdom.\\
% ...
$^{158}$Raymond and Beverly Sackler School of Physics and Astronomy, Tel Aviv University, Tel Aviv; Israel.\\
% ...
$^{164}$Department of Physics, University of Toronto, Toronto ON; Canada.\\
% ...
$^{a}$ Also at Borough of Manhattan Community College, City University of New York, NY; United States of America.\\
$^{b}$ Also at California State University, East Bay; United States of America.\\
$^{c}$ Also at Centre for High Performance Computing, CSIR Campus, Rosebank, Cape Town; South Africa.\\
$^{d}$ Also at CERN, Geneva; Switzerland.\\
$^{e}$ Also at CPPM, Aix-Marseille Universit\'e, CNRS/IN2P3, Marseille; France.\\
% ...
$^{*}$ Deceased

\end{flushleft}

 
%
%
\end{document}